\documentclass[mreferee]{gji}

\usepackage{graphicx}
\usepackage{epstopdf}
\title[Multifractal Omori Law]{Multifractal Omori Law for Earthquake Triggering: New Tests on 
the California, Japan and Worldwide Catalogs}

\begin{document}

\author[G.Ouillon, E. Ribeiro, D. Sornette]
       {G. Ouillon$^1$ \thanks{e-mail : lithophyse@free.fr}, 
        E. Ribeiro$^2$ \thanks{e-mail : shinigami@tele2.fr} 
        and D. Sornette$^{3,4}$ \thanks{e-mail : dsornette@ethz.ch}\\
       $^1$ Lithophyse, 1 rue de la croix, 06300 Nice, France,\\
       $^2$  Laboratoire de Physique de la Mati\`{e}re Condens\'{e}e, CNRS UMR 6622,\\
             Universit\'{e} de Nice-Sophia Antipolis, Parc Valrose, 06108 Nice, France,\\
       $^3$ D-MTEC, ETH Zurich, Kreuzplatz 5, CH-8032 Zurich, Switzerland,\\
       $^4$ Department of Earth and Space Sciences and
            Institute of Geophysics and Planetary Physics,\\ 
            University of California, Los Angeles, California 90095-1567}

\maketitle

\begin{summary}
The Multifractal Stress-Activated (MSA) model is a statistical model of
triggered seismicity based on mechanical and thermodynamic
principles. It predicts that, above a triggering magnitude cut-off $M_0$,
the exponent $p$ of the Omori law for
the seismic decay of aftershocks is a linear increasing function $p(M)
=a M+b$ of the main shock magnitude $M$.
We previously reported empirical support for this prediction, using
the Southern California SCEC catalog. Here, we confirm this law
using an updated, longer version of the same catalog, as well as new methods
to estimate $p$. One of this methods is the newly defined Scaling Function
Analysis, adapted from the wavelet transform. This method is able to measure
a singularity ($p$-value), erasing the possible regular part of a time series.
The Scaling Function Analysis also proves particularly efficient to
reveal the coexistence of several
types of relaxation laws (typical Omori sequences and short-lived swarms sequences)
which can be mixed within the same catalog.
The same methods are used on data from the worlwide Harvard CMT and show results
compatible with those of Southern California. For the Japanese JMA catalog, 
we still observe a linear dependence of $p$ on $M$, yet with a smaller slope.
The scaling function analysis shows however that results for this catalog
may be biased by numerous swarm sequences, despite our efforts to remove them 
before the analysis.
\end{summary}


\setcounter{secnumdepth}{4} 

\newcommand{\be}{\begin{equation}}
\newcommand{\ee}{\end{equation}}
\newcommand{\ba}{\begin{eqnarray}}
\newcommand{\ea}{\end{eqnarray}}
\newenvironment{technical}{\begin{quotation}\small}{\end{quotation}}


\section{Introduction}

The popular concept of triggered seismicity reflects the growing 
consensus that earthquakes interact through a variety of fields 
(elastic strain, ductile and plastic strains, fluid flow, dynamical shaking
and so on). The concept of triggered seismicity was first introduced from
mechanical considerations, by looking at the correlations between the
spatial stress change induced by a given event
(generally referred to as a main shock), and the spatial location of the
subsequent seismicity that appeared to be temporally correlated with
the main event (the so-called aftershocks)
(King et al. 1994; Stein 2003). Complementarily,
purely statistical models have been introduced to take account of
the fact that the main event is not the sole event to trigger some
others, but that aftershocks may also trigger their own aftershocks and
so on. Those models, of which the ETAS (Epidemic Type of Aftershock Sequences) model
(Kagan and Knopoff 1981; Ogata 1988) is a standard
representative with good explanatory power (Saichev and Sornette 2006), unfold the
cascading structure of earthquake sequences. This class of models show that real-looking
seismic catalogs can be generated by using a parsimonious set of parameters
specifying the Gutenberg-Richter distribution of magnitudes, the
Omori-Utsu law for aftershocks and the productivity law of the average
number of triggered events as a function of the magnitude of the triggering earthquake.

Very few efforts have been devoted to bridge these two approaches, so that
a statistical mechanics of seismicity based on physical principles could be built
(see (Sornette 1991; Miltenberger et al. 1993; Sornette et al. 1994) early attempts).
Dieterich (1994) has considered both the spatial
complexity of stress increments due to a main event and one possible physical
mechanism that may be the cause of the time-delay in the aftershock triggering,
namely state-and-rate friction. Dieterich's model predicts that
aftershocks sequences decay with time as $t^{-p}$ with $p \simeq 1$
independently of the main shock magnitude, a value which is often
observed but only for sequences with a sufficiently large number of
aftershocks triggered by large earthquakes, typically for main events of
magnitude 6 or larger. Dieterich's model has in particular the drawback of neglecting
the stress changes due to the triggered events themselves and cannot
be considered as a consistent theory of triggered seismicity.  

Recently, two of us (Ouillon and Sornette 2005; Sornette and Ouillon 2005)
have proposed a simple physical model of self-consistent earthquake triggering, the
Multifractal Stress-Activated (MSA) model, which takes into account
the whole deformation history due to seismicity. This model assumes that
rupture at any scale is a thermally activated process in which stress
modifies the energy barriers.
This formulation is compatible with all known models of
earthquake nucleation (see Ouillon and Sornette 2005 for a review), 
and in particular contains the state-and-rate
friction mechanism as a particular case. At any given place in the domain, the seismicity rate
$\lambda$ is given by $\lambda(t) = \lambda_0 \exp(\sigma(t)/\sigma_T)$, where $\sigma(t)$
is the local stress at time $t$ and $\sigma_T=kT/V$ is an activation stress
defined in terms of the activation volume $V$ and an effective temperature $T$
($k$ is the Boltzmann constant). 
Among others, Ciliberto et al. (2001) and Saichev and Sornette (2005)
have shown that the presence of frozen heterogeneities, always present in rocks and
in the crust, has the effect of renormalizing and amplifying the temperature of the rupture activation
processes  through
the cascade of micro-damage to the macro-rupture, while conserving the same Arrhenius
structure of the activation process. The prefactor
$\lambda_0$ depends on the loading rate and the local strength. The domain is considered as 
elasto-visco-plastic with a large Maxwell time $\tau_M$.
For $t < \tau_M$, the model assumes that the local stress relaxes according to 
$h(t) =  h_0/(t+c)^{1+\theta}$, where $c$ is is a small regularizing time scale.
The local stress $\sigma(t)$ depends
on the loading rate at the boundaries of the domain and on the stress fluctuations induced by all previous
events that occurred within that domain. At any place, any component $s$ of the stress fluctuations 
due to previous events is
considered to follow a power-law distribution $P(s) ds = C/(s^2+s_0^2)^{(1+\mu)/2} ds$.
For $\mu(1+\theta) \simeq 1$, Ouillon and Sornette (2005) found that 
(i) a magnitude $M$ event will be followed by a sequence of aftershocks
which takes the form of an Omori-Utsu law with exponent $p$, (ii)
this exponent $p$ depends linearly on the magnitude $M$ of the main
event and (iii) there exists a lower magnitude cut-off $M_0$ for main shocks below
which they do not trigger (considering that triggering implies a positive value of $p$). 
In contrast with the phenomenological statistical models such as
the ETAS model, the MSA model is based on firm mechanical and thermodynamical principles.

Ouillon and Sornette (2005) have tested this prediction on the SCEC catalog
over the period from 1932 to 2003. Using a superposed epoch procedure to stack
aftershocks series triggered by events within a given magnitude range,
they found that indeed the $p$-value increases with the magnitude $M$ of
the main event according to $p(M) = a M + b = a (M - M_{0})$, where
$a  = 0.10, b  = 0.37, M_{0}  = -3.7$. Performing the same analysis
on synthetic catalogs generated by the ETAS model for which $p$ is
by construction independent of $M$ did not show an increasing $p(M)$,
suggesting that the results
obtained on the SCEC catalog reveal a genuine multifractality which is
not biased by the method of analysis.

Here, we reassess the parameters $a$ and $b$ for Southern California, using an updated 
and more recent version of the catalog, and extend the analysis to other areas in the world 
(the worlwide Harvard CMT catalog and the Japanese JMA catalog), to put to test again
the theory and to check whether the parameters $a$ and $b$ are universal or 
on the contrary vary systematically from one catalog to the other, perhaps
revealing meaningful physical differences between the seismicity of different
regions. The methodology we use to measure values of $p$ are different from the one
in Ouillon and Sornette (2005),  based on the construction
of binned approximations of stacked time series. Here, we introduce a new method 
specifically designed to take account of the possible contamination of the singular signature
of the Omori law by a regular and non-stationnary background rate contribution that may
originate from several different origins described in section \ref{SFA}.

\section{Methodology of the multifractal analysis}

\subsection{Step 1: selection of aftershocks}

The method used here to construct stacked aftershocks time series is 
slightly different from the one used in (Ouillon and Sornette 2005), 
especially concerning the way we take account of the time dependence of  the
magnitude threshold $M_c(t)$ of completeness of earthquake catalogs.

All earthquakes in the catalog are considered successively as potential main shocks. 
For each event, we examine the seismicity following it over a period of $T=1$ year and
within a distance $R=2L$, where $L$ is the rupture length of
the main shock, which is determined empirically from
the magnitude using Wells and Coppersmith (1994)'s relationship. The same relationship
is used for all catalogs as no such relationship has been developed
specifically for the Japanese JMA catalog. 
Concerning the Harvard CMT catalog, it can be expected that a relationship relating
magnitudes to rupture length would be a weighted mixture
of different relationships holding in different parts of the world, with variations resulting
from local tectonic properties. We will see below
that our new method, the Scaling Function Analysis, is actually 
devised to take account of the uncertainties resulting from the use of approximate
length-magnitude relationships.
If the radius $R$ is smaller than the spatial location accuracy 
$\Delta$ (which is assumed here for simplicity in a first 
approach to be a constant for all events in a given catalog), we set $R=\Delta$. 
If an event has previously been tagged as an aftershock of a larger event,
then it is removed from the list of potential main shocks, as its own
aftershocks series could be contaminated by the influence of the previous, 
larger event. Even if an event has been removed from the list of main shocks,
we look for its potential aftershocks and tag them as well if necessary (yet they
are themselves excluded from the stacked time series).

Aftershock time series are then sorted according to the magnitude of the main event, and
stacked using a superposed epoch procedure within given main shock magnitude ranges. 
We choose main shock magnitude intervals to vary by half-unit magnitude steps, 
such a magnitude step being probably an upper-bound for the magnitude uncertainties.

This methodology to build aftershocks stacked series is straightforward when
the magnitude threshold $M_c(t)$ of completeness
is constant with time, which is the case for the Harvard catalog, for example.
For the SCEC and JMA catalog, we take into account the variation of $M_c(t)$ as follows.
Individual aftershock times series are considered in the stack only if the
magnitude of the main event, occurring at time $t_0$, is larger than
$M_c(t_0)$. If this main event obeys that criterion, only its
aftershocks above $M_c(t_0)$ are considered in the series. This methodology allows us
to use the maximum amount of data with sufficient accuracy to build a single set of
stacked time series of aftershock decay rates. Ouillon and Sornette (2005) used a slightly
different strategy accounting for 
the variation of $M_c$ with time by dividing the SCEC catalog
into subcatalogs covering different time intervals over which the catalog was
considered as complete above a given constant magnitude threshold. This led
Ouillon and Sornette (2005) to analyze four such subcatalogs separately.


\subsection{Step 2: fitting procedure of the stacked time series}

Once aftershocks time series have been selected, stacked, and sorted according to the main shock
magnitude, we fit the binned data with the following law:
\be
N(t) = A \cdot t^{-p} + B~,
\label{fitapb}
\ee
which includes an Omori-like power-law term and a constant background rate $B$.
Here, $N(t)$ is the rate of triggered seismicity at time $t$ after a main shock that occured at $t=0$. 
The time axis is binned in intervals according to a
geometrical series so that the width of the time intervals grows exponentially
with time. We then simply count the number of aftershocks contained
within each bin, then divide this number by the linear size of the
interval to obtain the rate $N$. The fitting parameters $A, B, p$ are then obtained by
a standard grid search.

As the linear density of bins decreases as the inverse of time, each bin receives a weight
proportional to time, balancing the weight of data points along the time axis. 
In our binning, the linear size of two consecutive intervals increases by a factor $r > 1$.
Since the choice of $r$ is arbitrary, it is important to check for the robustness
of the results with respect to $r$. We thus performed fits on time
series binned with $20$ different values of $r$, from $r=1.1$ to $r=3$ by
step of $0.1$. We then checked whether the fitted parameters $A$, $B$ and
$p$ were stable with $r$. We observed that the inverted parameters do not depend much
on $r$, so that we computed the average values and standard
deviations of all fitting parameters over the $20$ $r$ values. For some rare cases,
we obtained $p$-values departing clearly from the average (generally for the largest or smallest values
of $r$) - we thus excluded them to perform a new estimate of $p$. 
In order to provide reliable fits, we excluded
the early times of the stacked series, where aftershock catalogs appear to be incomplete (Kagan 2004). 
Finally, a $p$-value (and its uncertainty) determined within the main shock magnitude interval
$[M_1;M_2]$ was thus associated with the magnitude $\frac{M_1+M_2}{2}$. Our
approach extends that of
Ouillon and Sornette (2005) who performed fits on the same kind of data using
only a single value $r=1.2$.


For each magnitude range, we thus have $20$ different binned time series corresponding to
different values of $r$. For the sake of clarity, we only plot the binned 
aftershocks time series whose $p$-value is the closest
to the average $p$-value obtained over the $20$ different values $r$
for that magnitude range. Its fit using Eq. \ref{fitapb} will be plotted as well. 


\subsection{\label{SFA} Step 3: scaling function analysis}

The method presented above to fit binned data uses a magnitude-dependent 
spatio-temporal window within which aftershocks are selected. Consider a main event $E_1$ whose
linear rupture size is $L$.
The present methodology assumes that any event located within a distance $2L$ of $E_1$
and occurring no more than $1$ year after it is one of its aftershocks. Conversely, any event located at the same distance but which occurred after only just a little more than $1$ year after the main shock
is not considered as its aftershock but as a potential main shock $E_2$, with its own aftershocks sequence which can be used for stacking. Actually, any size for the time window to select
aftershocks is quite arbitrary and will not remove the possibility 
that the  aftershocks sequence of event $E_2$ may still be contaminated by the sequence triggered by event $E_1$, especially if $M(E_1) > M(E_2)$. Since the formula of Wells and Coppersmith [1994] does not strictly apply to each event in a given catalog, 
one can imagine many other scenarios of such a contamination that may also originate in the underestimation of $L$. 
A step towards taking into account this problem is to rewrite expression (\ref{fitapb})
for the time evolution of the sequence triggered by $E_2$ as
\be
N(t) = A \cdot t^{-p} + B(t)~,
\label{omnonstatback}
\ee
where $B(t)$ is a non-stationnary function that describes both the constant background seismicity rate 
and the decay of the sequence(s) triggered by $E_1$ (and possibly other events occuring prior to $E_2$). Here,  $t$ is the time elapsed since the event $E_2$ occurred, as we want to characterize the sequence which follows that event. As the event $E_1$ occurred before the event $E_2$, $B(t)$ is not singular at $t=0$. It is thus a regular contribution to $N(t)$, which we expect do decay rather slowly, 
so that it can be approximated by a polynomial of low degree $n_B$. 
We thus rewrite Eq. \ref{omnonstatback} as
\be
N(t) = A \cdot t^{-p} + \sum_{i=0}^{n_{B}} b_i t^i~,
\label{fitapbi}
\ee
where the sum on the right-hand side now stands for $B(t)$. We have a priori no information on the
precise value of $n_B$. For $n_B=0$, we recover the constant background term $B$ of expression (\ref{fitapb}). 
On the other hand, $n_B$ might be arbitrarily large in which case the coefficients $b_i$'s can be expected to 
decrease sufficiently fast with the order $i$ to ensure convergence, so that 
only the few first terms
of the sum will contribute significantly to $B(t)$. Their number will depend on the fluctuations of the 
seismicity rate at times prior to the event $E_2$.
The effect of this polynomial trend is to slow down the apparent time decay
of the aftershocks sequence triggered by event $E_2$, hence possibly leading to the determination of a spurious small $p$-value. This could be a candidate explanation for Ouillon and Sornette (2005)'s report of small values of $p$'s  for main events with small magnitudes. One could argue that their stacked aftershocks time series might be contaminated by the occurrence of previous, much larger events
(as well as of previous, smaller but numerous events).

In order to address this question, that is, to take account of the possible time-dependence of $B$, two  strategies are possible:
\begin{enumerate}
\item[(i)]  The $20$ different binned time series can be fitted using Eq.~(\ref{fitapbi}) with the unknowns being $A$, $n_B$  and the $b_i's$, 
\item[(ii)] One can use weights in the fitting procedure of the original data 
so that the polynomial trend is removed. One is  then left with a simple
determination of $A$ and $p$ alone. 
\end{enumerate}
We have implemented the second strategy in the form of what we refer to as the
 ``scaling function analysis''. The Appendix
 describes in details this method that we have developed, inspired by the
 pioneering work of Bacri et al. (1993), and presents several tests performed on synthetic  time series to illustrate its performance and the sensitivity of the results to the parameters.

\section{Results}

\subsection{The Southern California catalog}

\subsubsection{Selection of the data}

Ouillon and Sornette (2005) have analyzed
the magnitude-dependence of the $p$-value for aftershocks sequences
in Southern California. However, since we have here developed different methods to
build binned stacked series and to fit those series, it is instructive to reprocess the 
Southern California data in order to 1) test the robustness of Ouillon and Sornette (2005)'s
previous results and 2) provide a benchmark against which to compare the results obtained
with the other catalogs (Japan and Harvard). This also provides a training ground
for the new scaling function analysis method.

The SCEC catalog we use is the same as in (Ouillon and Sornette 2005), except that it now spans a larger time interval ($1932-2006$ inclusive). The magnitude completeness threshold is taken with the same time dependence as in (Ouillon and Sornette 2005): $M_0=3.0$ from $1932$ to $1975$, $M_0=2.5$ from $1975$ to $1992$,
$M_0=2.0$ from $1992$ to $1994$, and $M_0=1.5$ since $1994$. We assume a value $\Delta=5$~km
for  the spatial location accuracy (instead of $10$~km in Ouillon and Sornette (2005)).
This parameterization allows us to decluster 
the whole catalog and build a catalog of aftershocks, as previously explained.

\subsubsection{An anomalous zone revealed by the Scaling Function Analysis}

The obtained binned stacked series are very similar to
those presented by Ouillon and Sornette (2005). However, the scaling function analysis 
reveals deviations from a pure 
power-law scaling of the aftershock sequences, which take different shapes
for different magnitude ranges, as we now describe.

Let us first consider Fig.~\ref{SCEC_raw_4_45}, which shows the binned stacked series obtained for 
main shock magnitudes in the interval $[4;4.5]$. 
The many data points represent the binned series for all values of the binning
factor $r$.
The aftershock decay rate does not appear to be a pure power-law, 
and displays rather large fluctuations. A first scaling regime seems to
hold from $10^{-5}$~year to $4 \cdot 10^{-4}$~year, followed by a second scaling regime
up to $5 \cdot 10^{-3}$~year, then a third scaling regime which progressively
fades into the background rate. Note the similarity of this time series with the synthetic
one shown in Fig.~\ref{synthgammapowback} in the Appendix, which is the sum of three different contributions (a gamma law, 
a power law, and a constant background term).
Fig.~\ref{SCEC_raw_4_45_analysis} shows the scaling function analysis coefficient
(SFAC) of the corresponding set of aftershocks. 
The two solid lines correspond respectively (from top to bottom) to $n_B=0$
and $n_B=3$.

The first important observation is that the shapes of the SFAC 
as a function of scale are independent of $n_B$. This means that the term $B(t)$ in (\ref{omnonstatback})
is certainly quite close to a constant. Secondly, we clearly observe 
that a first power-law scaling regime holds for time scales within
$[5.10^{-5};5.10^{-3}]$ (for $n_B=0$ and similarly with the same exponent for $n_B=3$). The exponent being $\simeq 0.6$, this suggests a 
$p$-value equal to $p=0.4$.
Each curve then goes through a maximum, followed by a decay, and then increases again. 
This behavior is strikingly similar to that shown in Fig.~\ref{synthgammapowbackanalysis} in the Appendix. This suggests that the time series shown in Fig. \ref{SCEC_raw_4_45_analysis} may be
a mixture of several different contributions, such as gamma and power laws.
This simple example shows that the scaling function analysis
provides a clear evidence of a mixture of aftershock sequences with different nature within the same
stacked series -- a fact that has never been considered in previous studies of the same or of other catalogs.

We thus tried to identify in the SCEC catalog those events that may be responsible for the non-Omori behavior revealed by the scaling function analysis. After many trials, we were able to locate a very small
spatial domain in which many short-lived sequences occur. This zone is located within $[-115.6^\circ;-115.45^\circ]$
in longitude and $[32.8^\circ;33.1^\circ]$ in latitude. This zone corresponds to the Imperial Valley area, known to produce a significant amount of earthquake swarms (Scholz 2002). 
In the remaining of the SCEC catalog analysis, we decided to exclude any sequence triggered by a shock  in this small zone. The impact of excluding the Imperial Valley Area is illustrated in
Fig.~\ref{SCEC_raw_4_45_analysis} with the two dashed lines, which can be compared with the
two continuous lines. Excluding the the Imperial Valley Area 
significantly changes the scaling properties, and one can now measure an
exponent $p=0.80$ for time scales larger than $10^{-3}years$.

\subsubsection{Results on the cleaned SCEC catalog}

Following our identification of the anomalous Imperial Valley zone, we removed
all events in the aftershock catalog associated with this zone, 
and launched again our analysis of the binned stacked sequences using direct
fits as well as the Scaling Function Analysis.

\begin{figure}
\begin{center}
\includegraphics[width=8cm]{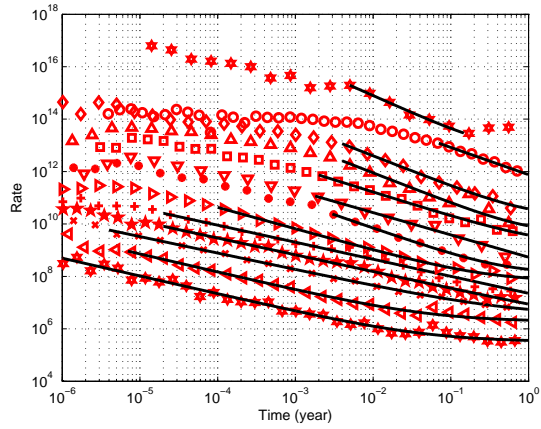}
\caption{\label{SCEC_bined_stacks} Binned stacked series of aftershock sequences in the SCEC catalog (after removing the events in the Imperial Valley
zone) for various magnitude ranges. Magnitude ranges are, from bottom to top: $[1.5;2],  [2;2.5],  [2.5;3], [3;3.5],  [3.5;4], [4;4.5], [4.5;5]$, $ [5;5.5], [5.5;6], [6;6.5], [6.5;7], [7;7.5], [7.5;8]$. The solid lines show the fits to individual  time series with formula (\ref{fitapb}).
All curves have been shifted along the vertical axis for the sake of clarity.}
\end{center}
\end{figure}

\begin{figure}
\begin{center}
\includegraphics[width=8cm]{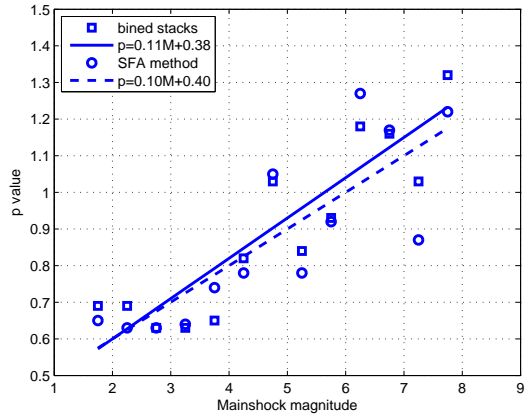}
\caption{\label{p_M_SCEC} P(M) values obtained for the SCEC catalog with fits of binned time series 
(squares - second column
of Table 1) and Scaling Function Analysis (circles - fifth column of Table 1). 
Continuous and dashed lines stand for their respective linear fits.}
\end{center}
\end{figure}

Figure \ref{SCEC_bined_stacks} shows the binned stacked series for the SCEC catalog. 
Each series corresponds to a given magnitude range. For each magnitude range, 
for the sake of clarity, we chose to plot only one  binned time series, corresponding
to a given $r$-value. The $r$-value we choose is the one for which the obtained $p$-value is the closest to  the average $p$-value over all $r$ values for that magnitude range. The solid lines show the fits of the corresponding series with formula (\ref{fitapb}).
For each magnitude range, the average $p$-values and their standard deviations are given in Table 1.
Figure \ref{p_M_SCEC} shows the corresponding dependence $p(M)$ of the $p$-value as a function of the magnitude of the main shocks. The dependence $p(M)$ is well fitted by the linear law $p(M)=0.11M+0.38$. This relationship is very close to the dependence $p(M)=0.10M+0.37$, reported by Ouillon and Sornette (2005).
 Despite the differences in the catalogs and in the general methodology, we conclude
 that the results are very stable and confirm a significant dependence of the exponent
 of the Omori law as a function of the magnitudes of the main shocks.
 
\begin{figure}
\begin{center}
\includegraphics[width=8cm]{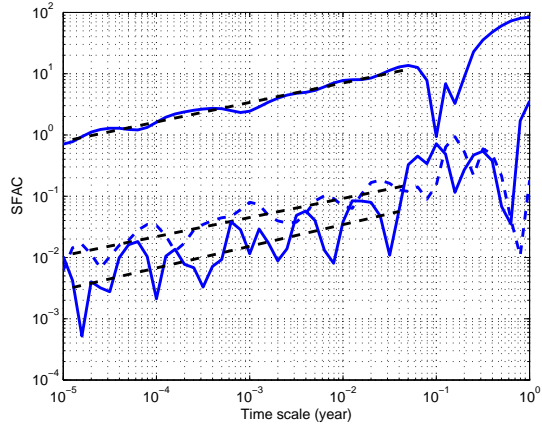}
\caption{\label{SCEC_15_2} SCEC - SFA method: main shock magnitudes M within $[1.5;2]$. Scaling breaks down due to the occurrence of a burst.  The upper solid curve corresponds to $n_B=0$ and $n_D=0$, the dashed curve corresponds to $n_B=3$ and $n_D=0$, while the lower solid curve corresponds to $n_B=0$ and $n_D=10$. }
\end{center}
\end{figure}

\begin{figure}
\begin{center}
\includegraphics[width=8cm]{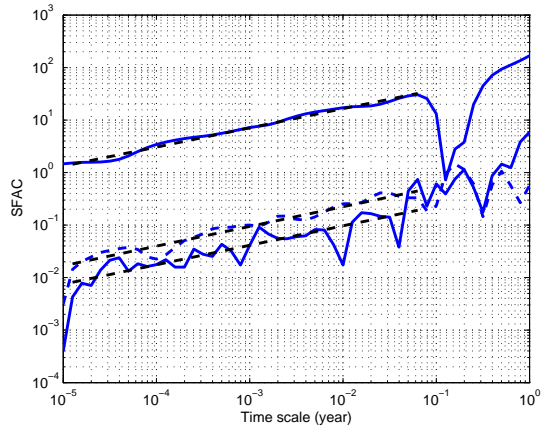}
\caption{\label{SCEC_2_25} Same as Fig.~\ref{SCEC_15_2} for 
M within $[2;2.5]$. Scaling breaks due to the occurrence of a burst. }
\end{center}
\end{figure}

\begin{figure}
\begin{center}
\includegraphics[width=8cm]{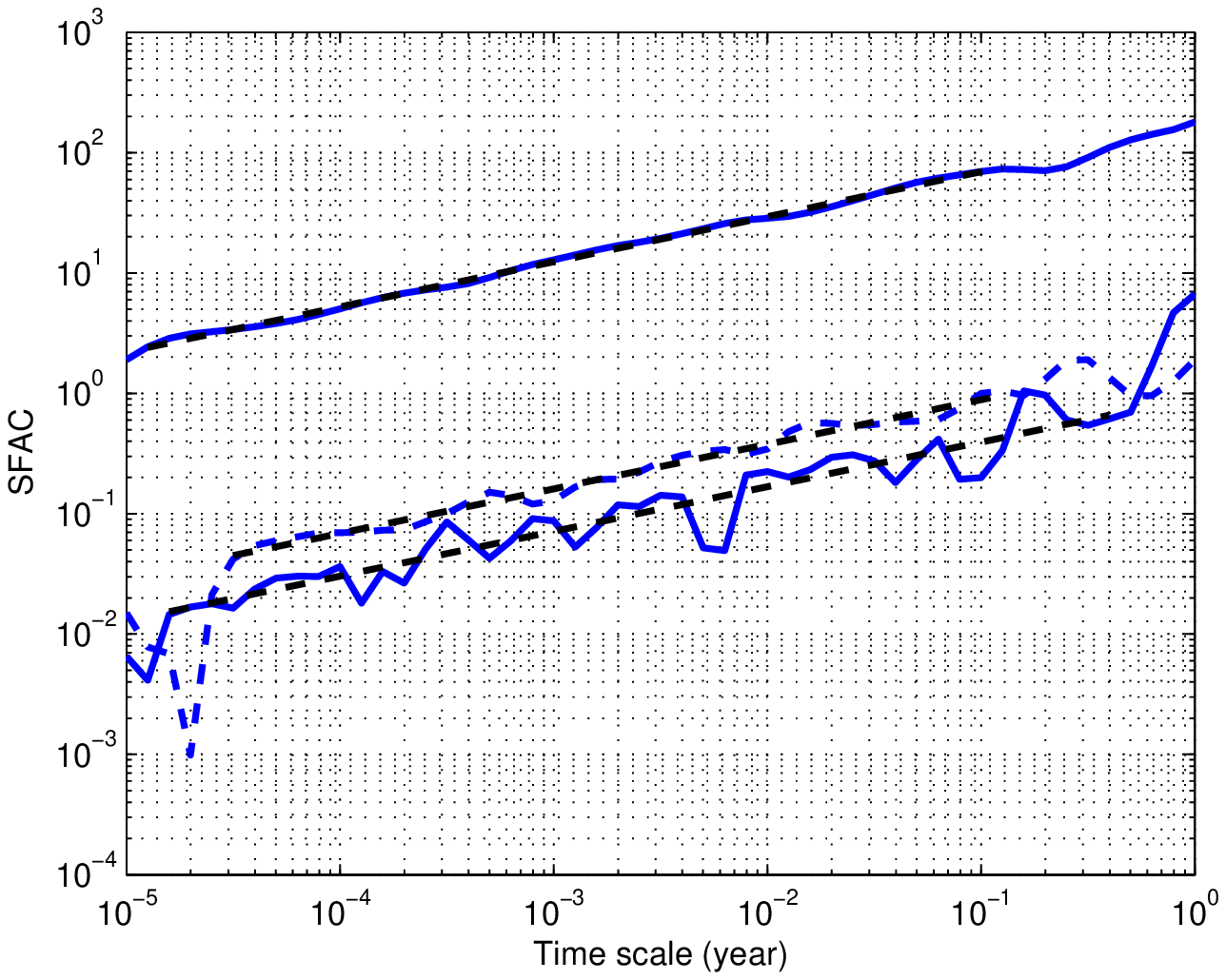}
\caption{\label{SCEC_25_3}Same as Fig.~\ref{SCEC_15_2} for 
 M within $[2.5;3]$.}
\end{center}
\end{figure}

\begin{figure}
\begin{center}
\includegraphics[width=8cm]{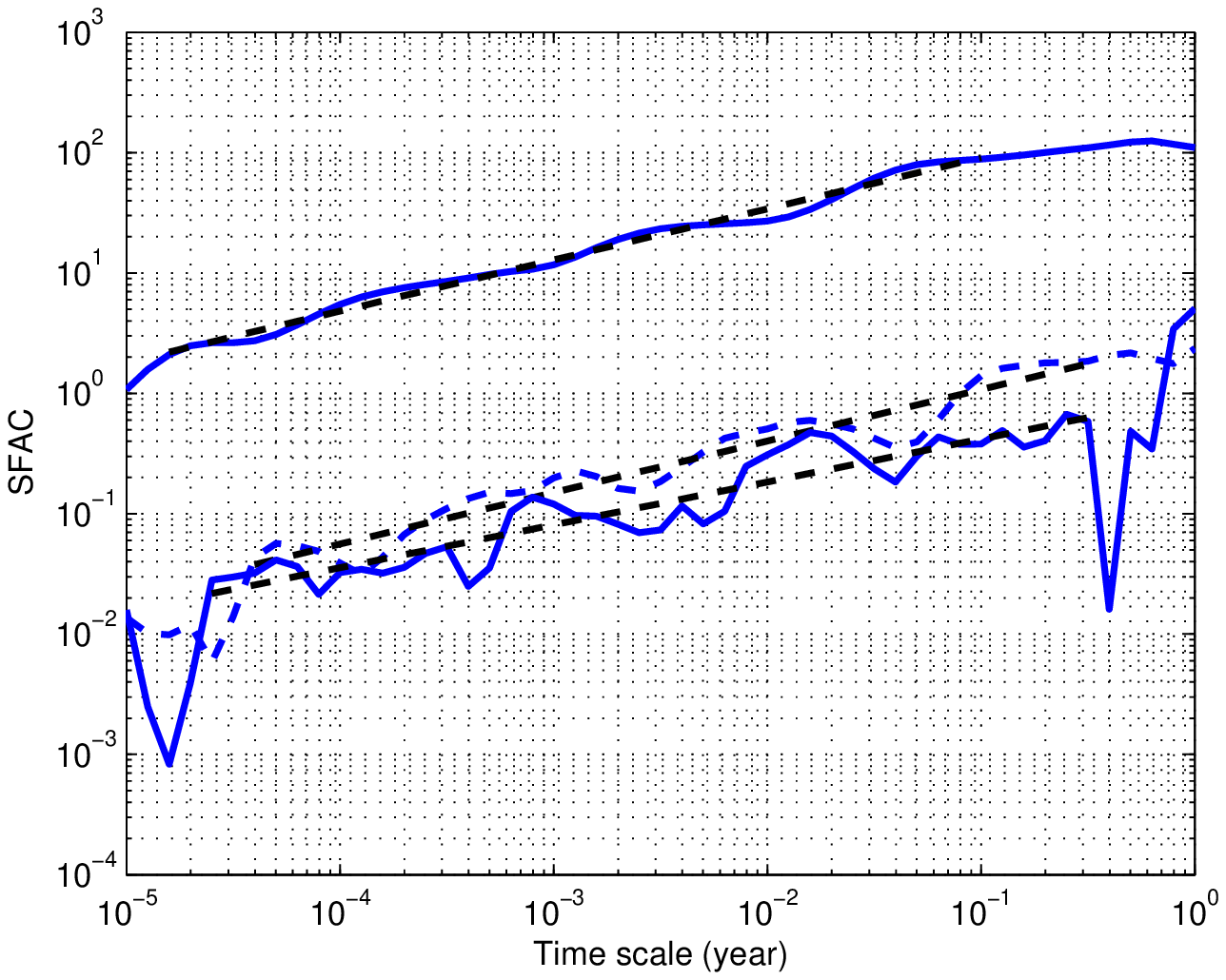}
\caption{\label{SCEC_3_35} Same as Fig.~\ref{SCEC_15_2} for 
M within $[3;3.5]$.}
\end{center}
\end{figure}

\begin{figure}
\begin{center}
\includegraphics[width=8cm]{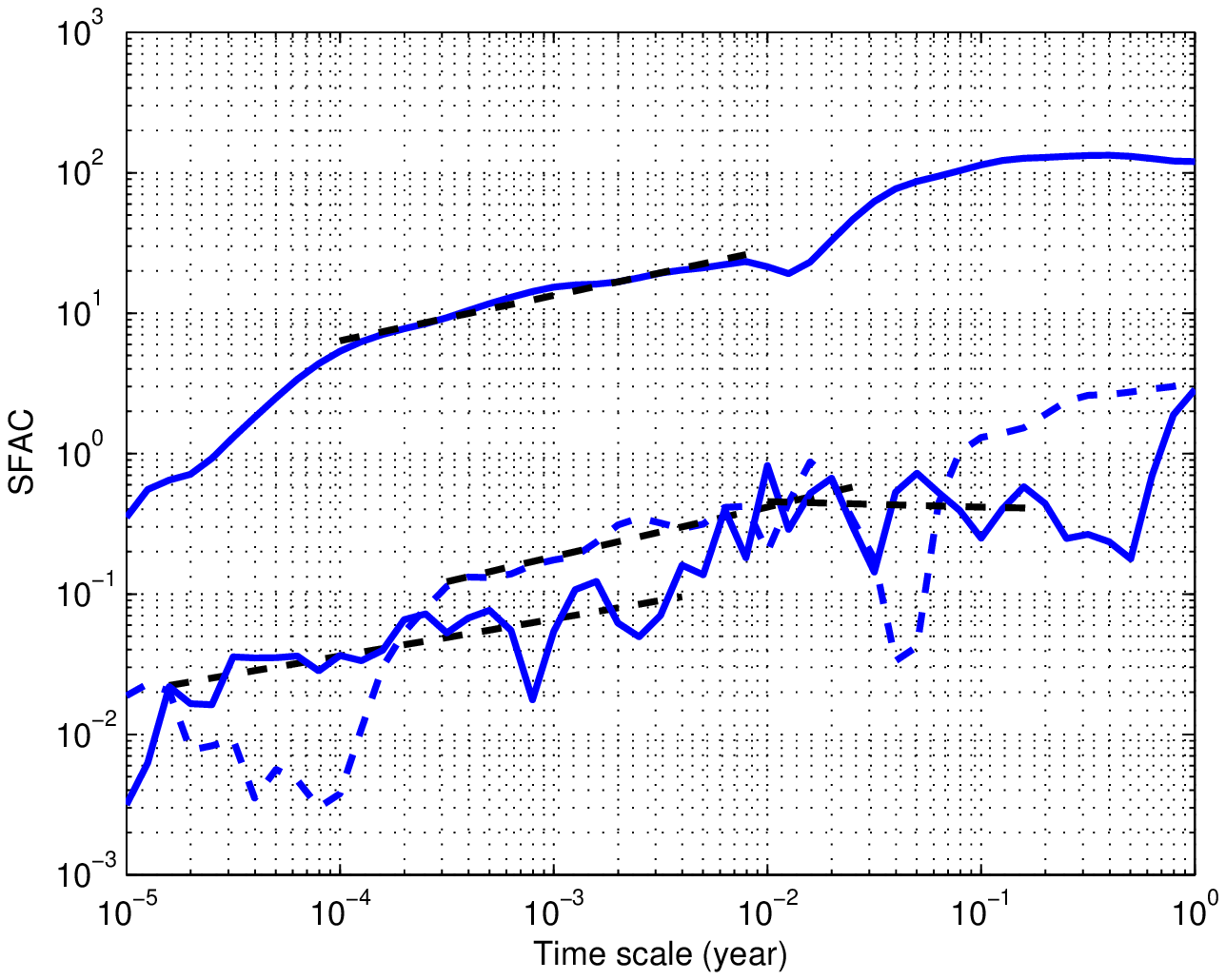}
\caption{\label{SCEC_35_4} Same as Fig.~\ref{SCEC_15_2} for 
 M within $[3.5;4]$. Scaling breaks due to
the occurrence of a burst at about $5 \cdot 10^{-3}$.}
\end{center}
\end{figure}

\begin{figure}
\begin{center}
\includegraphics[width=8cm]{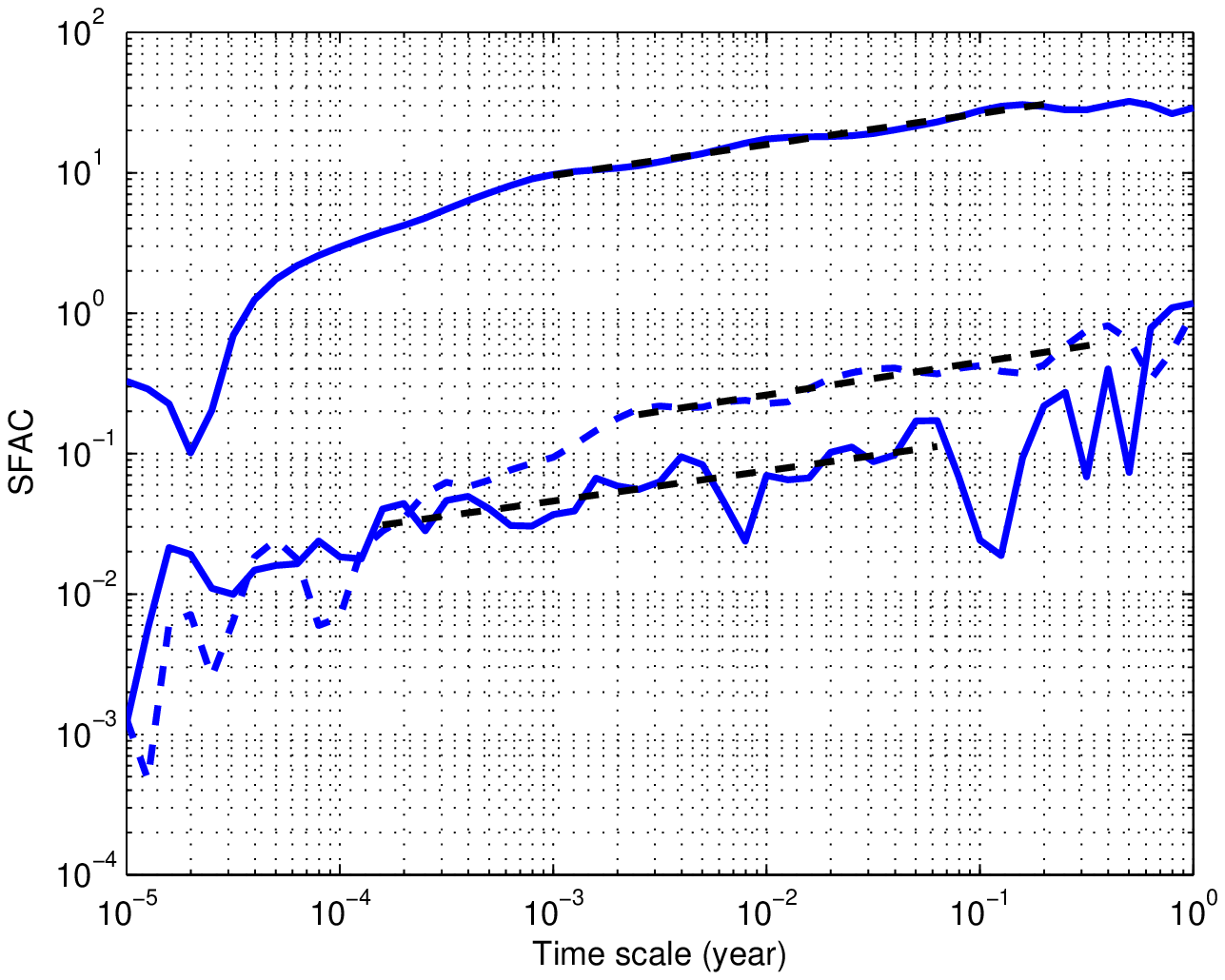}
\caption{\label{SCEC_4_45} Same as Fig.~\ref{SCEC_15_2} for 
 M within $[4;4.5]$.}
\end{center}
\end{figure}

\begin{figure}
\begin{center}
\includegraphics[width=8cm]{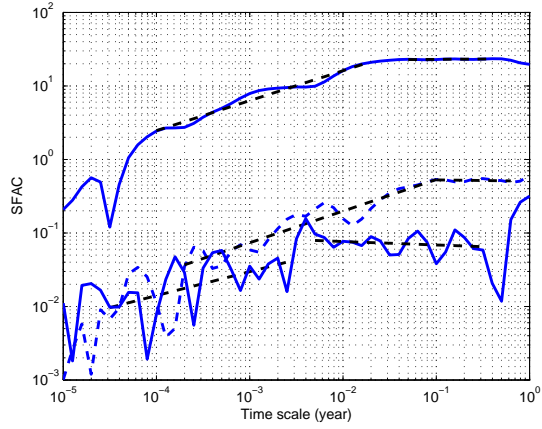}
\caption{\label{SCEC_45_5} Same as Fig.~\ref{SCEC_15_2} for 
 M within $[4.5;5]$. The first scaling range is due to the roll-off.}
\end{center}
\end{figure}

\begin{figure}
\begin{center}
\includegraphics[width=8cm]{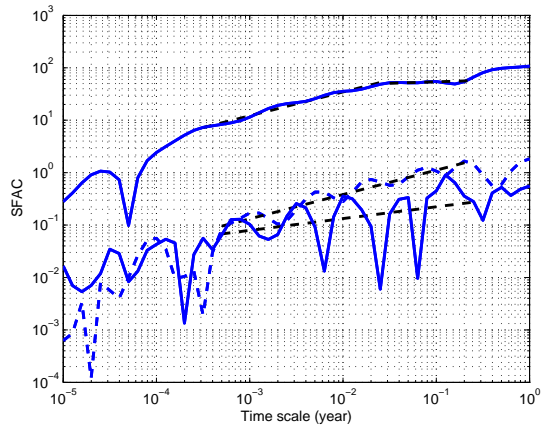}
\caption{\label{SCEC_5_55} Same as Fig.~\ref{SCEC_15_2} for 
 M within $[5;5.5]$. The existence of a roll-off imposes 
 to choose $n_D=10$ (lower solid line) as the relevant SFAC dependence.}
\end{center}
\end{figure}

\begin{figure}
\begin{center}
\includegraphics[width=8cm]{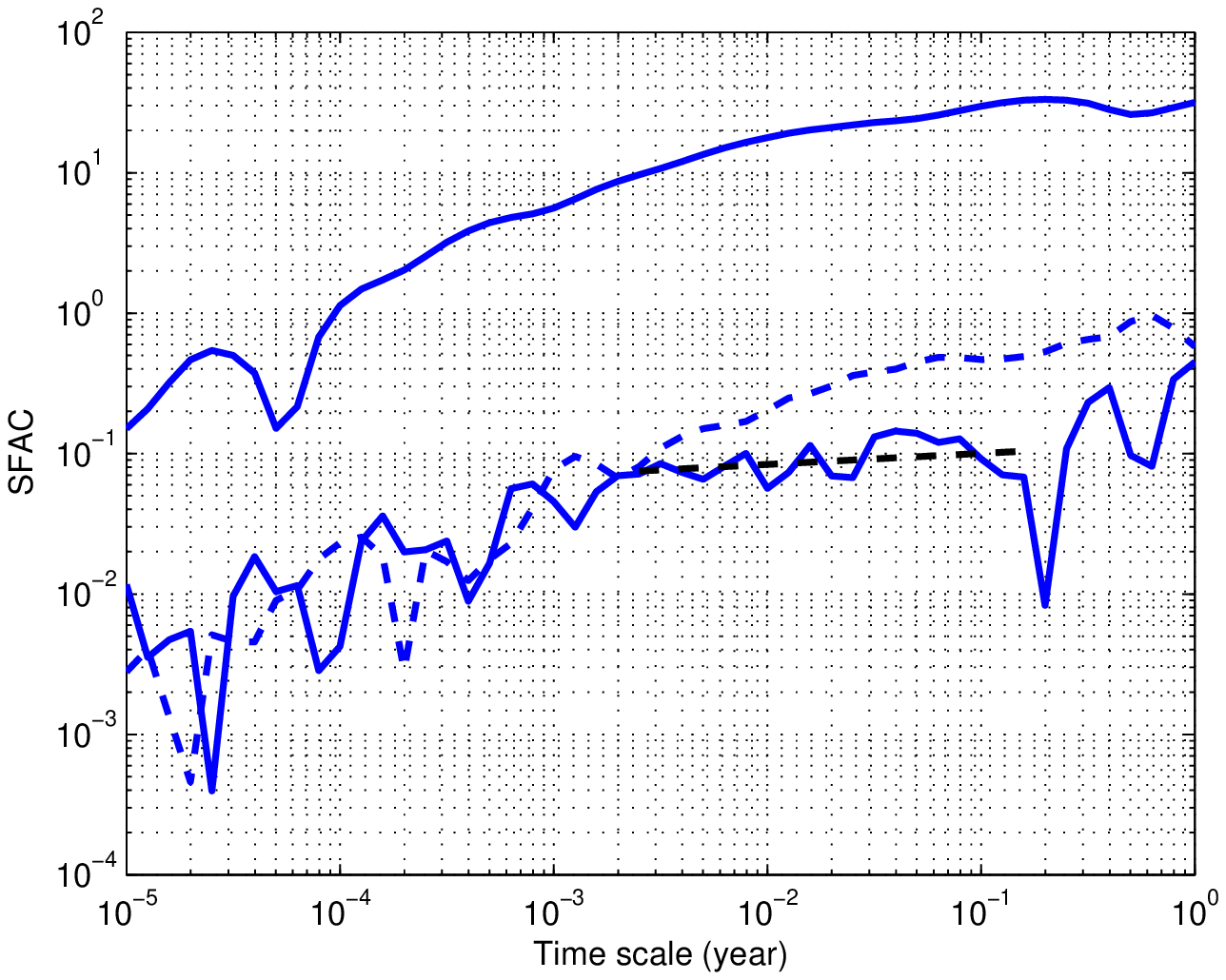}
\caption{\label{SCEC_55_6} Same as Fig.~\ref{SCEC_15_2} for 
M within $[5.5;6]$.}
\end{center}
\end{figure}

\begin{figure}
\begin{center}
\includegraphics[width=8cm]{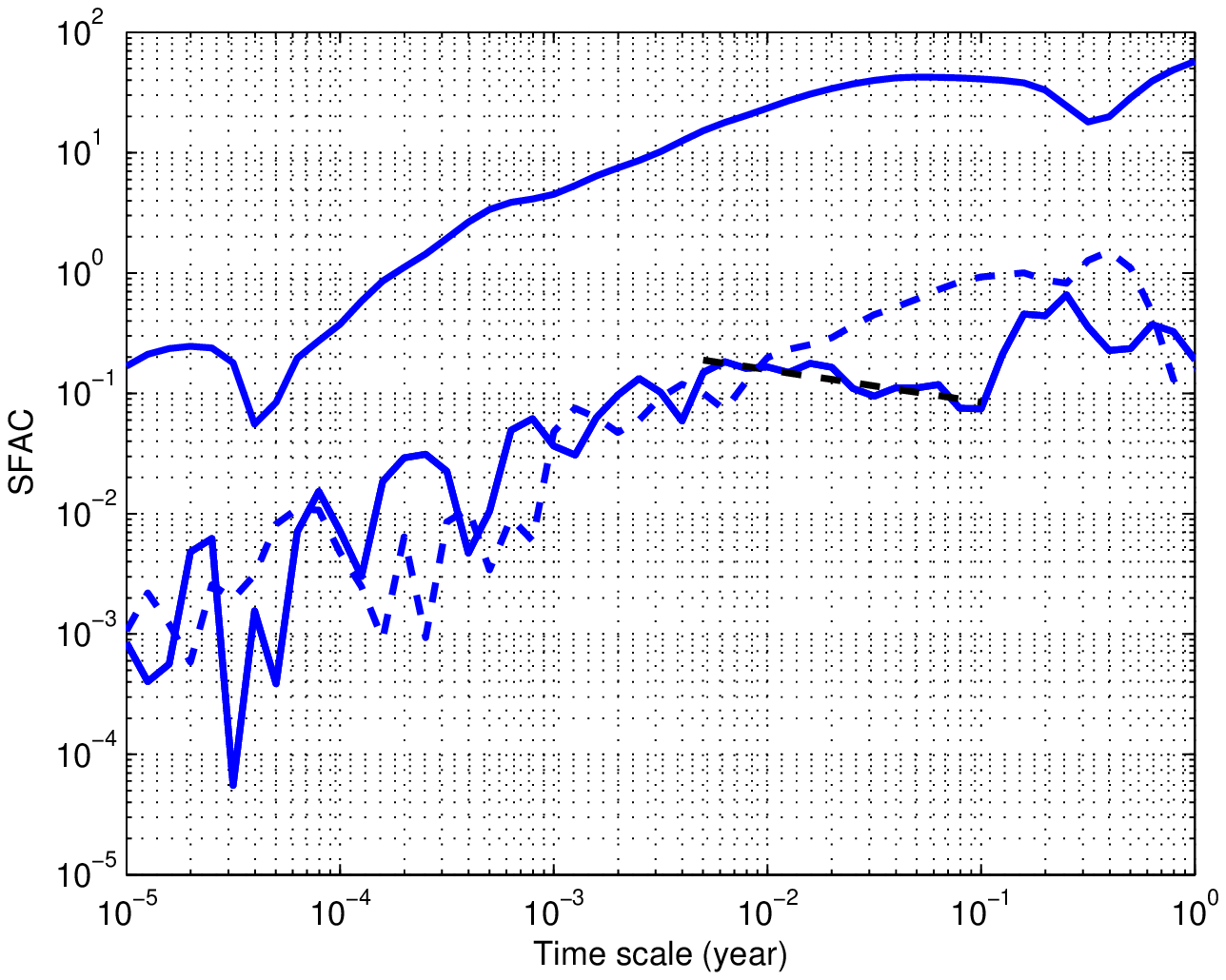}
\caption{\label{SCEC_6_65} Same as Fig.~\ref{SCEC_15_2} for 
M within $[6;6.5]$. The scaling range is limited by
the roll-off at small scales and by a burst at about $2 \cdot 10^{-1}year$.}
\end{center}
\end{figure}

\begin{figure}
\begin{center}
\includegraphics[width=8cm]{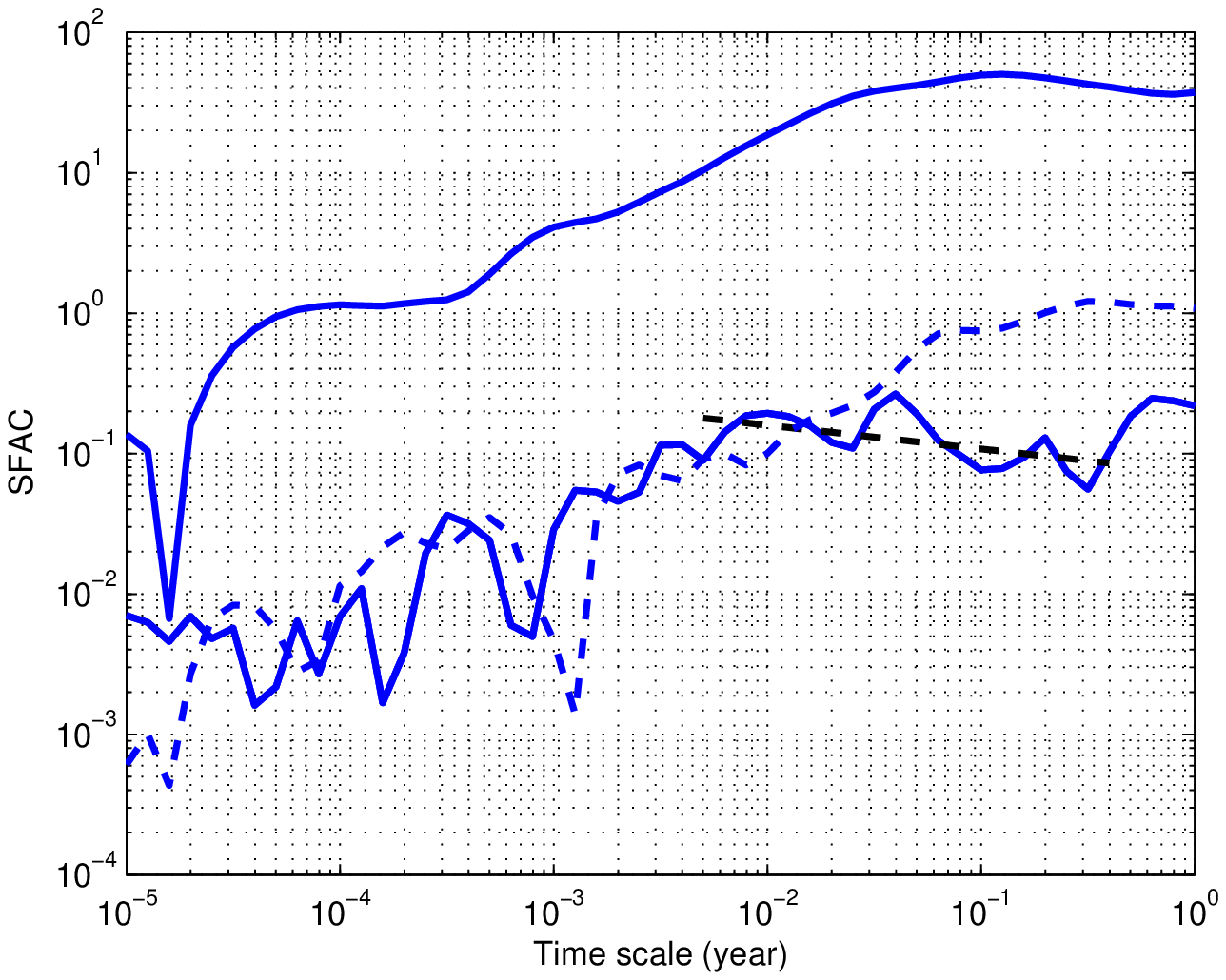}
\caption{\label{SCEC_65_7} Same as Fig.~\ref{SCEC_15_2} for  
 M within $[6.5;7]$.}
\end{center}
\end{figure}

\begin{figure}
\begin{center}
\includegraphics[width=8cm]{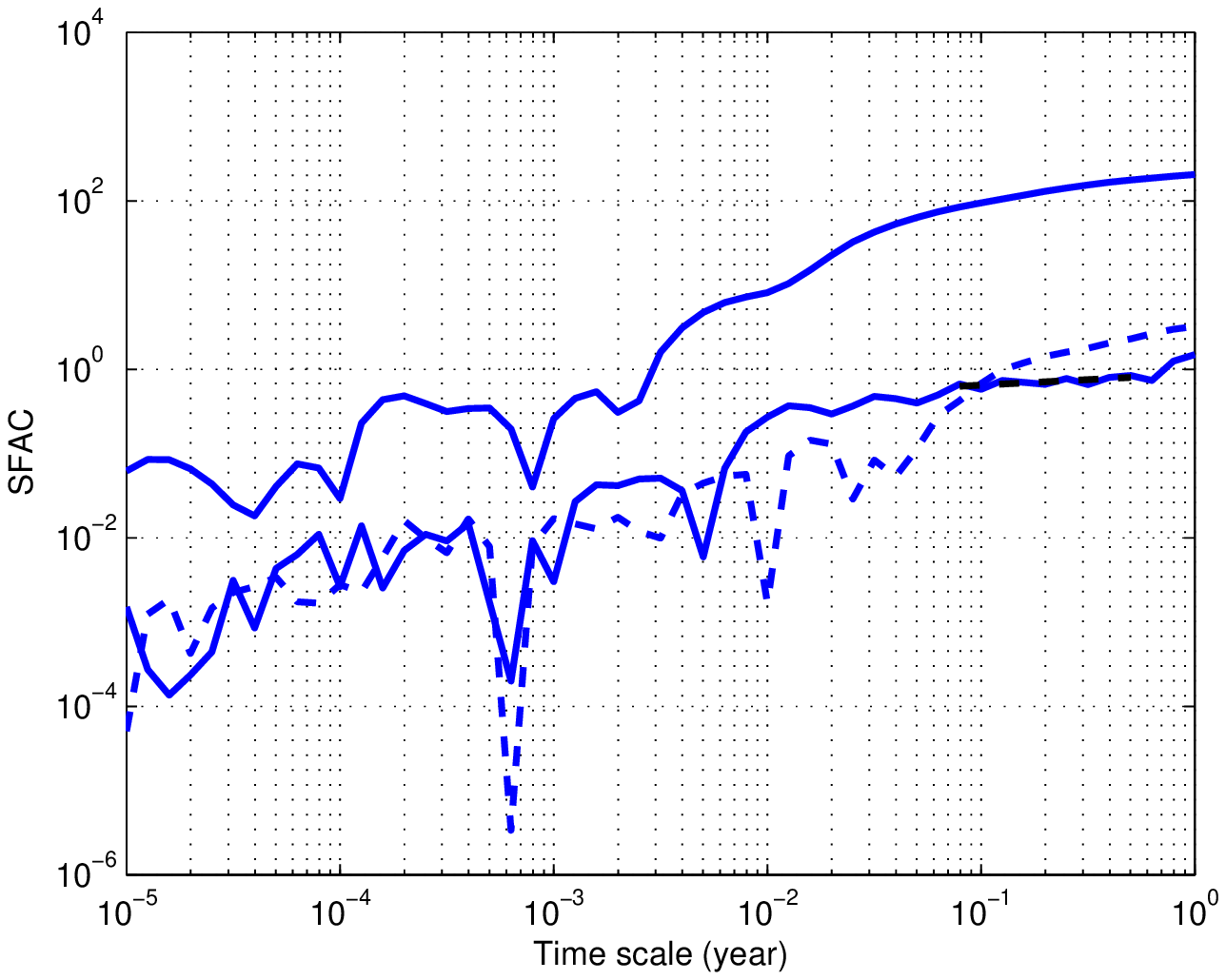}
\caption{\label{SCEC_7_75} Same as Fig.~\ref{SCEC_15_2} for  M within $[7;7.5]$.}
\end{center}
\end{figure}

\begin{figure}
\begin{center}
\includegraphics[width=8cm]{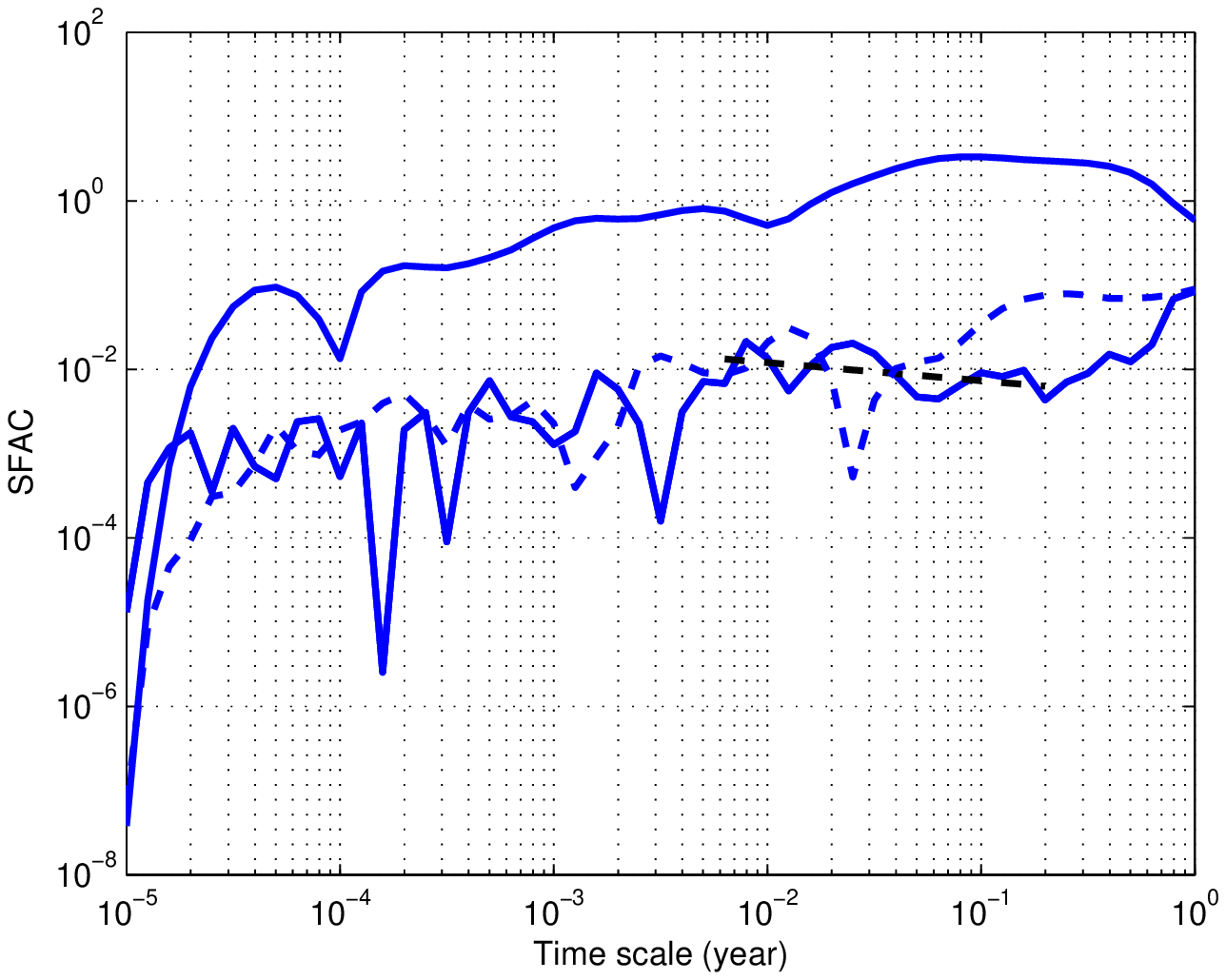}
\caption{\label{SCEC_75_8} Same as Fig.~\ref{SCEC_15_2} for M within $[7.5;8]$.}
\end{center}
\end{figure}

The scaling function analysis coefficients (SFAC) as a function of scale are displayed in Figures \ref{SCEC_15_2}-\ref{SCEC_75_8}. In each of these figures, we analyze the binned stacked aftershock time series for main shocks in a small magnitude interval, and vary the two parameters 
$n_B$ (which controls the ability of the SFA to filter non-Omori dependence) and $n_D$ (which controls the weight put to early times in the staked aftershock sequence, the larger $n_D$ is, the more are the early times removed from the analysis). Typically, we consider the following values: $n_B=0$ and $3$ and $n_D=0$ and $10$. In the set of figures  \ref{SCEC_15_2}-\ref{SCEC_75_8}, the upper solid curve corresponds to $n_B=0$ and $n_D=0$, the dashed curve
corresponds to $n_B=3$ and $n_D=0$, while the lower solid curve corresponds to $n_B=0$ and $n_D=10$. 
One can check that the value of $n_B$ has very little influence on the shape of the curves,
suggesting that the contribution of the background rate is practically constant with time. This in turn validates our aftershock selection
procedure. The straight dashed lines show the power-law fit of the SFAC as a function of time scale.
In some cases, different scaling regimes hold over different scale intervals, so that more than one fit is proposed
for the same time series (see for instance Fig. \ref{SCEC_35_4} and \ref{SCEC_45_5}). Note that
the fitting interval has a lower bound at small scales due to the roll-off effect observed in the time domain. At large time scales, several features
of the time series define the upper boundary of the fitting interval. The first feature is of course the finite size of the time series, as already discussed above. The other property is related to the occurrence of secondary aftershock sequences, that appear as localized bursts in the time series and distort it. For example,  consider the time series corresponding to the main shock magnitude range $[1.5;2]$ in Fig. \ref{SCEC_bined_stacks}, for which one can observe the
occurrence of a burst at a time of about $6 \cdot 10^{-2} year$. This corresponds to a break in the power law scaling of
the SFAC at time scales of about $10^{-1}$~year. We thus only retained the $p$-values measured using time scales before such  bursts occur. As the magnitude of the main shocks increases, the roll-off at small time scales extends to larger and larger time scales, so that the measure of $p$ proves impossible when $n_D=0$.
This is the reason why we consider the $p$-value measured with $n_D=10$ and $n_B=0$ as more reliable, especially for the large main shock magnitudes. Table 1 summarizes all our results obtained 
for the $p$-value using the SFA method. Notice that they agree very well with those obtained with the direct binning and fitting  approach.
Figure \ref{p_M_SCEC} shows the $p$-values obtained with $n_D=10$ as a function of $M$. 
A linear fit gives $p(M)=0.10M+0.40$, in excellent agreement with the results obtained using the direct fit to the binned stacked  series.

\subsection{JMA catalog}

The JMA catalog used here extends over a period from May 1923 to
January 2001 inclusive. We restricted our analysis to the zone
($+130^\circ$E to $+145^\circ$E in longitude and $30^\circ$N to $45^\circ$N in latitude), 
so that its northern and eastern boundaries fit with those of the
catalog, while the southern and eastern boundaries fit
with the geographic extension of the main japanese islands. This choice 
selects the earthquakes with the best spatial location accuracy, close
to the inland stations of the seismic network. In our analysis, 
the main shocks are taken from this zone and in the upper $70$~km, while we 
take into account their aftershocks which occur outside and at all depths.

Our detailed analysis of the aftershock time series at spatial scales down to
$20$~km reveals a couple of zones where large as well as small main
events are not followed by the standard Omori power-law relaxation of seismicity.
The results concerning these zones will be presented elsewhere. Here, we
simply removed the corresponding events from the analysis. The
geographical boundaries of these two anomalous zones are 
$[130.25^\circ{\rm E};130.375^\circ{\rm E}] \times [32.625^\circ{\rm N};32.75^\circ{\rm N}]$ 
for the first zone, and $[138.75^\circ{\rm E};139.5^\circ{\rm E}] \times 
[33^\circ{\rm N};35^\circ{\rm N}]$ for the second one (the so-called Izu
islands area). This last zone is well-known to be the locus of
earthquakes swarms, which may explain the observed anomalous aftershock relaxation.
We have been conservative in the definition of this zone along the latitude
dimension so as to avoid possible contamination in the data analysis which 
would undermine the needed precise quantification of the $p$-values.

The completeness of the JMA catalog is not constant in time, as the quality
of the seismic network increased more recently. We computed the
distribution of event sizes year by year, and used in a standard way
(Kagan 2003) the range over which the Gutenberg-Richter law is
reasonably well-obeyed to infer the lower magnitude of completeness. For our
analysis, we smooth out the time dependence of the magnitude threshold $M_c$
above which the JMA catalog can be considered complete from roughly
$M_c(1923)=6$, to $M_c(1930-1960)=5$, $M_c(1960-1990)=4.5$ with a
final progressive decrease to $M_c=2.5$ for the most recent past. This
time-dependence of the threshold $M_c(t)$ will be used for the selection
of main shocks and aftershocks. The assumed value of events location uncertainty $\Delta$ 
has been set to $10$~km. 

\subsubsection{Binned stacked times series}

\begin{figure}
\begin{center}
\includegraphics[width=8cm]{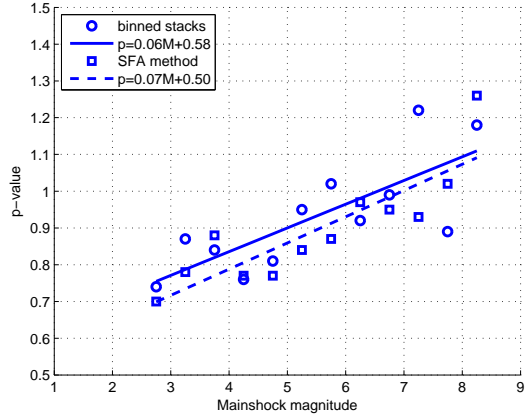}
\caption{\label{p_M_JMA} Exponents $p(M)$ of the Omori law obtained for the JMA catalog with
different methods (stacked binned method and SFA),
 with the corresponding fits:  binned time series  (squares - second column
of Table 2) and Scaling Function Analysis Coefficients (circles - fifth column of Table 2). 
Continuous and dashed lines correspond to their respective linear fits.}
\end{center}
\end{figure}

For the JMA catalog, $12$ magnitude intervals were used from $[2.5;3]$ to
$[8;8.5]$). Figure \ref{JMA_bined_stacks} shows the $12$ individual stacked
aftershocks time series and their fits (using a value for the binning factor $r$ determined
as described above for the SCEC catalog).
Figure \ref{p_M_JMA} plots the exponent $p$ averaged over the $20$ values of $r$
as a function of the middle value of the corresponding magnitude interval.
These values are also given in Table 2. A linear fit gives
$p=0.06M + 0.58$ (shown by the solid straight line in Fig.\ref{p_M_JMA}).
The $p$-value thus seems much less dependent on 
the main shock magnitude $M$ than for the SCEC catalog.

\subsubsection{SFA method}

\begin{figure}
\begin{center}
\includegraphics[width=8cm]{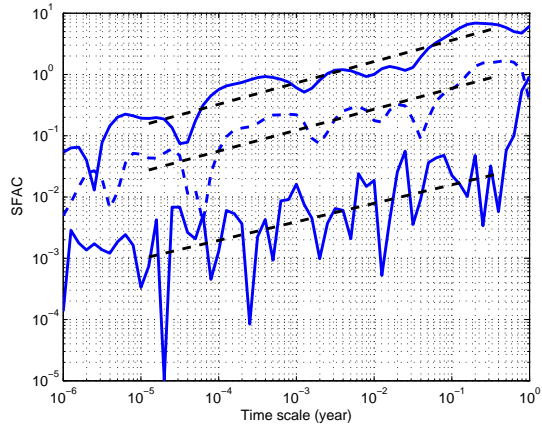}
\caption{\label{JMA_25_3} JMA - SFA method: main shock magnitudes M within $[2.5;3]$.
The upper solid curve corresponds to $n_B=0$ and $n_D=0$, the dashed curve corresponds to $n_B=3$ and $n_D=0$, while the lower solid curve corresponds to $n_B=0$ and $n_D=10$.
}
\end{center}
\end{figure}

\begin{figure}
\begin{center}
\includegraphics[width=8cm]{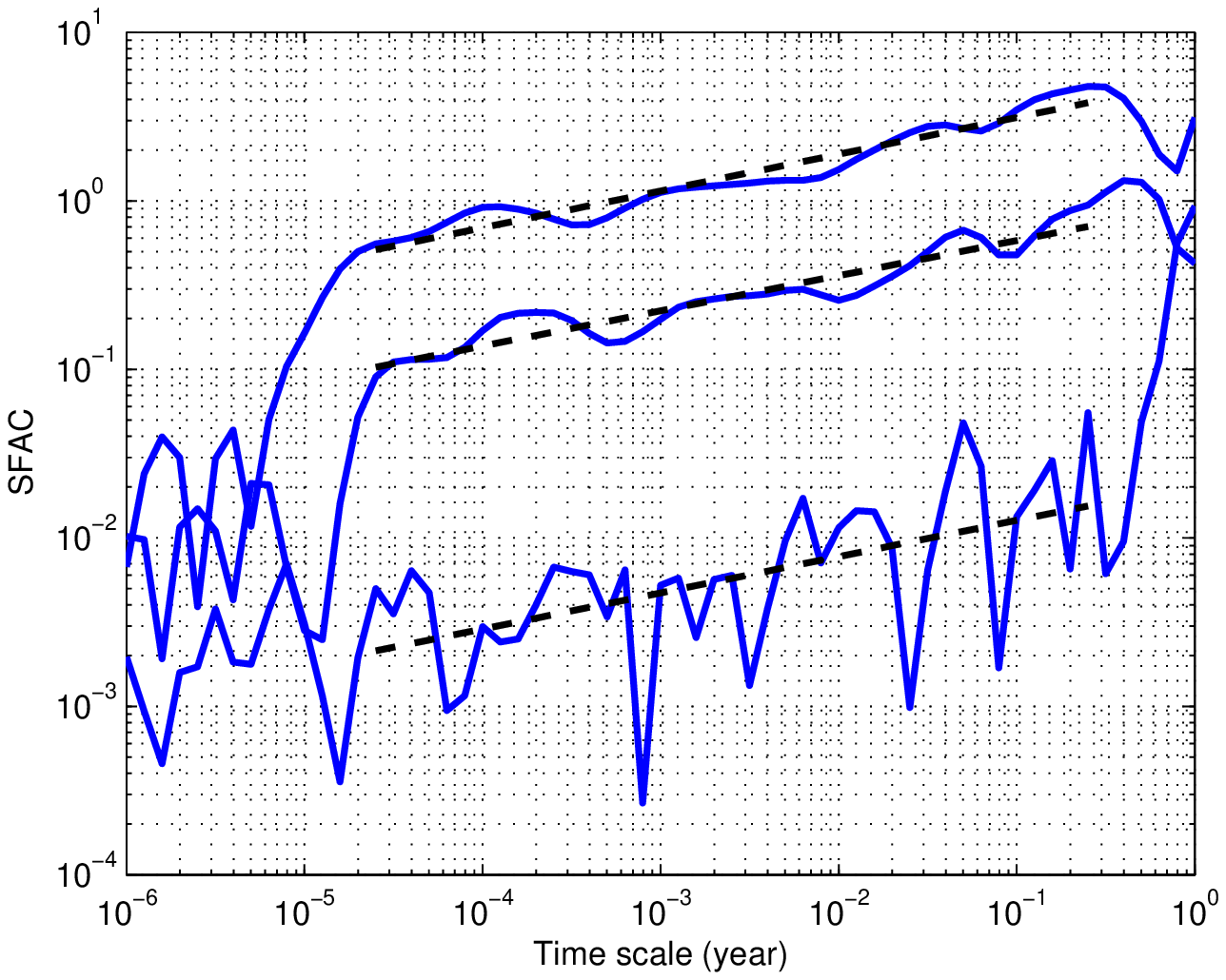}
\caption{\label{JMA_3_35} Same as Fig.\ref{JMA_25_3} for M within $[3;3.5]$.}
\end{center}
\end{figure}

\begin{figure}
\begin{center}
\includegraphics[width=8cm]{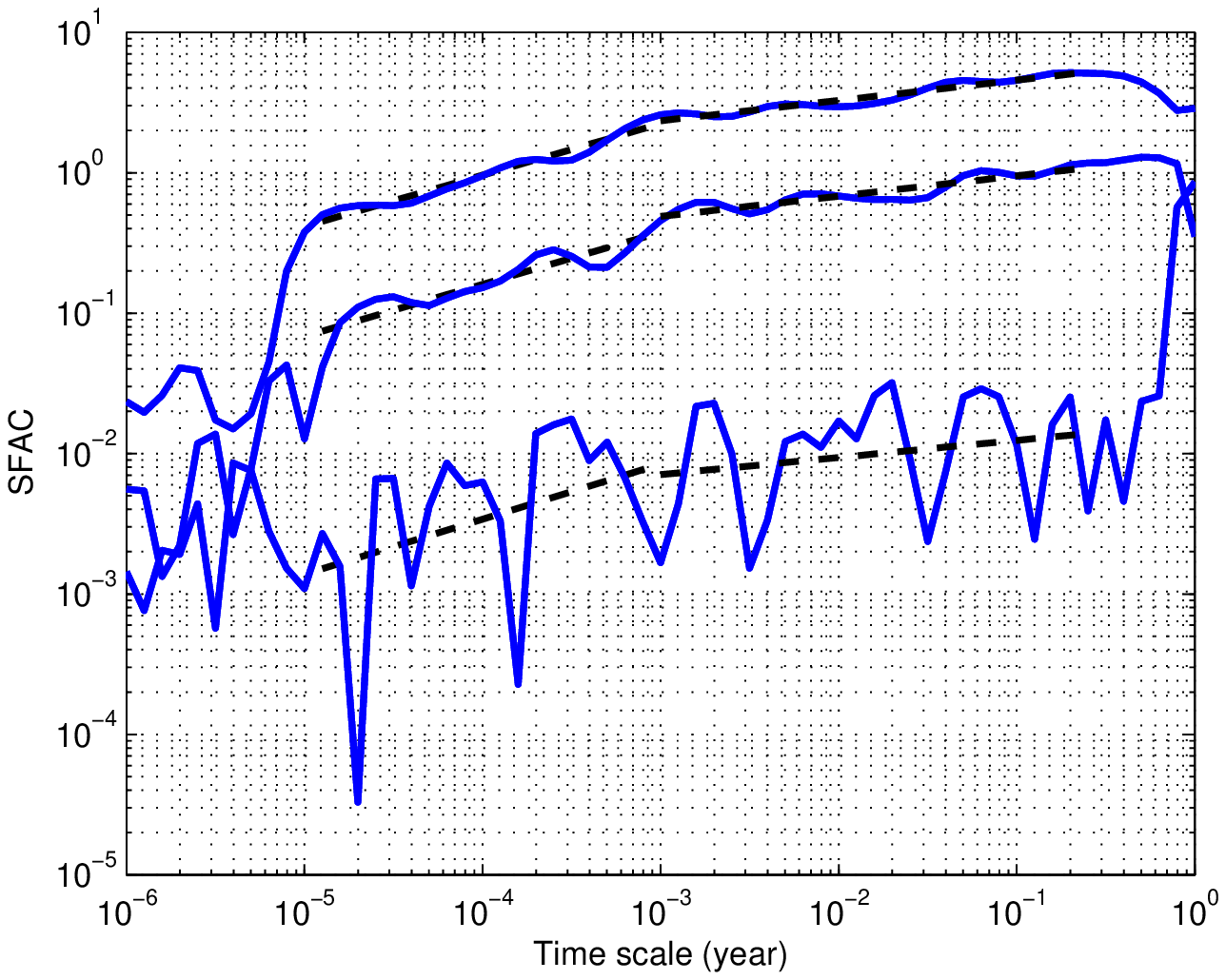}
\caption{\label{JMA_35_4} Same as Fig.\ref{JMA_25_3} for  M within $[3.5;4]$. Scaling breaks due to
the occurrence of a burst at about $5 \cdot 10^{-3}$.}
\end{center}
\end{figure}

\begin{figure}
\begin{center}
\includegraphics[width=8cm]{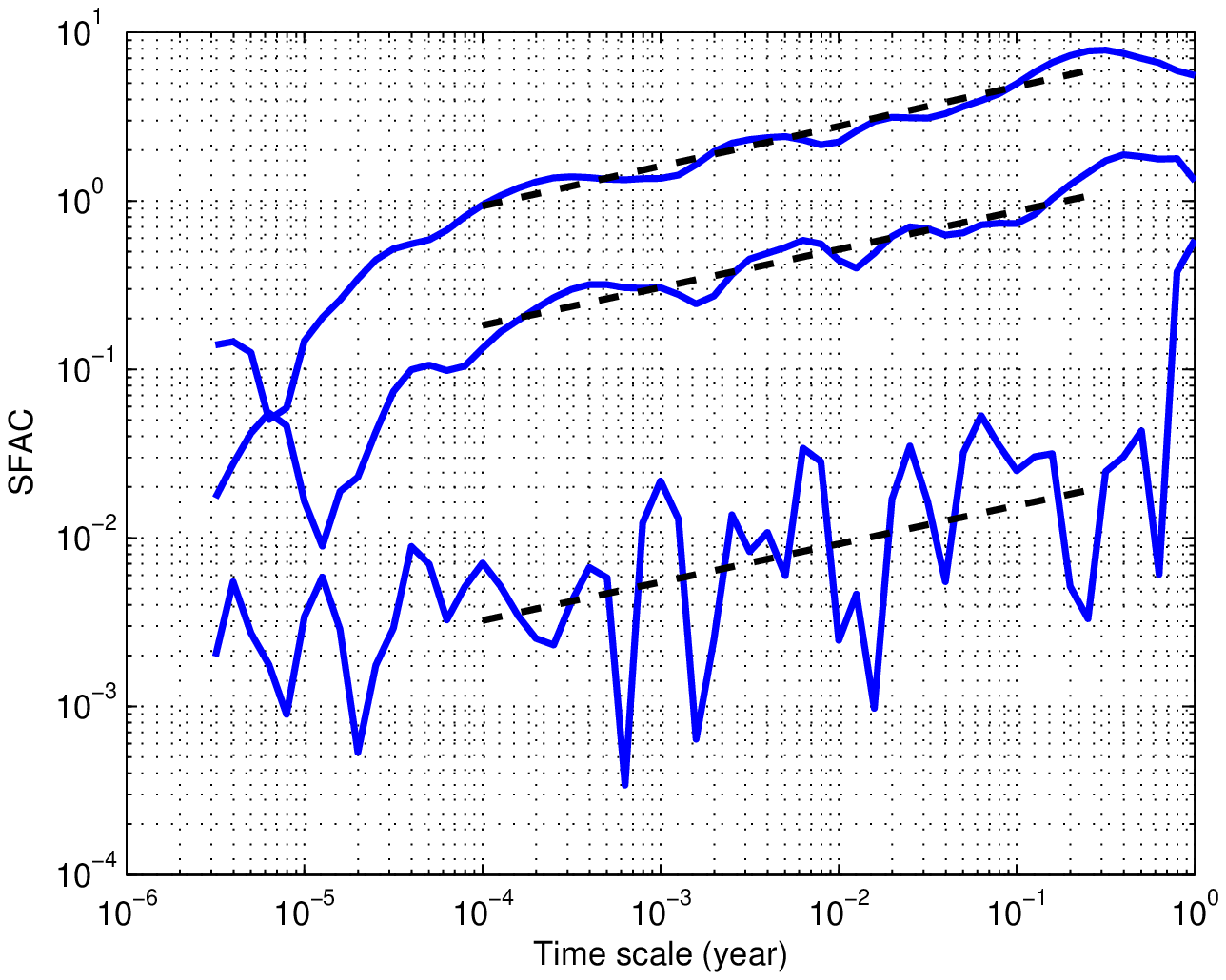}
\caption{\label{JMA_4_45} Same as Fig.\ref{JMA_25_3} for M within $[4;4.5]$.}
\end{center}
\end{figure}

\begin{figure}
\begin{center}
\includegraphics[width=8cm]{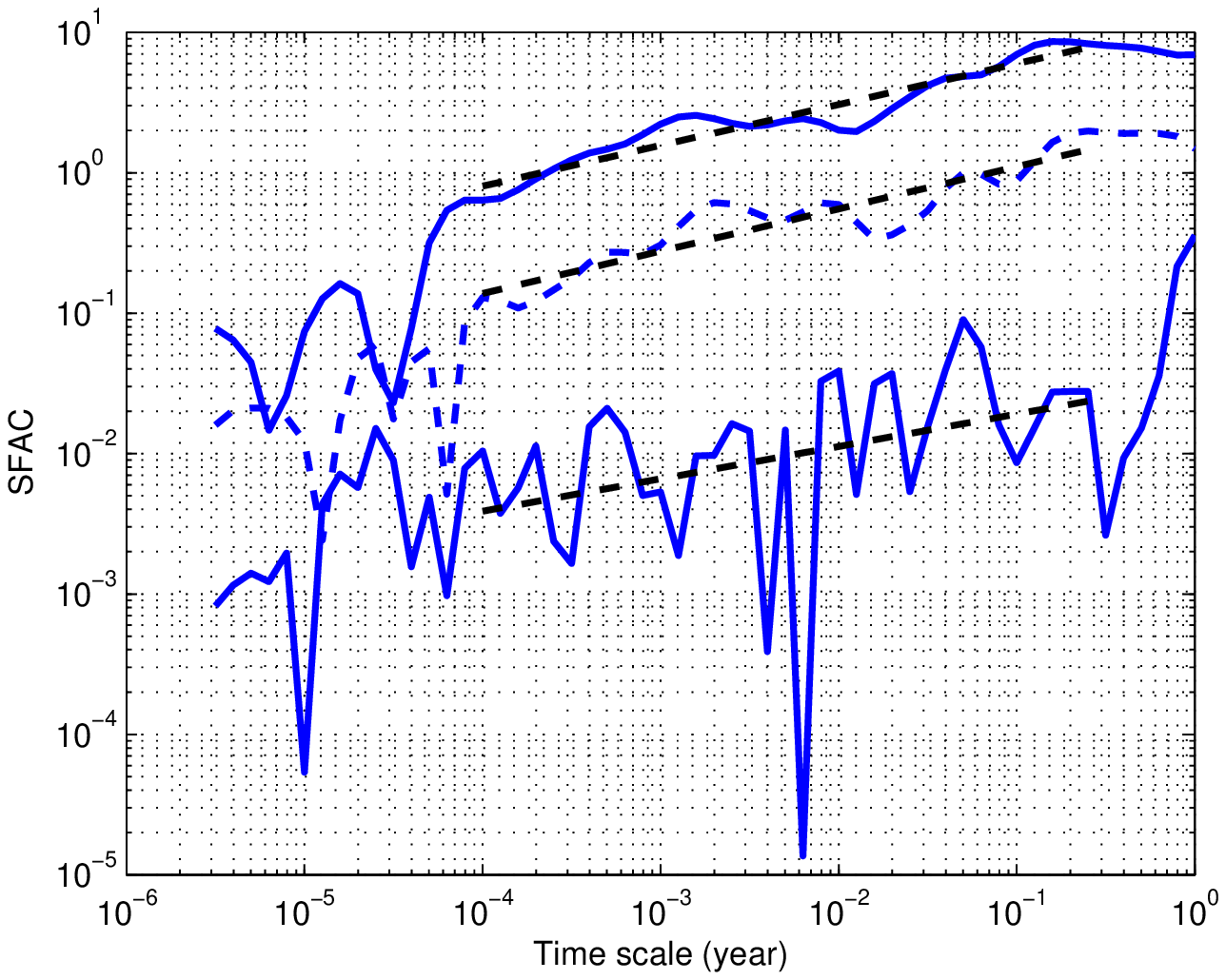}
\caption{\label{JMA_45_5} Same as Fig.\ref{JMA_25_3} for M within $[4.5;5]$. }
\end{center}
\end{figure}

\begin{figure}
\begin{center}
\includegraphics[width=8cm]{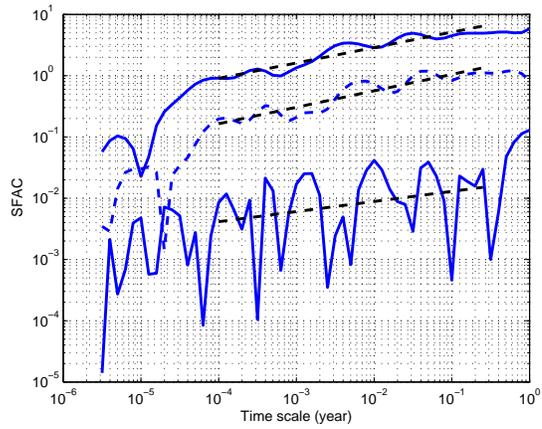}
\caption{\label{JMA_5_55} Same as Fig.\ref{JMA_25_3} for  M within $[5;5.5]$. The
presence of the roll-off implies
 to choose $n_D=10$ (lower solid line).}
\end{center}
\end{figure}

\begin{figure}
\begin{center}
\includegraphics[width=8cm]{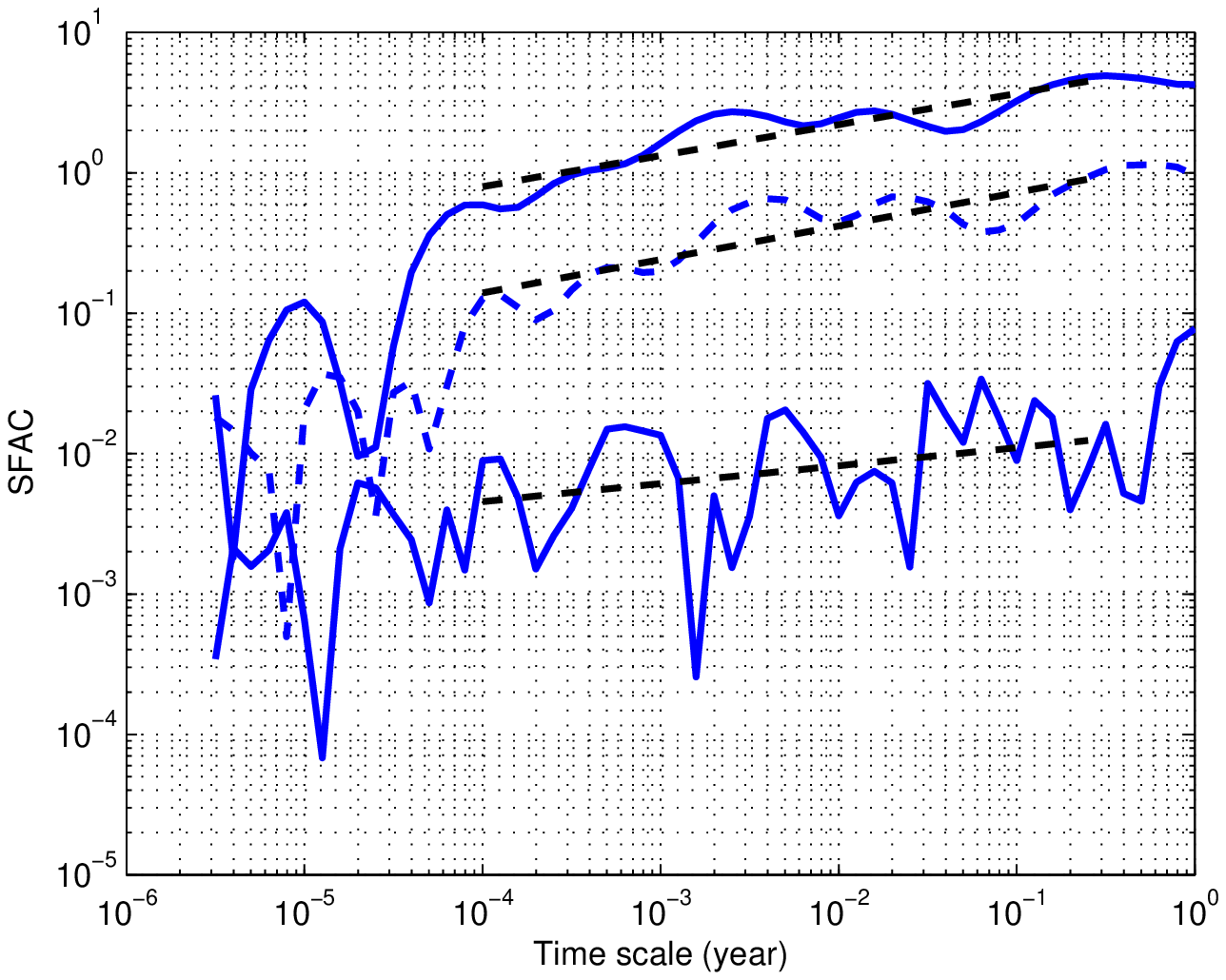}
\caption{\label{JMA_55_6} Same as Fig.\ref{JMA_25_3} for M within $[5.5;6]$.}
\end{center}
\end{figure}

\begin{figure}
\begin{center}
\includegraphics[width=8cm]{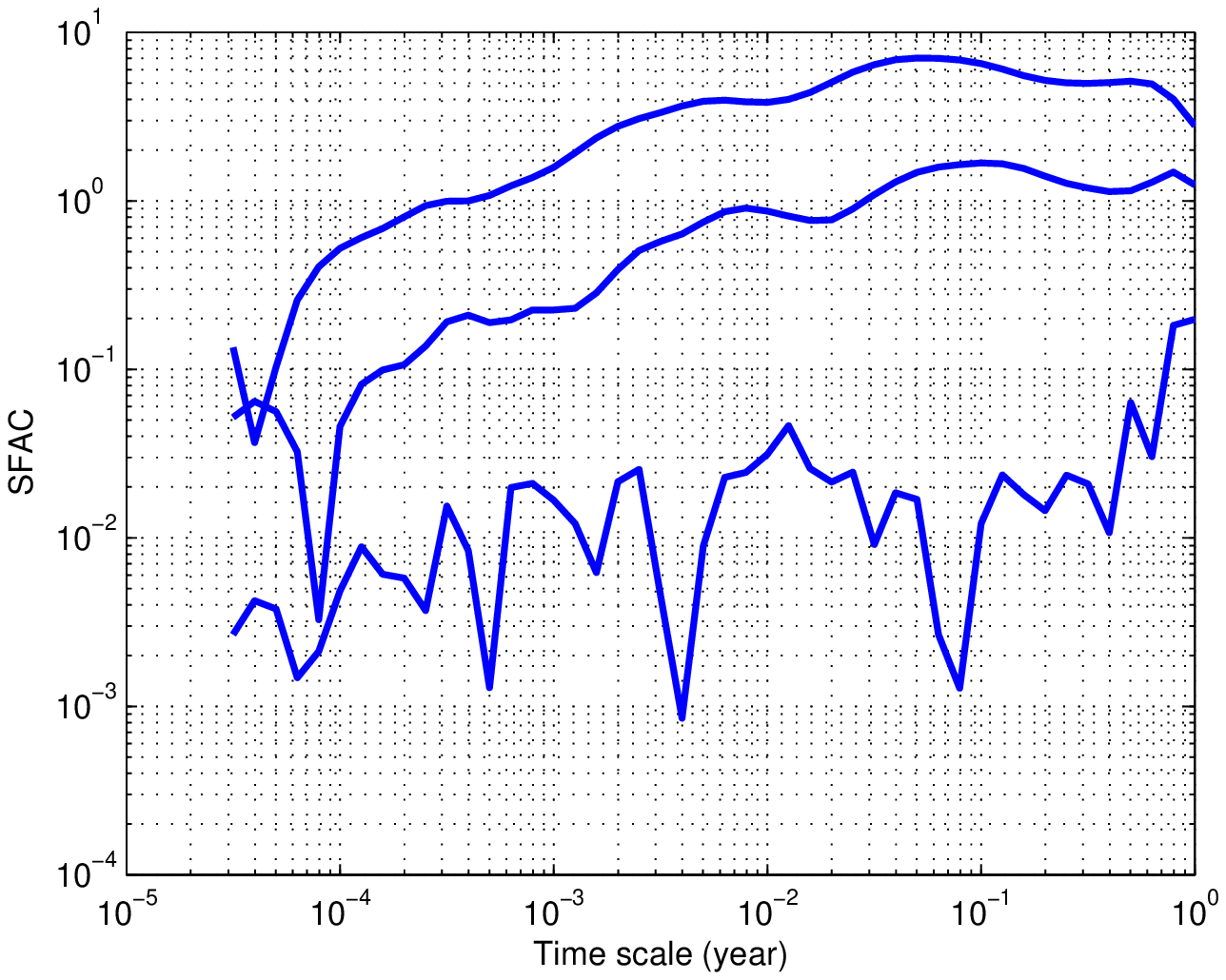}
\caption{\label{JMA_6_65} Same as Fig.\ref{JMA_25_3} for M within $[6;6.5]$. The scaling range is limited by the roll-off at small scales and a burst at about $2 \cdot 10^{-1}$~year.}
\end{center}
\end{figure}

\begin{figure}
\begin{center}
\includegraphics[width=8cm]{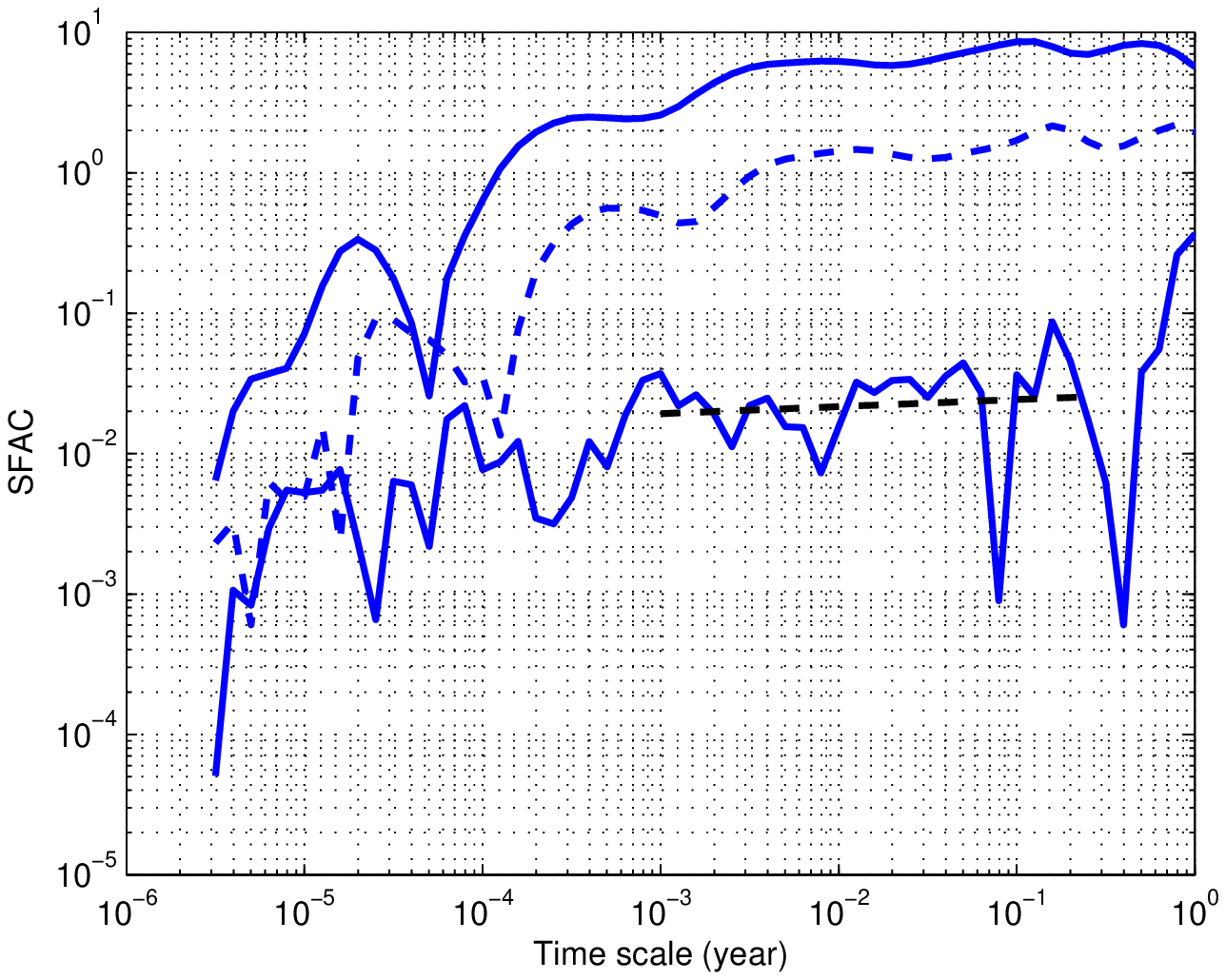}
\caption{\label{JMA_65_7} Same as Fig.\ref{JMA_25_3} for  M within $[6.5;7]$.}
\end{center}
\end{figure}

\begin{figure}
\begin{center}
\includegraphics[width=8cm]{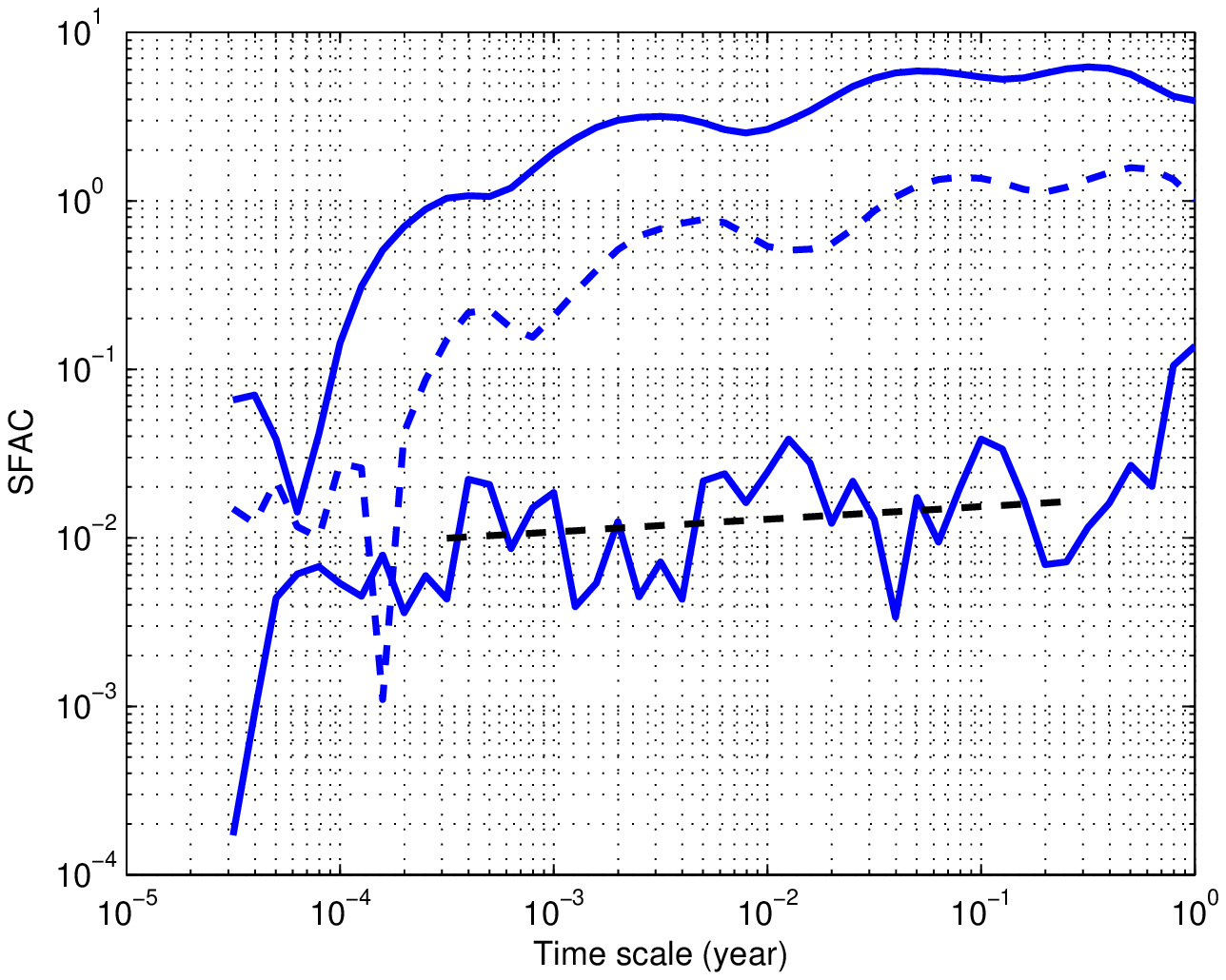}
\caption{\label{JMA_7_75} Same as Fig.\ref{JMA_25_3} for  M within $[7;7.5]$.}
\end{center}
\end{figure}

\begin{figure}
\begin{center}
\includegraphics[width=8cm]{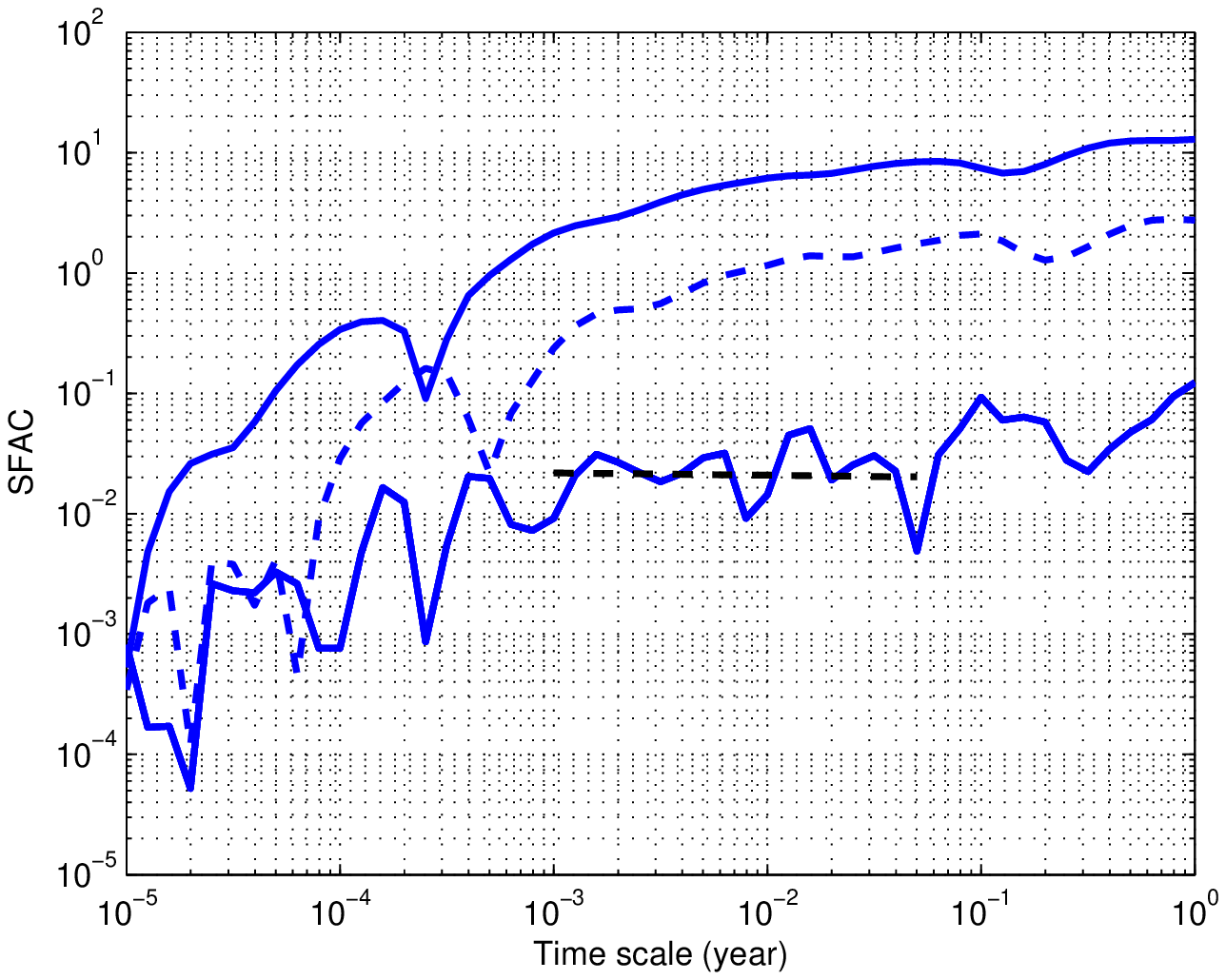}
\caption{\label{JMA_75_8} Same as Fig.\ref{JMA_25_3} for  M within $[7.5;8]$.}
\end{center}
\end{figure}

\begin{figure}
\begin{center}
\includegraphics[width=8cm]{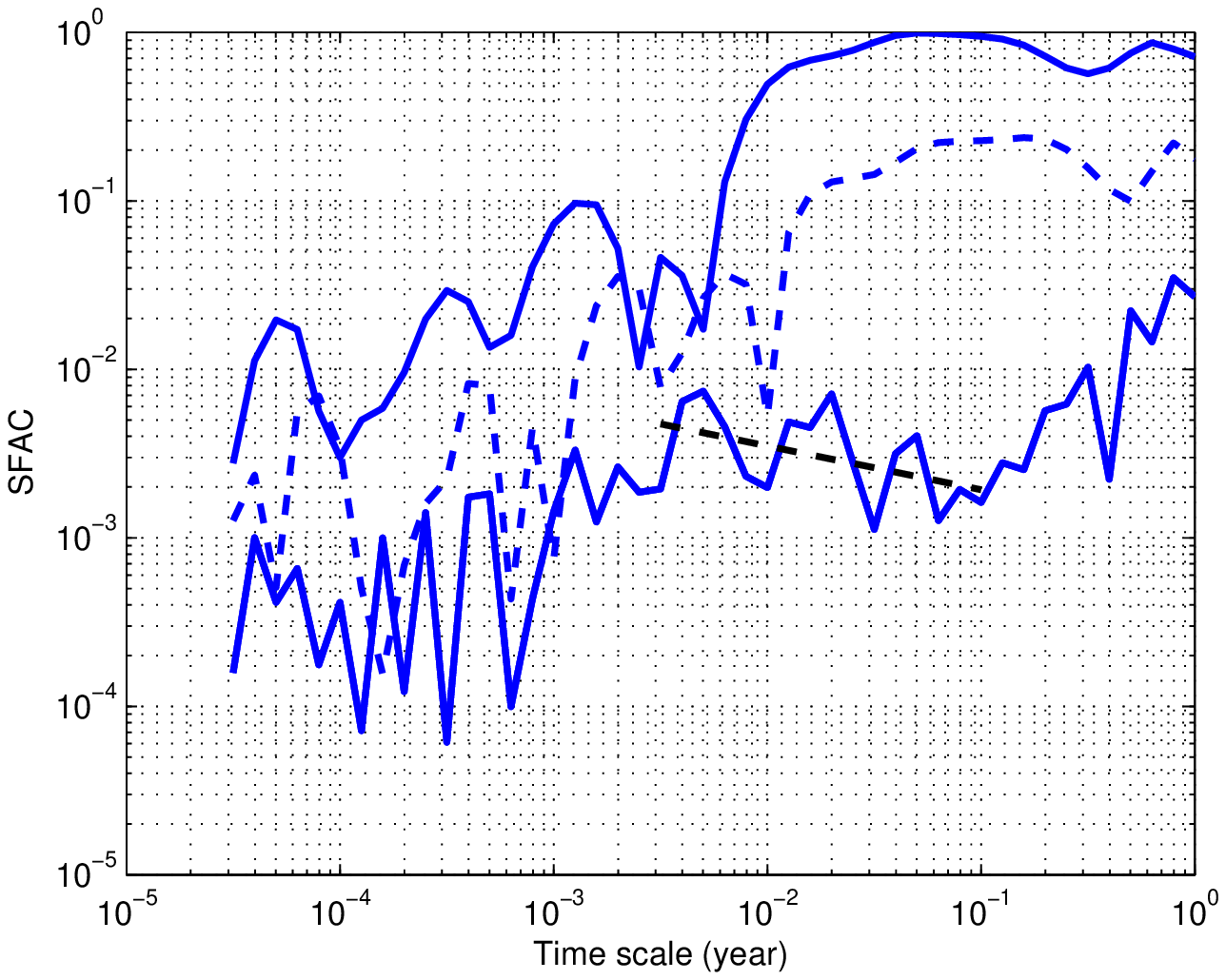}
\caption{\label{JMA_8_85} Same as Fig.\ref{JMA_25_3} for  M within $[8;8.5]$.}
\end{center}
\end{figure}

We also applied the SFA method to the same dataset. We checked that the resulting curves were not
dependent on the value of $n_B$, suggesting that the background term is constant. Fig. \ref{JMA_25_3}
to \ref{JMA_8_85} show the SFAC as a function of scale for different values of $(n_B,n_D)$: $(0,0)$ (upper solid
curve), $(3,0)$ (dashed curve), and $(0,10)$ (lower solid curve). One can observe that some of them exhibit a more complex 
scaling behavior than found for the SCEC catalog. This may reveal a complex mixture of sequences with different properties (see for example Fig.\ref{JMA_45_5} and \ref{JMA_55_6} which exhibit two characteristic time scales
of about $10^{-3}$~year and $10^{-1}$~year), despite our efforts to exclude zones that have
a large number of swarms. The characteristic scales disappear with $n_D=10$, but this may just be due to the strongly oscillating character of the filter and therefore of the SFAC which may mask its
local maxima.
Table 2 and Fig. \ref{p_M_JMA} report the corresponding measured exponents. There is a general agreement between the
$p$-values mesured using different sets of parameters or methods.
Using the set of $p$-values
corresponding to $n_B=0$ and $n_D=10$, we obtain the following dependence of the $p$-value 
as a function of the magnitude $M$ of the main shocks:  $p=0.07M+0.50$. Excluding the largest magnitude range leads to a weaker dependence: $p=0.05M+0.58$. Note that the dispersion of data points around the best fit line is much smaller for the  $p$-values obtained by the
SFA method. This thus confirms the weaker dependence of $p$ as a function of $M$ for the JMA catalog. Our SFA suggests that this weaker dependence may have to do with the presence of
many swarms in the Japanese catalogs. Our methodology has allowed us to diagnose the existence of mixtures of aftershock relaxation regimes, probably
swarms and standard Omori standard sequences.

\subsection{The Harvard CMT catalog}

The worldwide CMT Harvard catalog used here goes
from January $1976$ to August $2006$ inclusive. This catalog is
considered to be complete for events of magnitude $5.5$ or larger. 
We thus removed events below this threshold before searching for the aftershocks.
Due to the rather small number of events in this
catalog, we did not impose any limit on the depth of events. The assumed value
of location uncertainties has been set to $\Delta=10$~km. Note that instead of using the 
hypocenter location as we did for the two other catalogs, we considered the location
of the centroid, which is certainly closer to the center of the aftershock zone.  

\subsubsection{Binned stacks}

\begin{figure}
\begin{center}
\includegraphics[width=8cm]{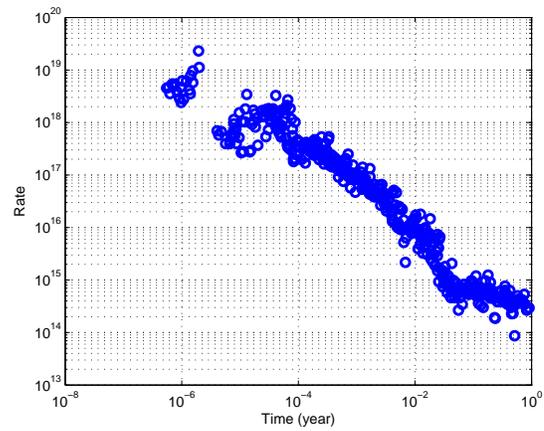}
\caption{\label{HAR_55_6_raw} Binned stacked time series of sequences triggered by main events with $M$ within
$[5.5;6]$. This plot features binned series corresponding to all $r$ values.}
\end{center}
\end{figure}

For the Harvard catalog, seven magnitude intervals were used from $[5.5;6]$ to
$[9;9.5]$ (the $[8.5;9]$ interval being empty). The binned stacked times series for the 
$[5.5;6]$ magnitude range is shown 
in Fig. \ref{HAR_55_6_raw}, using all values of the binning factor $r$. The underlying decay law is
obviously not of Omori-type, which suggest that it is the result from the superposition of different
distributions. We attribute the different behavior of the $([5.5;6])$ magnitude range to the fact that the corresponding times series contain many events occurring at mid-oceanic ridges, where many swarms are  known to occur. As very few events of magnitude $>6$ occur in this peculiar tectonic settings, swarms (from the mid-ocean ridges) do not contaminate too much the time series associated with larger magnitude main shocks.  We will see below that the SFA confirms this intuition,
and doesn't provide any evidence of a power law scaling for the $([5.5;6])$ magnitude range
while the other magnitude ranges (except the largest) give reliable estimates for the Omori
exponent $p$.

\begin{figure}
\begin{center}
\includegraphics[width=8cm]{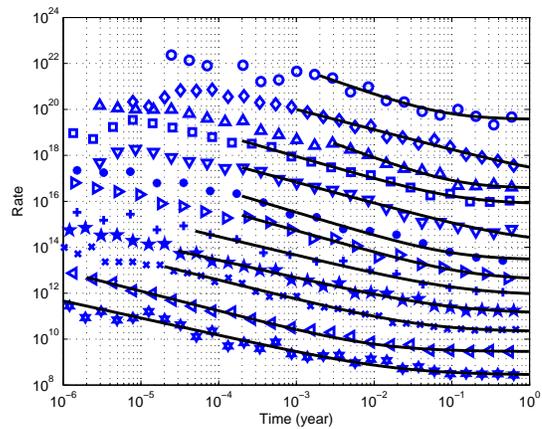}
\caption{\label{JMA_bined_stacks} Binned stacked series of aftershock sequences in the JMA catalog
for various magnitude ranges (from $[2.5;3]$ at the bottom to $[8;8.5]$ at the top by steps of $0.5$). 
The solid lines show the fits of formula (\ref{fitapb}) to the individual time series.
All curves have been shifted along the vertical axis for the sake of clarity.}
\end{center}
\end{figure}

Figure \ref{HAR_bined_stacks} shows the six remaining stacked
aftershocks time series and their fits (constructed as in Figs. \ref{SCEC_bined_stacks} and 
\ref{JMA_bined_stacks}).
One can clearly observe Omori-like behaviors. The corresponding $p$-values are reported
in Table 3 and in Fig. \ref{p_m_HAR} as a function of the main shock magnitudes $M$. The linear fit 
of the dependence of $p$ as a function of $M$ gives  $p(M)=0.16M-0.09$. The magnitude
dependence of $M$ is thus much larger than found in Southern California but we have to consider that the
magnitude range over which the fit is performed is much more restricted that for the SCEC catalog, leading to larger uncertainty. Note that the $[9;9.5]$ magnitude range displays an unusual small $p$ value of $0.69$. This may be due to the fact that we are still in the roll-off time range, or to the very limited  amount of data as only one main shock occurred in that magnitude range. For this reason, we excluded it in the plots and in the estimation of the $p(M)$ relationship.

\subsubsection{SFA method}

\begin{figure}
\begin{center}
\includegraphics[width=8cm]{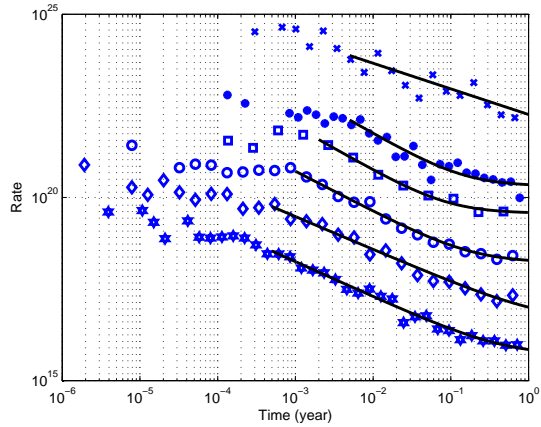}
\caption{\label{HAR_bined_stacks} Binned stacked series of aftershock sequences in the Harvard catalog
for various magnitude ranges. Magnitude ranges are, from bottom to top: $[6;6.5], [6.5;7], [7;7.5], 
[7.5;8], [8;8.5]$ and $[9;9.5]$.
The solid lines show the fits to the individual time series. The
 $[9;9.5]$ magnitude range displays an unusual small $p$ value of $0.69$ (see text for further discussion of this anomaly in comparison with the other magnitude ranges.
 All curves have been shifted along the vertical axis for the seek of clarity.}
\end{center}
\end{figure}

\begin{figure}
\begin{center}
\includegraphics[width=8cm]{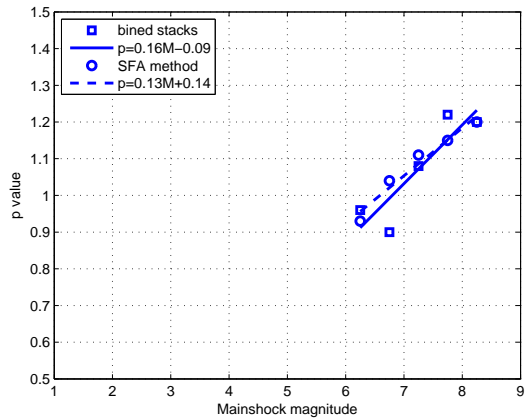}
\caption{\label{p_m_HAR} Dependence of the Omori exponent $p$ as a function of the 
main shock magnitude $M$ obtained for the HARVARD catalog with fits of the binned time series 
(squares - second column
of Table 3) and of the Scaling Function Analysis Coefficent (circles - third column of Table 3). 
Continuous and dashed lines stand for their respective linear fits.}
\end{center}
\end{figure}

\begin{figure}
\begin{center}
\includegraphics[width=8cm]{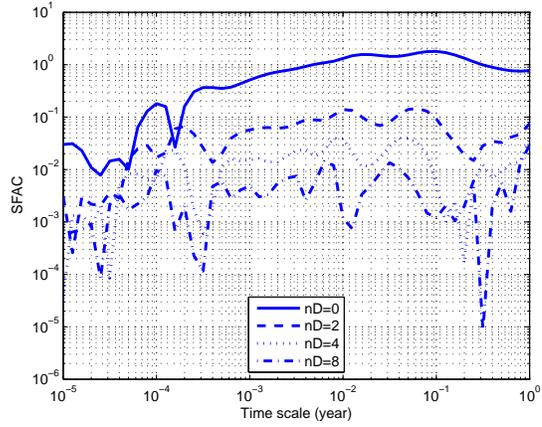}
\caption{\label{HAR_55_6} HAR - SFA method for main shock magnitudes M within $[5.5;6]$. Note the existence of two characteristic scales. The parameter $n_B$ is set to zero and the four
curves correspond to different values of $n_D$ as indicated in the insert panel.
}  
\end{center}
\end{figure}

\begin{figure}
\begin{center}
\includegraphics[width=8cm]{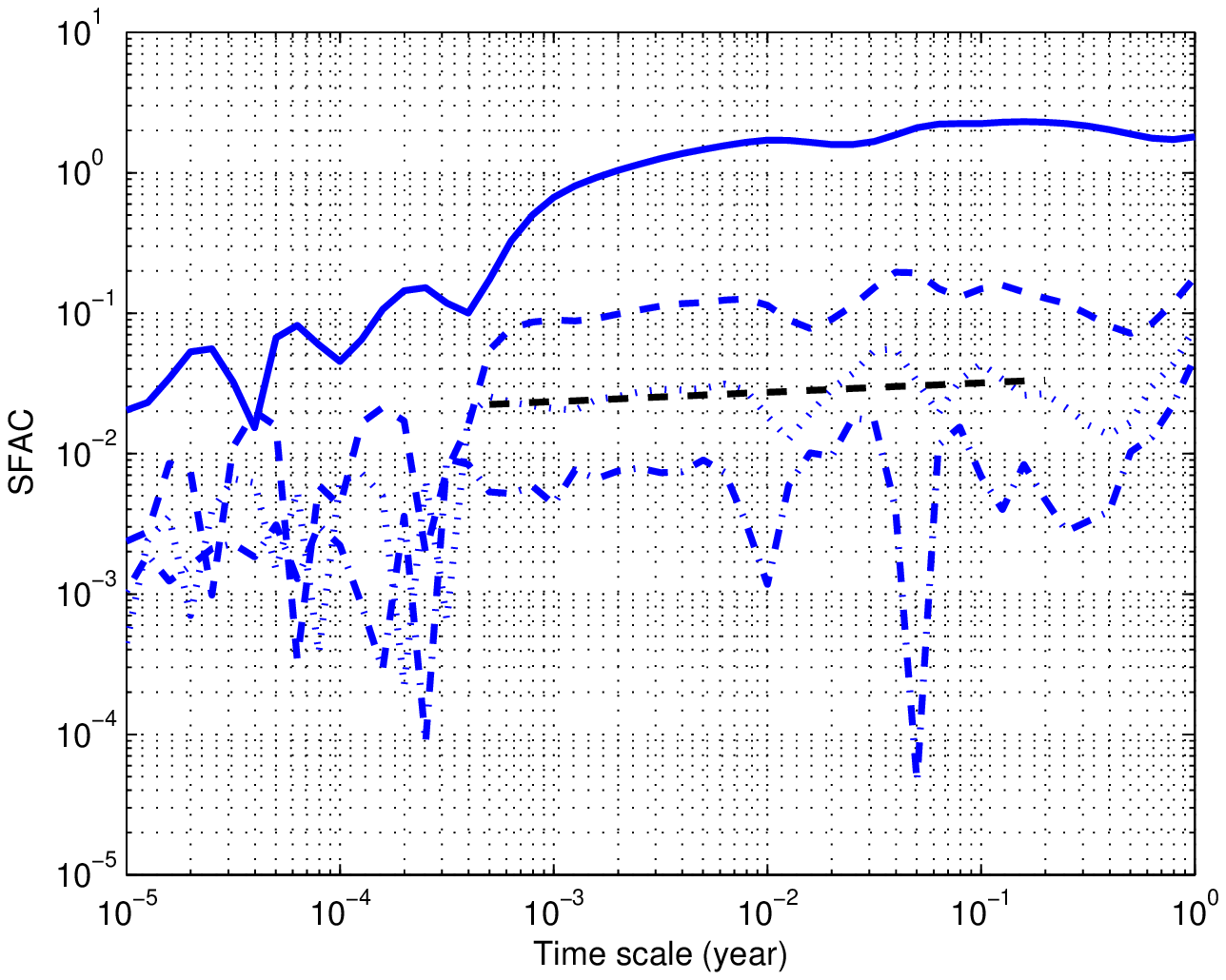}
\caption{\label{HAR_6_65} Same as Fig.\ref{HAR_55_6} for M within $[6;6.5]$. The scaling range is limited by the roll-off at small scales and a burst at about $2 \cdot 10^{-1}$~year.}
\end{center}
\end{figure}

\begin{figure}
\begin{center}
\includegraphics[width=8cm]{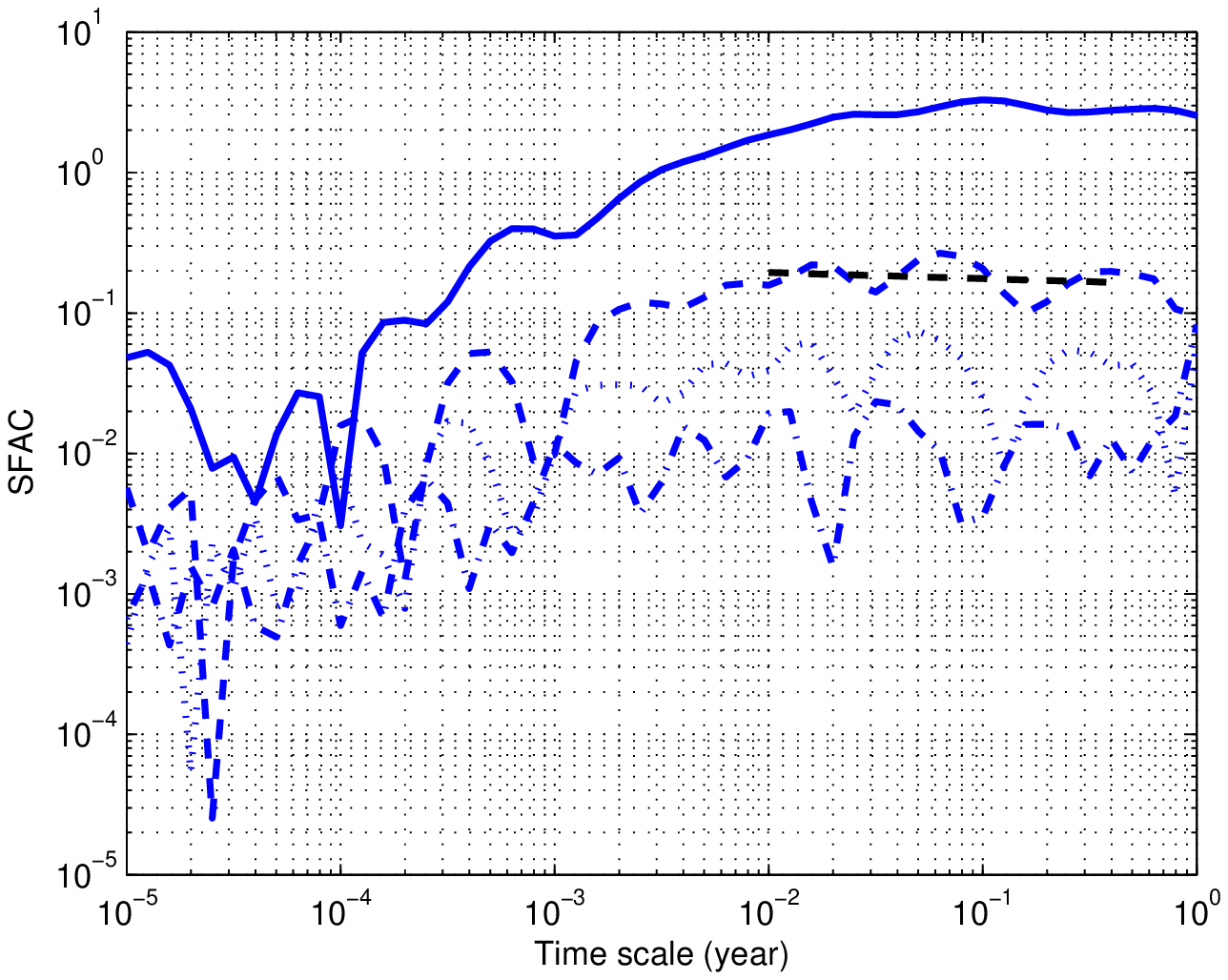}
\caption{\label{HAR_65_7} Same as Fig.\ref{HAR_55_6} for M within $[6.5;7]$.}
\end{center}
\end{figure}

\begin{figure}
\begin{center}
\includegraphics[width=8cm]{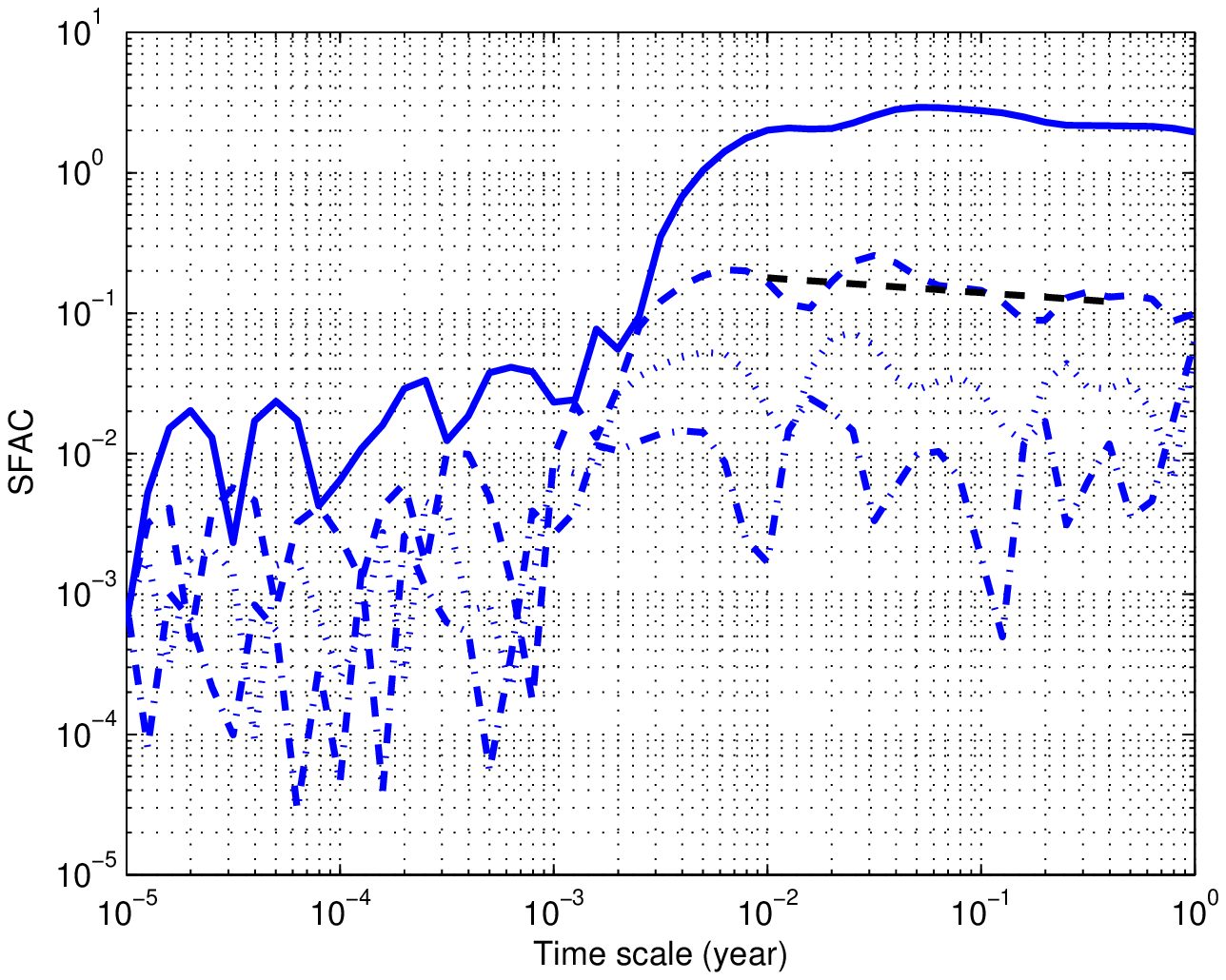}
\caption{\label{HAR_7_75} Same as Fig.\ref{HAR_55_6} for M within $[7;7.5]$.}
\end{center}
\end{figure}

\begin{figure}
\begin{center}
\includegraphics[width=8cm]{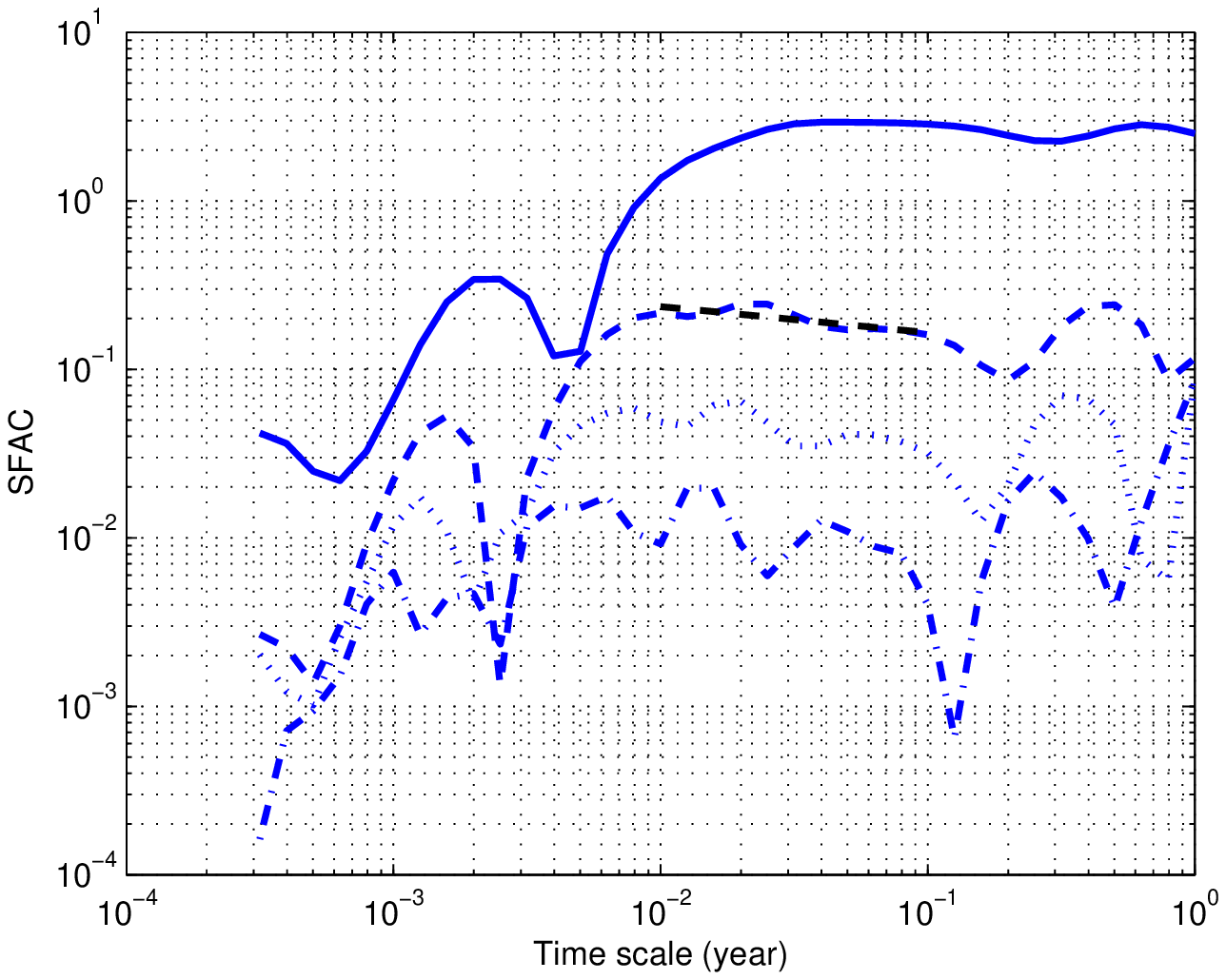}
\caption{\label{HAR_75_8} Same as Fig.\ref{HAR_55_6} for M within $[7.5;8]$.}
\end{center}
\end{figure}

\begin{figure}
\begin{center}
\includegraphics[width=8cm]{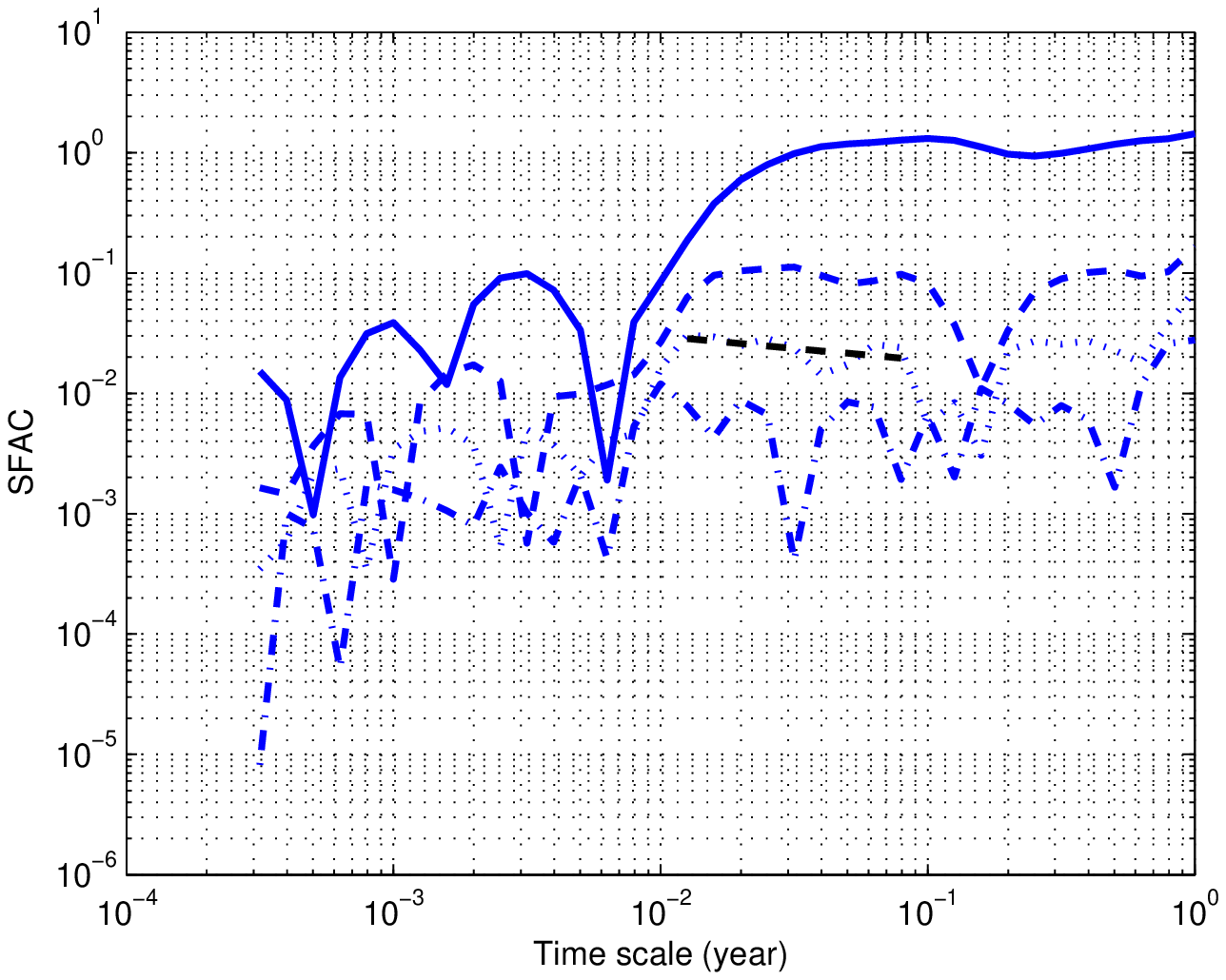}
\caption{\label{HAR_8_85} Same as Fig.\ref{HAR_55_6} for M within $[8;8.5]$.}
\end{center}
\end{figure}

\begin{figure}
\begin{center}
\includegraphics[width=8cm]{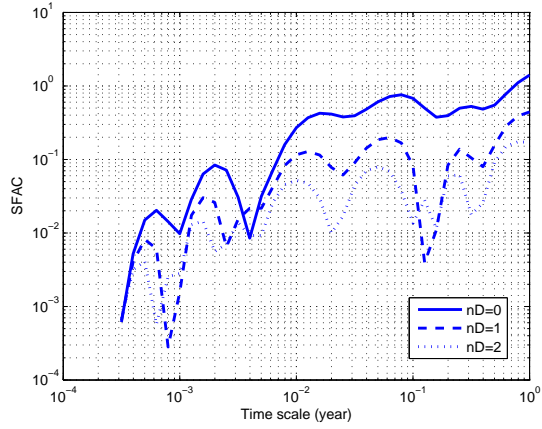}
\caption{\label{HAR_9_95} Same as Fig.\ref{HAR_55_6} for M within $[9;9.5]$. As the 
dependences of the SFAC as a function of scale are too strongly oscillating, we do not provide any fit.}
\end{center}
\end{figure}

Figures \ref{HAR_55_6} to \ref{HAR_9_95} present the dependence of the SFAC 
as a function of scale for the different main shock magnitude ranges.
Due to the incompleteness (roll off) effect and to the rather large value of 
magnitude $M_c=5.5$ of completeness, one can observe in Fig. \ref{HAR_bined_stacks}
that the power law scaling do not hold at scales smaller than about $10^{-3}$~year. This thus prevents
us from using the SFA method with $n_D=0$ to measure an accurate value of the Omori exponent $p$. We thus first checked that, using $n_D=0$, 
the shape of the SFAC curves is independent of $n_B$. We then set $n_B=0$ and considered different values of  $n_D=0,2,4$ and $8$. Larger values of $n_D$ lead to strongly oscillating SFAC as a function of scale, which are difficult to interpret.
Only one fit (straight dashed line) is shown in each figure, and 
the corresponding $p$-values are gathered in Table 3 and plotted in Fig. \ref{p_m_HAR}. We chose the fits with non-zero $n_D$ with a value such that 
the SFAC curve does not oscillate too much. We can visually check that the
chosen fit is compatible with other non-zero values of $n_D$, as well as with the extrapolation to scales where the SFAC is oscillating. Due the small amount of data, no $p$-value could be determined for the $[9;9.5]$ range, as the SFAC is strongly oscillating for any value of $n_D$ (see Fig. \ref{HAR_9_95}). Concerning the
smallest magnitude range ($[5.5;6]$), one can note the existence of two characteristic scales so that no
power law scaling holds. Those scales are of the order $10^{-2}$~year and $10^{-1}$~year. Fig. \ref{HAR_55_6}
should be compared with Fig. \ref{JMA_45_5} and \ref{JMA_55_6} for similar behaviors of the SFAC observed in the JMA catalog.
This strengthens our conjecture that the JMA catalog we used still contains numerous swarms that may alter the quality of our results.

Excluding the largest magnitude range, a linear fit of the dependence of the Omori exponent $p$ as a function of the main shock magnitude $M$ gives $p(M)=0.13M+0.14$. This fit is different from that 
 obtained with the binned stacking method, probably due to the limited magnitude range 
 available for the Harvard catalog. In any case, both methods confirm a strong magnitude
 dependence of the Omori exponent $p$. 
 
\section{Conclusion}

\begin{figure}
\begin{center}
\includegraphics[width=8cm]{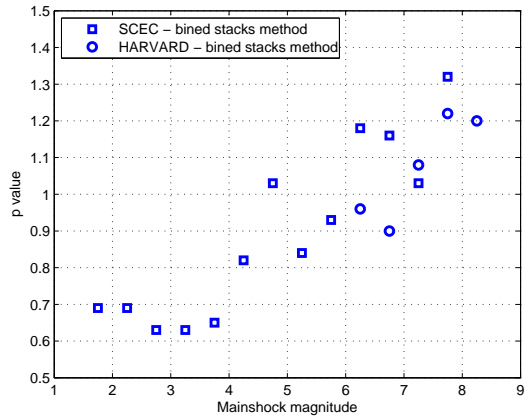}
\caption{\label{p_m_SCEC_HAR_bin} P(M) values obtained for the SCEC and 
Harvard catalogs with fits of binned time series.}
\end{center}
\end{figure}

\begin{figure}
\begin{center}
\includegraphics[width=8cm]{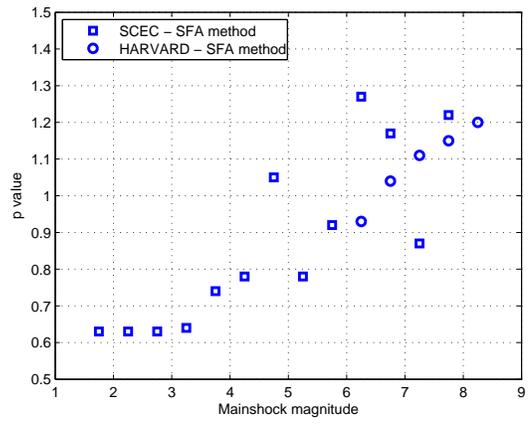}
\caption{\label{p_m_SCEC_HAR_sfa} Exponent $p(M)$ of the Omori law
obtained for the SCEC and Harvard catalogs with the SFA method.}
\end{center}
\end{figure}

We have introduced two methods to analyze the time-relaxation of aftershock sequences. One is based on standard binning methods, while the other one is based on the wavelet transform adapted to the present problem, leading to the Scaling Function Analysis (SFA) method.
We analyzed three different 
catalogs using a very simple declustering technique based on the definition of a magnitude-dependent 
space-time window for each
event. The SFA method showed that this declustering method was certainly sufficient as aftershock sequences
of small events are not contaminated by aftershock sequences triggered by previous larger events.
Both methods yield very similar results for each of the three catalogs, suggesting that our results 
are reliable.
The SFA method confirms the results of the binning method already presented by Ouillon and Sornette (2005), showing that the $p$-value of the Omori law 
increases linearly as a function 
of the magnitude of the main shock for the SCEC catalog. Those results are also in good agreement with the $p(M)$ dependence
measured for the Harvard CMT catalog (see Figs. \ref{p_m_SCEC_HAR_bin}  and \ref{p_m_SCEC_HAR_sfa} which present the results for both
catalogs and methods). The magnitude dependence of $p$ is much less obvious for the Japanese JMA catalog, but the SFA method 
clearly diagnosed that a rather significant number of swarm sequences are still mixed with more standard Omori-like sequences, so
that the obtained results should not be considered as representative of the latter. Overall, the extensive
analysis presented here strengthens the validity of the major prediction of the 
MSA model, namely that the relaxation rate of aftershock sequences is an Omori power law
with an exponent $p$ increasing significantly with the main shock magnitude. To the best
of our knowledge, the MSA model is the only one which predict this remarkable multifractal property.


\clearpage

{}

\clearpage

\begin{table}
\caption{$p$-values for the SCEC catalog obtained from fitting binned stacked sequences
with formula (\ref{fitapb}) (second column) and from using the
SFA method (third to fifth columns). $(n_B,n_D)$ correspond to the parameters used to define the mother
scaling function. $p(M)$ values in the second and fifth columns are plotted in Fig. \ref{p_M_SCEC}.}
\begin{center}
\begin{tabular}{lcccccc}
\hline
\\
Magnitude &    binned    & $(n_B,n_D)=(0,0)$ & $(n_B,n_D)=(3,0)$ & $(n_B,n_D)=(0,10)$ \\
\hline \\
$1.5-2.0$ &    $0.69 \pm 0.03$   &   $0.68$    &   $0.69$    &   $0.63$    \\
$2.0-2.5$ &    $0.69 \pm 0.02$   &   $0.63$    &   $0.63$    &   $0.63$    \\
$2.5-3.0$ &    $0.63 \pm 0.01$   &   $0.63$    &   $0.63$    &   $0.63$    \\
$3.0-3.5$ &    $0.63 \pm 0.02$   &   $0.58$    &   $0.57$    &   $0.64$    \\
$3.5-4.0$ &    $0.65 \pm 0.01$   &   $0.68$    &   $0.65$    &   $0.74$    \\
$4.0-4.5$ &    $0.82 \pm 0.02$   &   $0.78$    &   $0.77$    &   $0.78$    \\
$4.5-5.0$ &    $1.03 \pm 0.03$   &   $0.99$    &   $1.02$    &   $1.05$    \\
$5.0-5.5$ &    $0.84 \pm 0.04$   &   $0.94$    &   $0.54$    &   $0.78$    \\
$5.5-6.0$ &    $0.93 \pm 0.03$   &   no value  &  no value   &   $0.92$    \\
$6.0-6.5$ &    $1.18 \pm 0.05$   &   no value  &  no value   &   $1.27$    \\
$6.5-7.0$ &    $1.16 \pm 0.03$   &   no value  &  no value   &   $1.17$    \\
$7.0-7.5$ &    $1.03 \pm 0.02$   &   no value  &  no value   &   $0.87$    \\
$7.5-8.0$ &    $1.32 \pm 0.17$   &   no value  &  no value   &   $1.22$    \\
\\
\hline
\end{tabular}
\end{center}
\end{table}

\begin{table}
\caption{$p$-values for the JMA catalog obtained by fitting binned stacked sequences 
(second column) and the SFAC (third to fifth columns). $(n_B,n_D)$ correspond to the parameters used to define 
the mother scaling function. $p(M)$ values in the second and fifth columns are plotted in Fig. \ref{p_M_JMA}.}
\begin{center}
\begin{tabular}{lcccccc}
\hline
\\
Magnitude &    binned    & $(n_B,n_D)=(0,0)$ & $(n_B,n_D)=(1,0)$ & $(n_B,n_D)=(0,10)$ \\
\hline \\
$2.5-3.0$ &    $0.74 \pm 0.04$   &   $0.65$    &   $0.66$    &   $0.70$    \\
$3.0-3.5$ &    $0.87 \pm 0.06$  &   $0.78$    &   $0.79$    &   $0.78$    \\
$3.5-4.0$ &    $0.84 \pm 0.03$   &   $0.86$    &   $0.86$    &   $0.88$    \\
$4.0-4.5$ &    $0.76 \pm 0.05$   &   $0.76$    &   $0.77$    &   $0.77$    \\
$4.5-5.0$ &    $0.81 \pm 0.04$   &   $0.71$    &   $0.70$    &   $0.77$    \\
$5.0-5.5$ &    $0.95 \pm 0.04$   &   $0.75$    &   $0.73$    &   $0.84$    \\
$5.5-6.0$ &    $1.02 \pm 0.15$   &   $0.78$    &   $0.76$    &   $0.87$    \\
$6.0-6.5$ &    $0.92 \pm 0.04$   &   no value  &  no value   &   $0.97$    \\
$6.5-7.0$ &    $0.99 \pm 0.07$   &   no value  &  no value   &   $0.95$    \\
$7.0-7.5$ &    $1.22 \pm 0.07$   &   no value  &  no value   &   $0.93$    \\
$7.5-8.0$ &    $0.89 \pm 0.04$   &   no value  &  no value   &   $1.02$    \\
$8.0-8.5$ &    $1.18 \pm 0.13$   &   no value  &  no value   &   $1.26$    \\
\\
\hline
\end{tabular}
\end{center}
\end{table}

\begin{table}
\caption{$p$-values for the HARVARD catalog obtained by fitting the binned stacked sequences (second column) and the
SFAC (third column). $p(M)$ values are plotted in Fig. \ref{p_m_HAR}.}
\begin{center}
\begin{tabular}{lcccccc}
\hline
\\
Magnitude &    binned    &     SFA     \\
\hline \\
$5.5-6.0$ &   no value  &   no value  \\
$6.0-6.5$ &    $0.96 \pm 0.04$   &    $0.93$   \\
$6.5-7.0$ &    $0.90 \pm 0.04$   &    $1.04$   \\
$7.0-7.5$ &    $1.08 \pm 0.08$   &    $1.11$   \\
$7.5-8.0$ &    $1.22 \pm 0.08$   &    $1.15$   \\
$8.0-8.5$ &    $1.20 \pm 0.24$   &    $1.20$   \\
$8.5-9.0$ &   no value  &   no value  \\
$9.0-9.5$ &    $0.69 \pm 0.03$   &   no value  \\
\\
\hline
\end{tabular}
\end{center}
\end{table}

\clearpage

\begin{figure}
\begin{center}
\includegraphics[width=8cm]{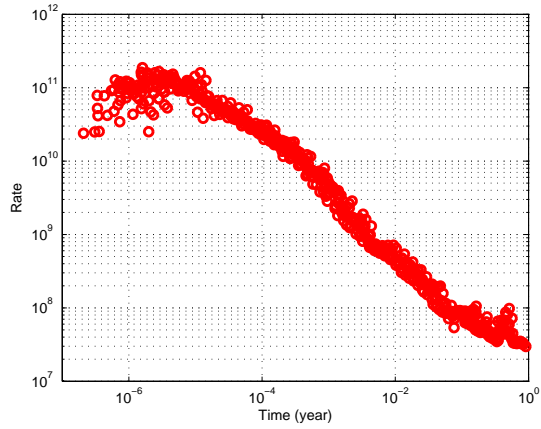}
\caption{\label{SCEC_raw_4_45} Binned stacked time series of sequences triggered by main events with 
magnitudes $M$ within the interval $[4;4.5]$ in the SCEC catalog. 
This plot shows all binned series corresponding to all the 20 binning factors from $r=1.1$ to $r=3$.}
\end{center}
\end{figure}

\begin{figure}
\begin{center}
\includegraphics[width=8cm]{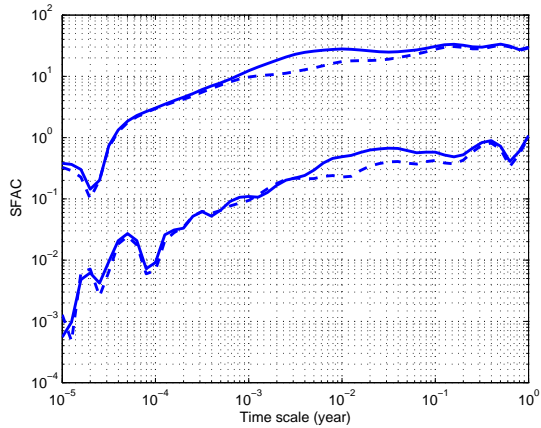}
\caption{\label{SCEC_raw_4_45_analysis} Scaling function analysis coefficient (SFAC) of the time series shown in Fig.~\ref{SCEC_raw_4_45}. The two top curves correspond to $n_B=0$, the bottom curves to $n_B=3$.
The solid curves refer to the data sets which include events in the Imperial Valley zone. The dashed curves correspond to the data sets excluding those events.}
\end{center}
\end{figure}

\clearpage

\clearpage

\clearpage

\clearpage

\clearpage

\clearpage

\clearpage

\clearpage

\clearpage

\clearpage

\clearpage

\clearpage

\clearpage

\clearpage

\clearpage

\clearpage

\clearpage

\clearpage

\clearpage

\clearpage

\clearpage

\clearpage


\appendix
\section{The scaling function analysis}

The scaling function analysis described in this Appendix develops a fitting procedure
that removes the impact of non-Omori law terms in expression (\ref{fitapbi})
as described by the polynomial expansion $\sum_{i=0}^{n_{B}} b_i t^i$ describing
a tectonic background contribution and the impact of aftershock sequences of 
main events preceding the main shock under investigation.

\subsection{Construction of the mother scaling functions (MSF)}

The first step in developing the scaling function analysis, which is inspired
from the well-known wavelet transform, is
to define a mother scaling function (hereafter MSF) that we shall name $\Psi$, and define
the associated scaling function analysis coefficient $C(s=1)$ of the rate function $N(t)$ by
\be
C(s=1) = \int_0^{\infty} \Psi(t) N(t) dt~.
\ee
We then define
a set of daughter scaling functions $\Psi(\frac{t}{s})$,  where $s$ is a time scale parameter 
(that should not be mistaken for the fluctuations of the stress described in the MSA model),
and compute the associated scaling function analysis coefficients (herafter SFAC):
\be
C(s) = \int_0^{\infty} \Psi \left(\frac{t}{s} \right) N(t) dt~.
\label{gnn;;tg}
\ee
In the analogy with a wavelet transform, the SFAC is nothing but the wavelet coefficient measured at the time location $t=0$.
If we now assume that $N(t) = A \cdot t^{-p} + B(t)$, we have (using a simple change of variable):
\be
C(s) = s^{1-p} A \int_0^{\infty} \Psi(t) t^{-p} dt + \int_0^{\infty} \Psi \left(\frac{t}{s} \right) B(t) dt~.
\label{scaleq}
\ee
For a given stacked series, the first integral is independent of $s$, so that
the variation of $C(s)$ with $s$ stems from two contributions: a power-law term with exponent $1-p$, plus a term depending on the shape of $B(t)$ and $\Psi(t)$.

We choose the MSF so as to respect the following constraints. First of all, $\Psi$ is designed to analyze the scaling properties
of aftershock sequences, {\it i.e} of sequences that are triggered shortly after a main shock. It
is thus rational to choose $\Psi$ so that it is defined only for positive times, and that its modulus 
decays rather quickly with time (so that it focuses on short-term rather than long-term scales).
Secondly, the power law scaling behavior of $C(s)$ results from the singularity of $N(t)$ at $t=0$.
In real catalogs, of course, the seismicity rate does not diverge at small times, and one rather observes a roll-off  of $N(t)$, mainly due to the incompleteness of the catalog.
The Omori law thus breaks down for too short times, so that the scaling analysis presented in Eq. \ref{scaleq} doesn't hold. In order to circumvent this effect, we impose that $\Psi(t=0)=0$, so that aftershocks
occurring at short times will have a negligible weight in the computation of $C(s)$, preserving the
announced scaling properties of the SFAC. Thirdly, in order to measure $p$ more easily, we impose that $\Psi$ should filter out polynomials, so that the second term in the right-hand side of Eq. \ref{scaleq} gives a vanishing contribution to $C(s)$. These three conditions are fulfilled with the
following construction
\be
\Psi(t) = \sum_{i=0}^{n_P} a_i t^i \exp \left( -a t^2 \right)~,
\ee
where the coefficients $a$ and $a_i, i=0, ..., n_P$ are determined as follows.
For all integer values $j=0,...,n_B$, we impose that the function $\Psi$ obeys the conditions
\be
\int_0^{\infty} t^j \Psi(t) dt = 0~,
\label{hnkkoef}
\ee
If we find the corresponding coefficients $a_i$'s, then our goal of removing the influence
of the non-stationary background and of previous main shocks will be fulfilled.
Expression (\ref{hnkkoef}) leads to
\be
\sum_{i=0}^{n_P} I_{i+j} a_i =0~,
\label{eqijai}
\ee
where
\be
I_m = \int_0^{\infty} t^m \exp \left(-a t^2 \right) dt = \frac{\Gamma[(m+1)/2]}{2a^{(m+1)/2}}~.
\ee
As equation \ref{eqijai} must hold for all $j$ values between $0$ and $n_B$, and as we also impose $\Psi(0)=0$, the set of conditions (\ref{eqijai})
defines a linear system of $n_B+2$ equations which can be solved to obtain the $n_P+1$ unknowns $a_i$. 
In order to obtain a non-degenerate solution, we impose $n_P=n_B+2$ and arbitrarily fix $a_{n_P}= \pm 1$.  The sign of $a_{n_P}$ is chosen so that the most extreme value of the MSF is positive. 
The MSF is then normalized so that its maximum value is $1$.
In order to fully define the MSF, we still have to specify the two parameters
$a$ and $n_B$. In the remaining of this paper, we shall fix $a=5$ yr$^{-2}$ (which ensures a good
temporal localization of $\Psi$). As for the parameter $n_B$, it requires a specific
discussion for each of the studied catalogs.

Figure \ref{MSFexemples} shows the shape of the function $\Psi$ for 
\begin{itemize}
\item $n_B=0$ $(n_P=2)$  (which filters out only constant background terms $B(t)=b_0$), 
\item $n_B=1$ $(n_P=3)$ (which filters out linear trends like  $B(t)=b_0+b_1t$), and 
\item $n_B=2$ $(n_P=4)$ (which filters out quadratic trends like $B(t)=b_0+b_1t+b_2t^2$). 
\end{itemize}
The higher the order of the polynomial that needs to be filtered out, 
the more oscillating is the MSF. It is noteworthy that the shape of the MSF is independent of
the precise shape of the function $B(t)$ (and of its coefficients $b_i$). Only the degree $n_B$ of the polynomial is needed to determine the corresponding MSF.

\begin{figure}
\begin{center}
\includegraphics[width=8cm]{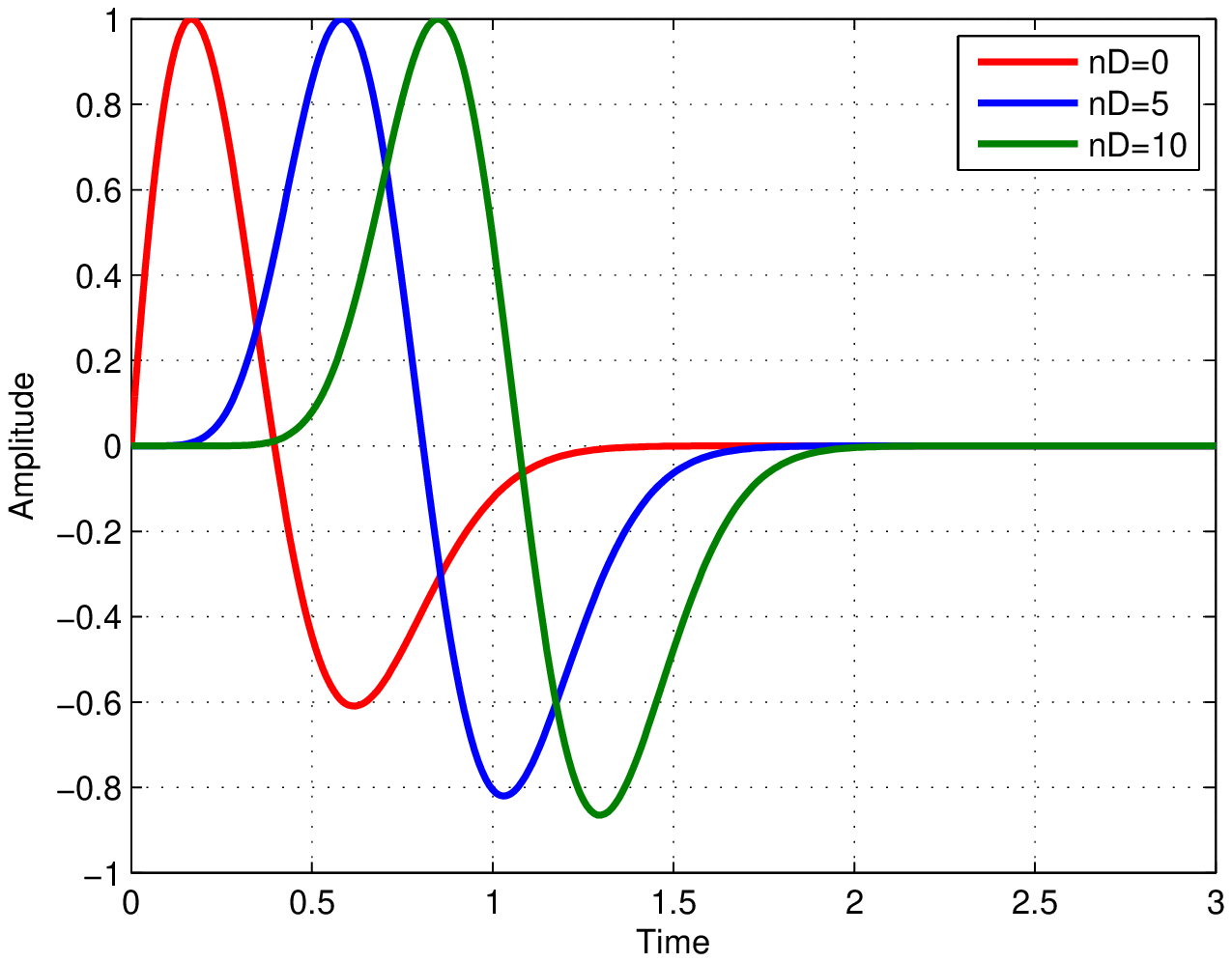}
\caption{\label{MSFexemples2} Plot of three different MSFs corresponding to different values of $n_D$ (using $n_B=0$).}
\end{center}
\end{figure}

Imposing $\Psi(0)=0$ decreases the influence of the incompleteness of real catalogs at short times
after main shocks. However, for large 
main shocks, the corresponding roll-off in the Omori law can extend over weeks or months after the main shock. The MSF we just
introduced may then prove unable to provide anything but spurious SFAC scaling estimations. We thus 
introduce additional constraints to build a suitable MSF with less sensitivity to the early times. Specifically, we impose in addition that all derivatives of $\Psi$ up to order $n_D$
vanish at $t=0$. To obtain a non-degenerate system of equations determining the coefficients of the expansion  $\Psi$, we have  $n_P=n_B+2+n_D$, and impose $a_i=0$ for $i=0,...,n_D$. Fig. \ref{MSFexemples2} shows the MSFs for $n_B=0$ and $n_D=0,5,10$. At short times, $\Psi$ takes negligible values over a time
interval whose width increases with $n_D$. We shall see in our analysis of real catalogs that 
this set of MSF will provide much better estimates of $p$ 
in a few peculiar situations.

Another advantage of using the scaling function analysis is that we do not need to bin
the time series of aftershock rates. Indeed, consider a given sequence of $N_{\rm aft}$ aftershocks occurring at successive times $t_1, ..., t_k, ..., t_{N_{aft}}$ after their  triggering main shock.
By definition, the aftershock rate is a sum of Dirac functions
\be
N(t) = \sum_{k=1}^{N_{aft}} \delta(t-t_k)~,
\ee
which yields the SFAC
\be
C(s) = \sum_{k=1}^{N_{aft}} \Psi \left(\frac{t_k}{s} \right)~,
\ee
according to the definition (\ref{gnn;;tg}).
The estimation of $C(s)$ is a simple discrete sum without any need
for some intermediate manipulation of the data.

\subsection{Scaling function analysis of synthetic cases}

We now apply the scaling function analysis to a variety of synthetic cases to demonstrate its efficiency. These synthetic tests will 
define benchmarks that will be used to interpret the results obtained for real catalogs.
We use $n_D=0$ to build the MSFs, except when explicitely mentioned.

\subsubsection{Omori law with a quadratic background term}

\begin{figure}
\begin{center}
\includegraphics[width=8cm]{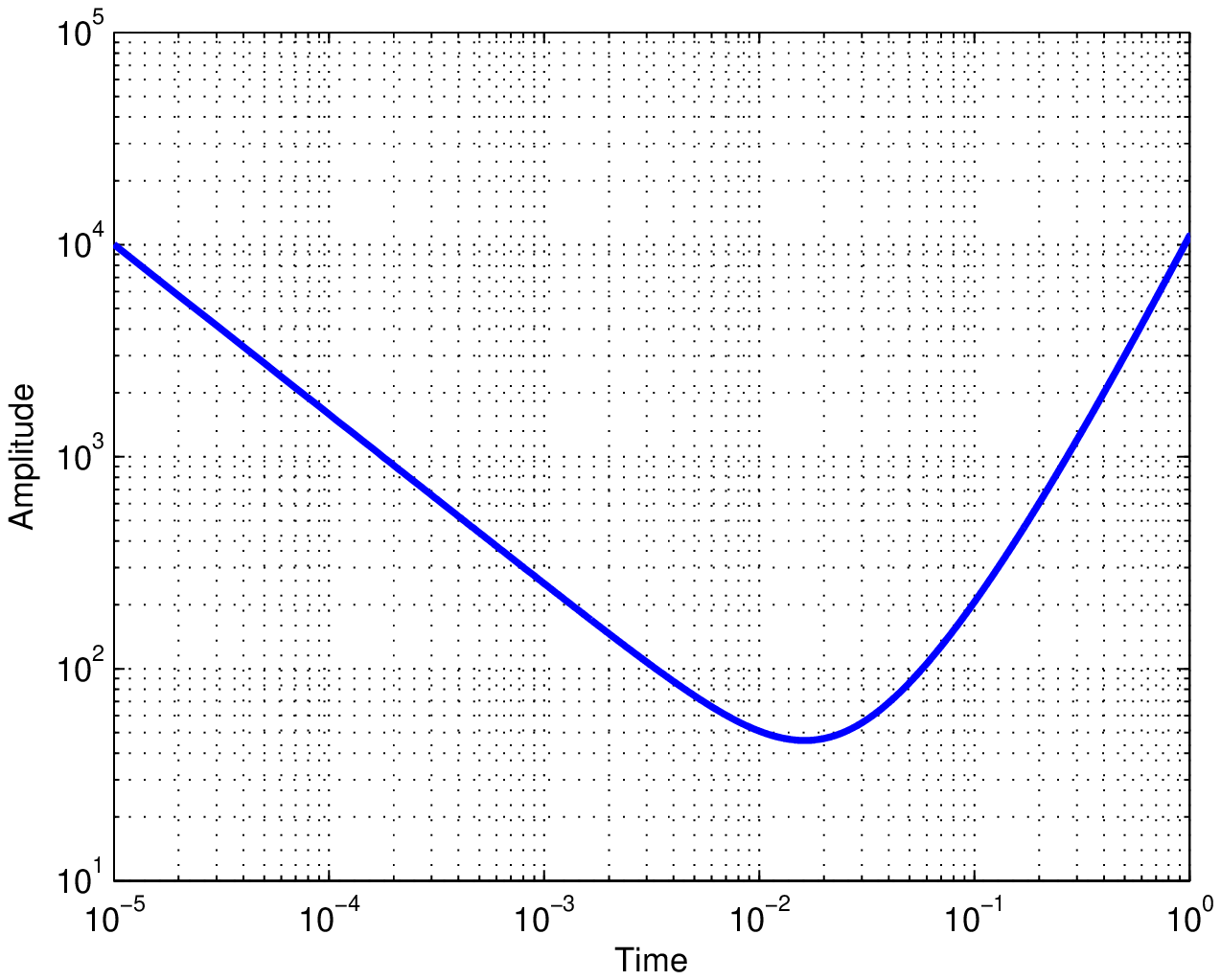}
\caption{\label{synthplpol} Synthetic time series $N(t) = t^{-0.8} + 10^3t + 10^4t^2$.}
\end{center}
\end{figure}

While not directly similar to a real case, the first example illustrates the power
of the Scaling Function Analysis. The synthetic time series that we choose to analyze is
generated with the following formula
\be
N(t) = t^{-0.8} + 10^3t + 10^4 t^2 
\label{hjbpwdc}
\ee
over the interval  $[10^{-5};1]$ and is plotted in Fig. \ref{synthplpol}. 
This interval (where the time unit is $1$ year)
is similar to those used for real time series analyzed in the text. The sampling rate
is $10^{-5}$. It first exhibits a power-law decay followed
by an explosive increase of $N(t)$. 

\begin{figure}
\begin{center}
\includegraphics[width=8cm]{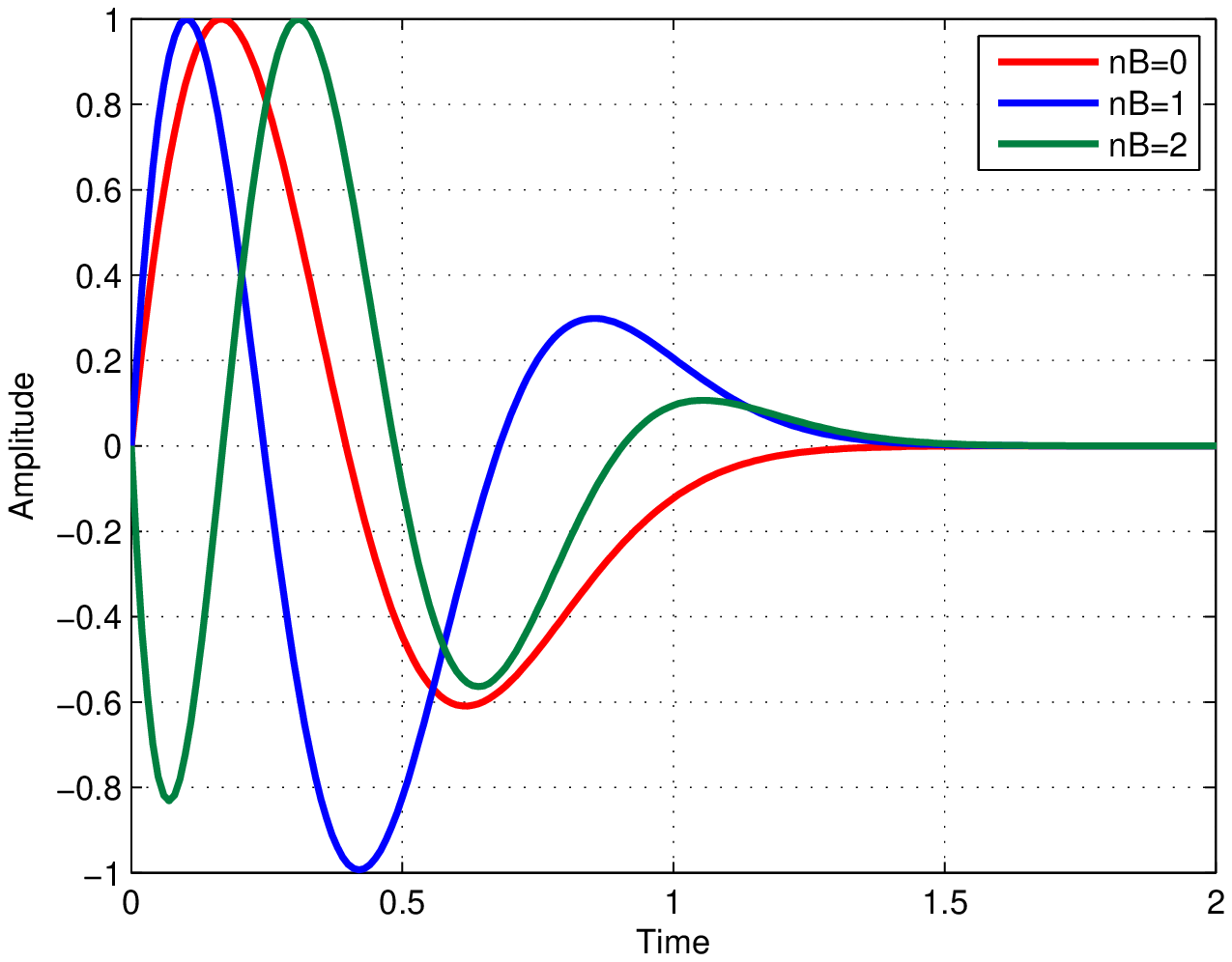}
\caption{\label{MSFexemples} Plot of three different MSFs corresponding to different values of $n_B$.}
\end{center}
\end{figure}

\begin{figure}
\begin{center}
\includegraphics[width=8cm]{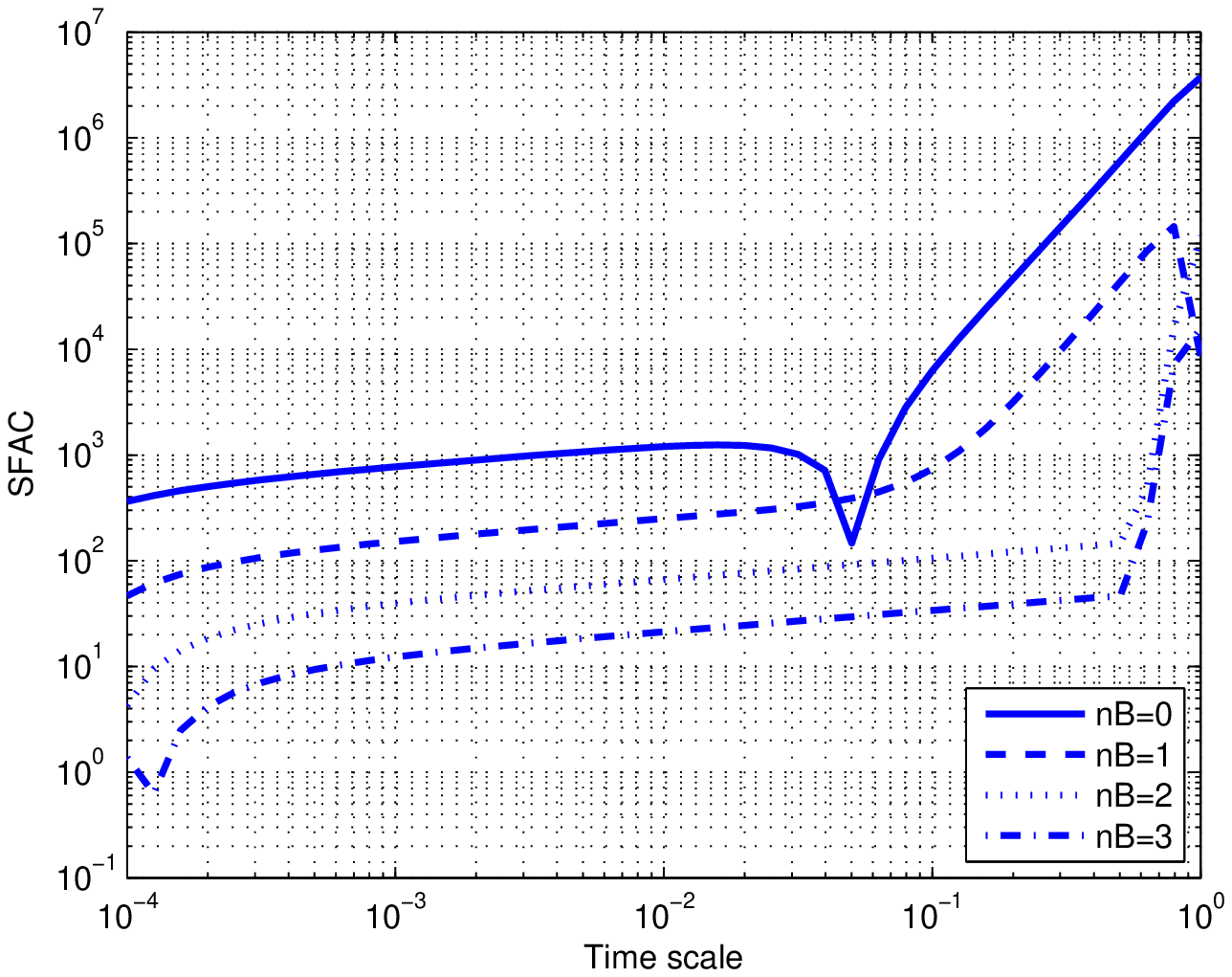}
\caption{\label{synthplpolanalysis} Scaling function analysis coefficient (SFAC) of the time series shown on Fig. \ref{synthplpol} as a function of the time scale $s$.
Each curve corresponds to a given value of $n_B$ used to build the corresponding MSF (mother scaling function).}
\end{center}
\end{figure}

In order to analyze the time series defined by
(\ref{hjbpwdc}), we used four different MSFs, each function corresponding to
a different value of $n_B$ ($0,1,2$ or $3$, the first three being represented in Fig.~\ref{MSFexemples}). 
The results are plotted on Fig. \ref{synthplpolanalysis}.

According to the previous section and expression (\ref{scaleq}), 
a linear behavior in the $log-log$ plot of Fig.~\ref{MSFexemples} reveals an underlying power-law with exponent $1-p$. For each curve, the power-law scaling is absent at
the smallest scales which are comparable with the sampling rate, reflecting 
signal digitization effect.
The powerlaw scaling also breaks down at the largest scales, as $N(t)$ is defined over a finite time range (a finite size effect), whereas the daughter scaling functions can take values significantly different from $0$ over a larger range.
For example, Fig.~\ref{MSFexemples} shows that the chosen MSF remains significant
in the interval $[0;1.5]$. 
\begin{enumerate}
\item For $n_B=0$, the MSF erases only the constant background contribution, which
is anyway absent in the present example for $N(t)$. As a consequence, a power-law scaling holds at small scales (up to about $10^{-2}$)
with an exponent close to $0.2$ (as expected from the prediction $1-p$ for $p=0.8$). Scaling then breaks down due to the existence of both the linear and 
quadratic contributions. At large scales, the exponent is close to $3$, which means that the corresponding $p$-value
is close to $-2$, which is exactly the signature of the quadratic term. 
\item For $n_B=1$, the linear trend is erased,
so that the power-law scaling now extends over a slightly larger range of time scales, with the same
exponent, but the quadratic
trend influence remains. 
\item For $n_B=2$, the influence of the quadratic trend should be also erased, which is indeed the case as the power-law trend
with exponent $0.2$ now extends up to a scale $s \approx 0.5$. 
\item If we now increase $n_B$ to $3$, we see that the scaling
range and exponent are the same, as there is indeed no contribution of higher degree to filter out (we obtain the
same results using scaling functions with even larger $n_B$ values). 
\end{enumerate}
Using this analysis,
we are thus able to retrieve that the degree of the polynomial background term is $n_B=2$, and that the Omori exponent is $p=1-0.2=0.8$.

\subsubsection{Gamma law with constant background term}

\begin{figure}
\begin{center}
\includegraphics[width=8cm]{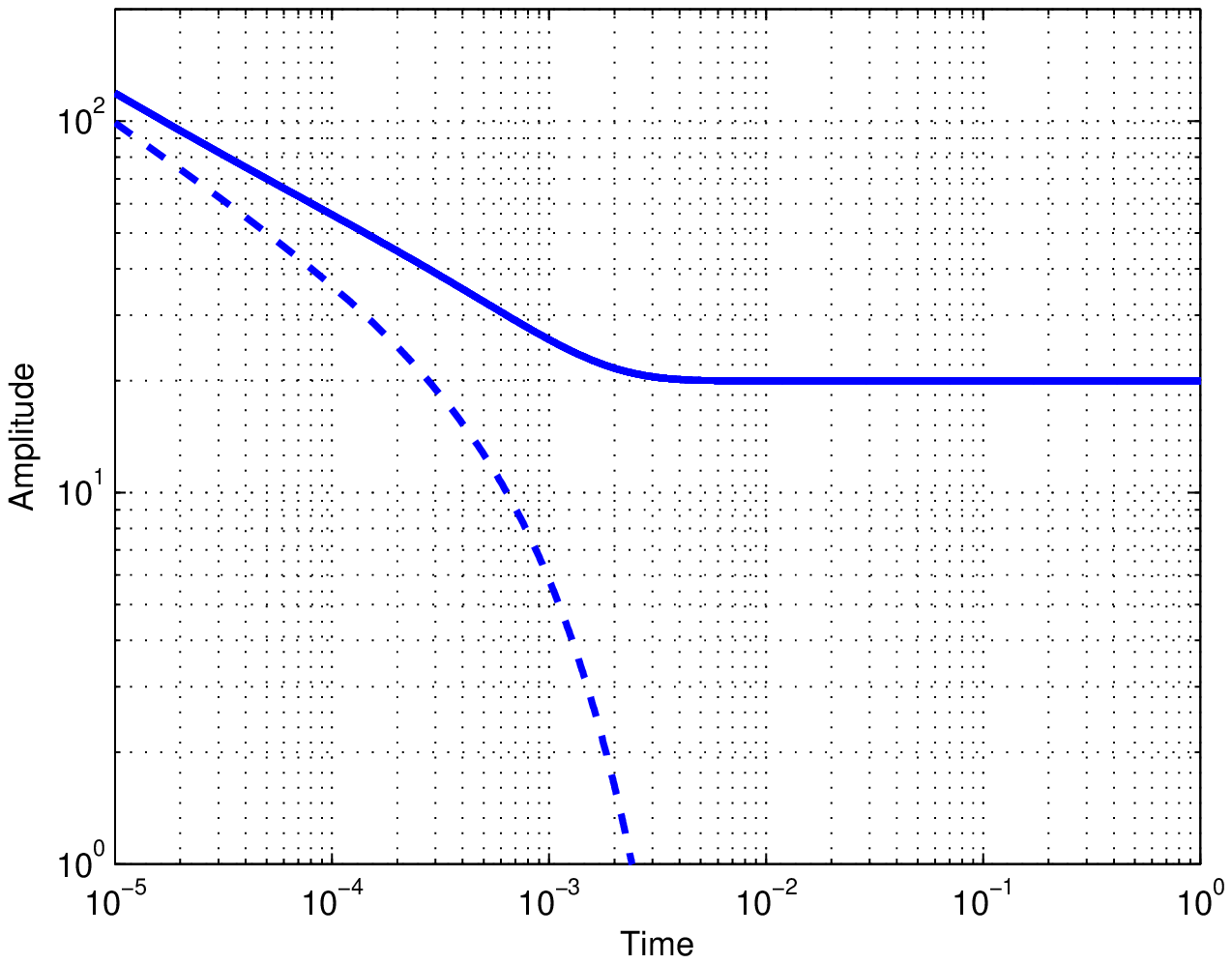}
\caption{\label{synthgamma} Gamma function (dashed line) defined by 
expression (\ref{jghmmb}) and Gamma function with an added constant background (continuous line).}
\end{center}
\end{figure}

\begin{figure}
\begin{center}
\includegraphics[width=8cm]{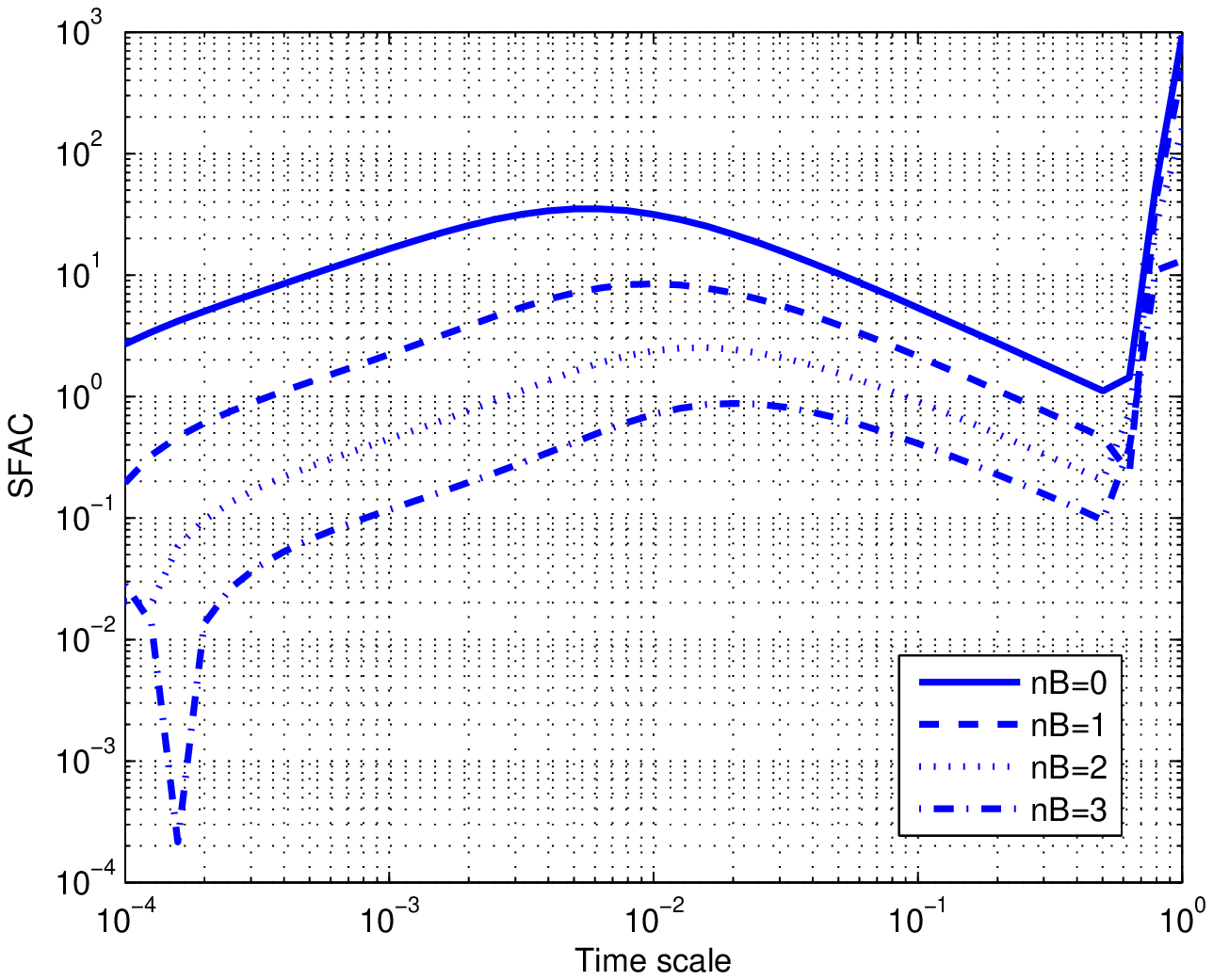}
\caption{\label{synthgammaanalysis} Scaling function analysis of the Gamma function with constant background term
shown in Fig. \ref{synthgamma}. Each curve corresponds to a given value of $n_B$ used to build the corresponding MSF.}
\end{center}
\end{figure}

Figure \ref{synthgamma} shows a dashed-line plot of the gamma function
\be
N(t) = t^{-0.4} \exp \left(-\frac{t}{\tau_0} \right)~,
\label{jghmmb}
\ee
It exhibits a power law behavior at small times, followed by an exponential roll-off at large times. 
This law could describe the time decay of swarms in
volcanic areas, for example, with $\tau_0$ being the characteristic duration of the swarm (here we took
$\tau_0=10^{-3}$).  The continuous line on the same figure shows the same function to which a constant
background term $B=20$ has been added. Note that this new time series could very easily be mistaken
for a pure Omori-law with a constant background. We performed a scaling function analysis of this last
time-series, and Fig. \ref{synthgammaanalysis} shows the obtained results using the same four scaling 
functions as above.

As the only polynomial trend in $N(t)$ is a constant term, all curves exhibit the same scaling behavior,
which results from two complementary effects. The first effect is that 
the gamma function can be described as an effective Omori-like power law with a tangent exponent $p$ that continuously increases
with time. Since the effective exponent is smaller than $1$ at small times and larger than $1$ at large times, the SFAC first increases and then decreases
with time scale. The second effect is of a different nature. Fig. \ref{synthgamma} illustrates that the Gamma
function takes values significantly different from zero within a finite interval spanning roughly $[0;10^{-2}]$.
As the time scale increases, the associated SFAC will thus increase as the daughter scaling function 
progressively enters a kind of resonance with this finite-size feature. The maximum resonance is obtained when the
scale of the daughter scaling function is of the order of $10^{-2}$. Further increasing
the time scale, the resonance amplitude decreases, leading to a decreasing SFAC. The interplay between those two effects leads to a reasonably well-defined maximum of the dependence of the SFAC as a function of the scale $s$, providing a rough estimate of $\tau_0$. The drawback is that the left side of the power-law scaling behavior in Fig. \ref{synthgammaanalysis} 
is distorted and doesn't provide an accurate measure of $p$ (in the present example, the measured $p$ value is $0.2$, compared with the true value $p=0.4$).
Overall, we conclude that the scaling function analysis clearly reveals the existence of a characteristic scale which precludes
the existence of a genuine Omori scaling over the whole range of time. In this sense, the scaling function analysis provides a useful diagnostic.

\subsubsection{Mix of gamma law, Omori law and constant background term}

\begin{figure}
\begin{center}
\includegraphics[width=8cm]{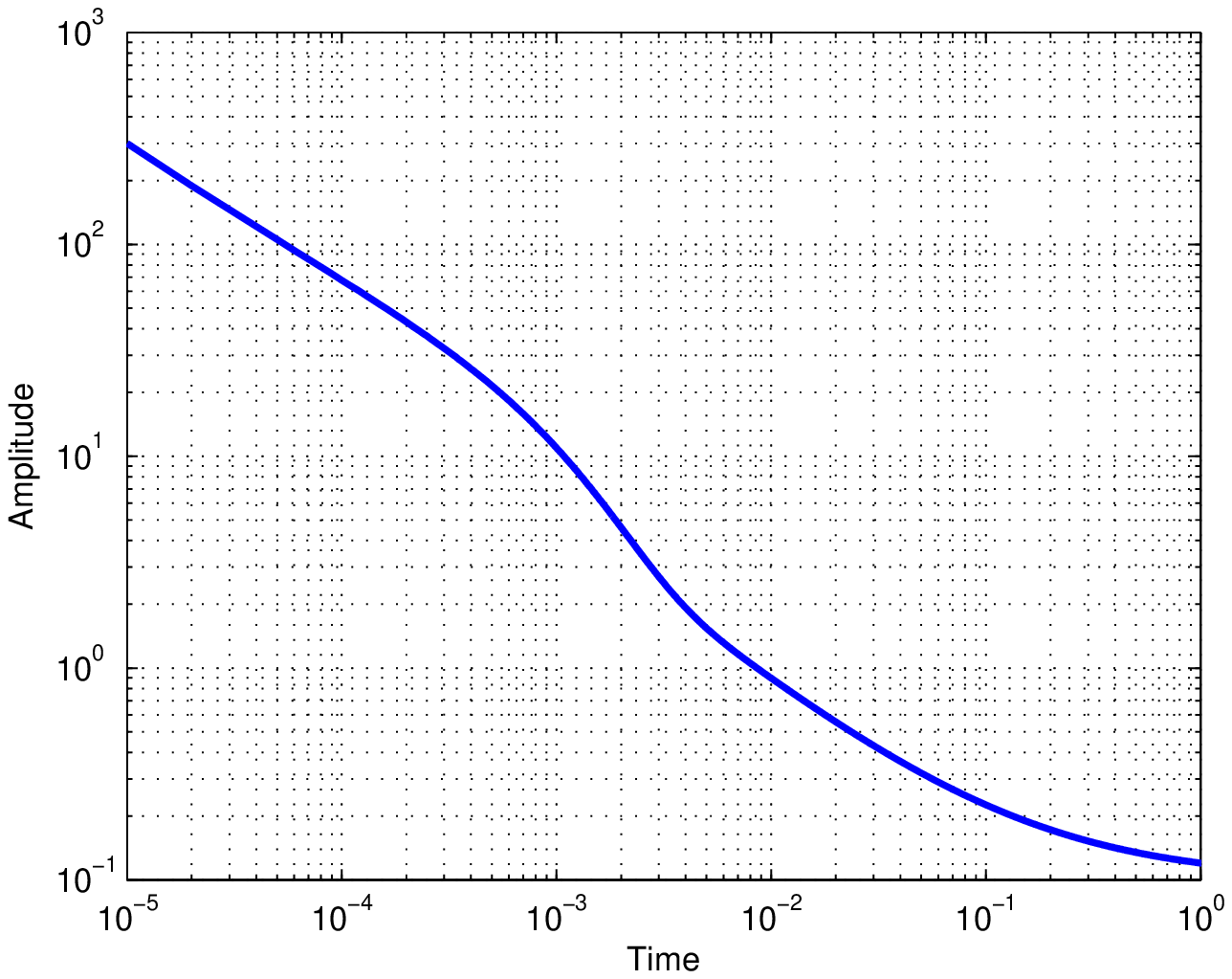}
\caption{\label{synthgammapowback} Time series defined as the sum of a Gamma function, an Omori-law and a constant
background term.}
\end{center}
\end{figure}

The next synthetic example we wish to present is a sum of an Omori-like power law, a gamma-law and a 
constant background term:
\be
N(t) = 0.02~ t^{-0.8} + t^{-0.4} \exp \left(-\frac{t}{\tau_0} \right) + 0.1~,
\label{mjhinpopr}
\ee
with $\tau_0=10^{-3}$. This function can describe the mixture of pure Omori-like sequences with
swarm sequences in the presence of a constant background noise within the same data set. This function is plotted in 
Fig. \ref{synthgammapowback} and displays a very complex time behavior, that is sometimes observed in real 
time series (see Fig. \ref{SCEC_raw_4_45}). When observing such time series, one
generally tries to fit it with an Omori-law, considering that its fluctuations in
$log-log$ scale are just of statistical nature.

\begin{figure}
\begin{center}
\includegraphics[width=8cm]{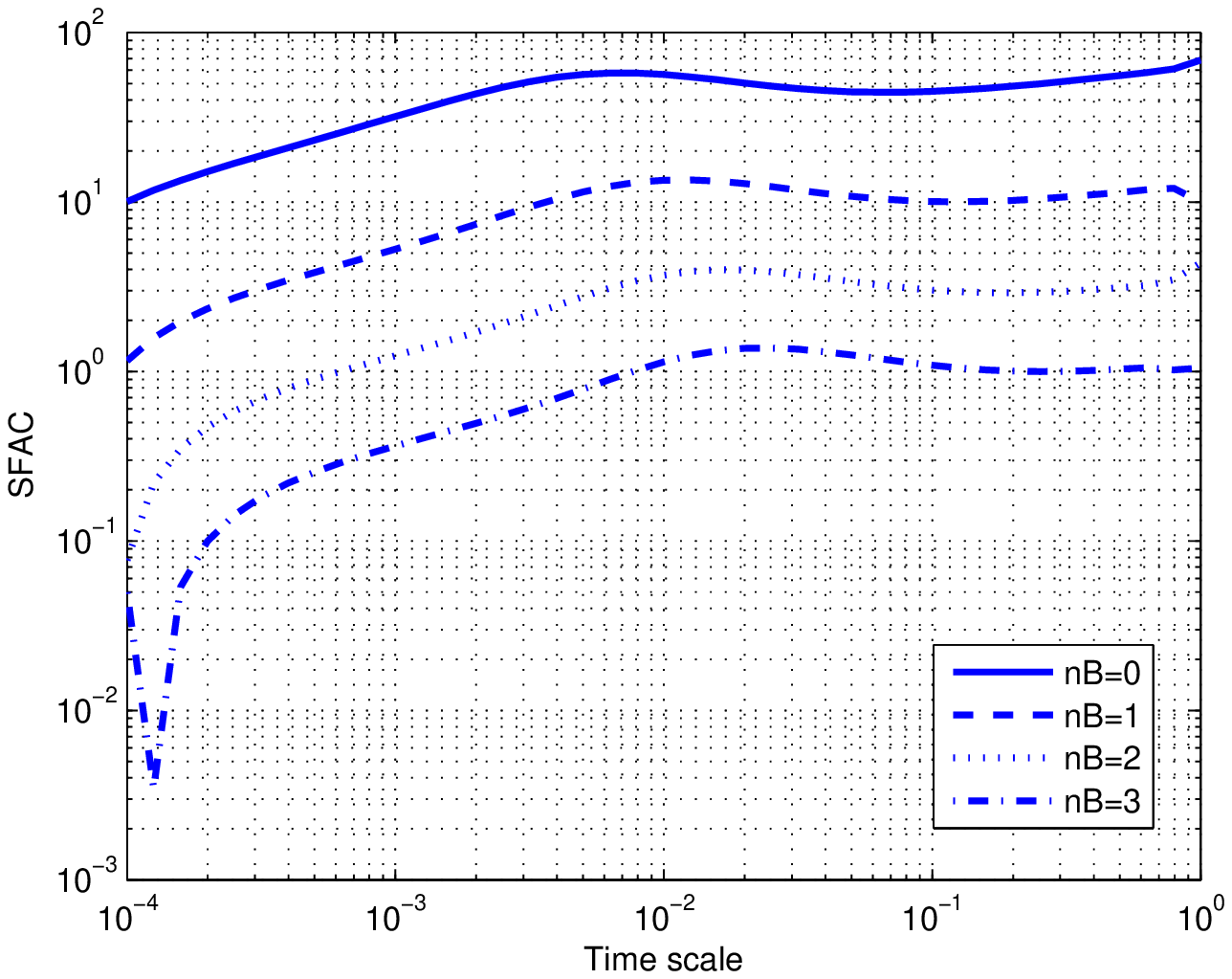}
\caption{\label{synthgammapowbackanalysis} Scaling function analysis of the time series
shown on Fig. \ref{synthgammapowback}. Each curve corresponds to a given value of $n_B$ used to build the corresponding MSF.}
\end{center}
\end{figure}

Using the same approach as before, Fig. \ref{synthgammapowbackanalysis} shows the results of the scaling function analysis on this function (\ref{mjhinpopr}).
The obtained trend for small time-scales is the same whatever the chosen value
for $n_B$, and is compatible with a power-law with an exponent close to $0.5$ 
(corresponding to $p=0.5$). The difference from the real exponent $p=0.4$ is due to the same effects as in the case of the single gamma law discussed above. All curves then 
go through a maximum, and then decrease. This reveals the existence of a characteristic scale (which is
$\tau_0=10^{-3}$ for expression (\ref{mjhinpopr})). Then, for time scales larger than $10^{-1}$, all curves increase  again. This behavior is due to the fact that, at
such time scales, the gamma function is now negligible compared with the Omori-like contribution, 
and the SFAC exhibits a positive slope compatible with the true exponent $p=0.8$ of the Omori law. 
As $n_B$ increases, the maximum is shifted to
larger and larger time scales, which implies that the positive slope to the right
of this maximum which is associated with the Omori law
can be observed only at larger and larger scales. As the time scales are limited
by the time range of $N(t)$, the slope corresponding to the Omori component can
not always be measured with sufficient accuracy for the larger $n_B$ values. However, we qualitatively
find the same shape for all values of $n_B$.

\subsubsection{The modified Omori-Utsu law with constant background term}

\begin{figure}
\begin{center}
\includegraphics[width=8cm]{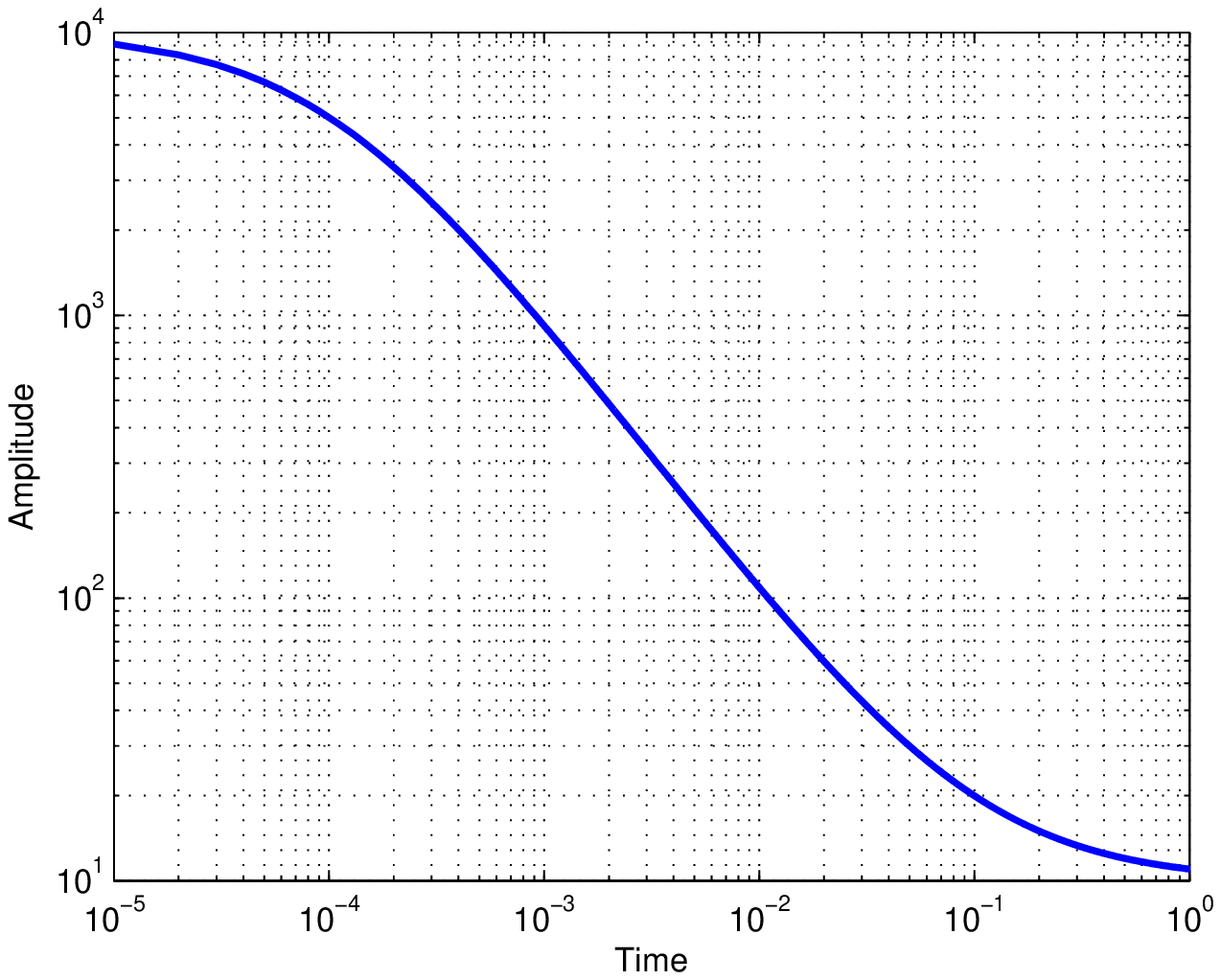}
\caption{\label{synthmodomori} Modified Omori law defined by (\ref{khkihts}) with a constant background term, for $p=1$ and $\tau_0=10^{-4}$.}
\end{center}
\end{figure}

The modified Omori-Utsu law has been introduced as a convenient way to model the nearly constant
seismicity rate after a large event at short time scales. We thus considered the following
decay function:
\be
N(t) = (t+\tau_0)^{-p} + 10 ~,
\label{khkihts}
\ee
which is shown on Fig. \ref{synthmodomori} for $p=1$ and $\tau_0=10^{-4}$. 

\begin{figure}
\begin{center}
\includegraphics[width=8cm]{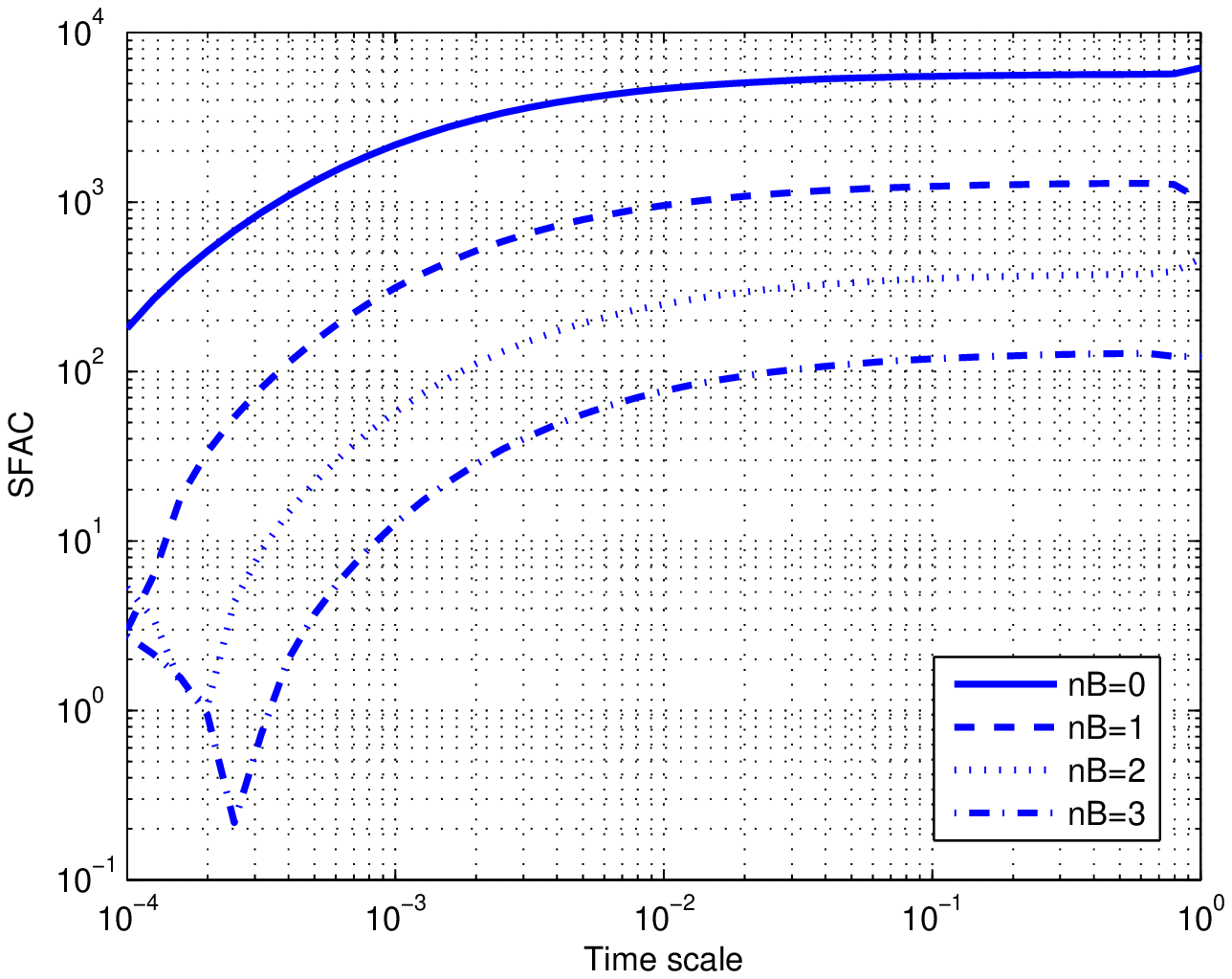}
\caption{\label{synthmodomorianalysis} Scaling function analysis of the time series
shown on Fig. \ref{synthmodomori}. Each curve corresponds to a given value of $n_B$ used to build the 
corresponding MSF.}
\end{center}
\end{figure}

Results of the scaling function analysis are shown in Fig. \ref{synthmodomorianalysis}.
The SFACs first increase non-linearly (in $log-log$ scales) up to a
scale of about $10^{-1}$, then behave as power-laws with the associated exponent $1-p=0$. Note that
the transition from non-powerlaw to powerlaw scaling is very smooth and thus offers a very small time
scale range to estimate $p$, despite the fact that $\tau_0$ is small.

\begin{figure}
\begin{center}
\includegraphics[width=8cm]{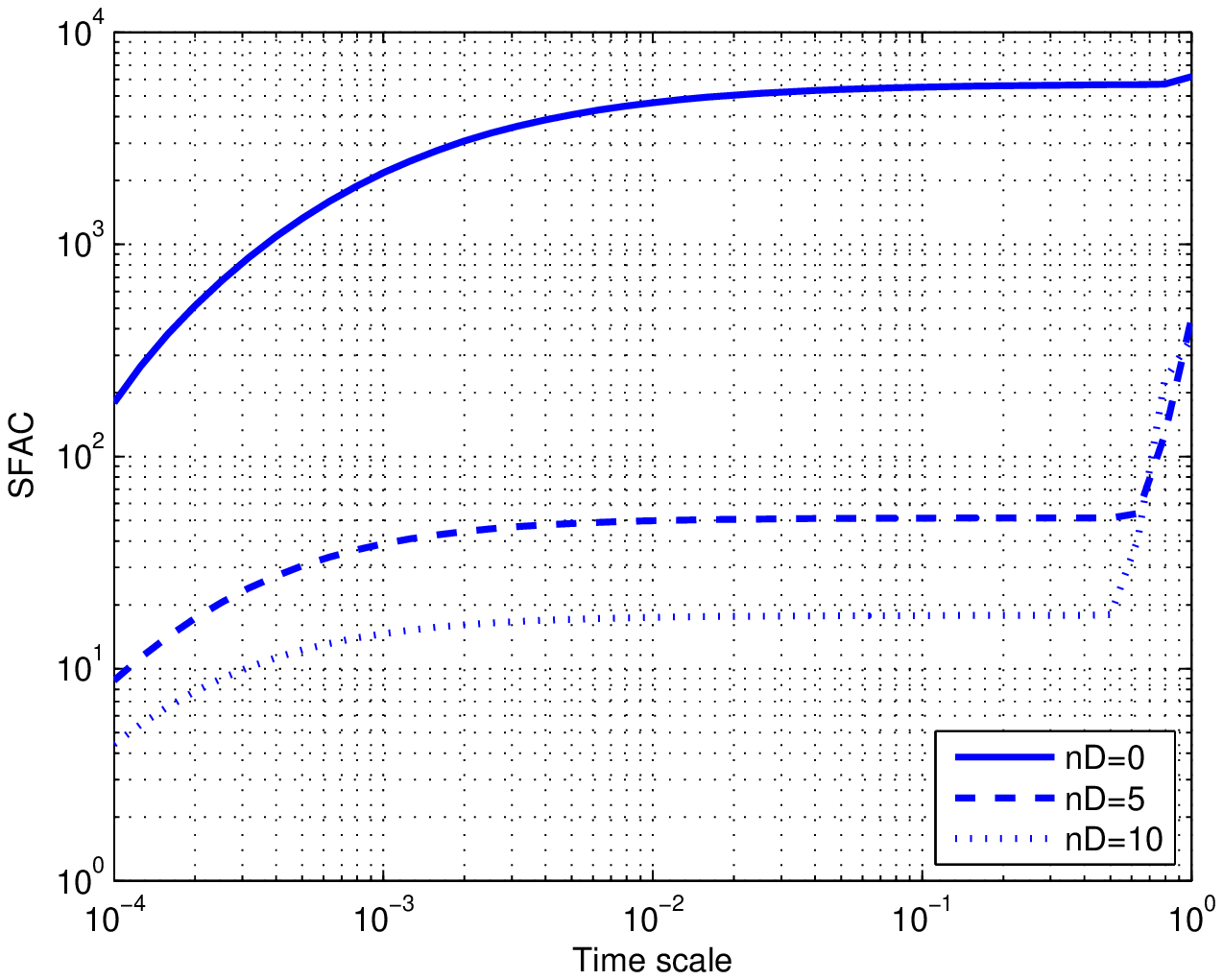}
\caption{\label{synthmodomorianalysis2} Scaling function analysis of the time series
shown in Fig. \ref{synthmodomori}. Each curve corresponds to a given value of $n_D$ used to build the 
corresponding MSF: $n_D$ is the number of orders of derivatives of $\Psi$ that vanish at $t=0$.}
\end{center}
\end{figure}

We also performed a SFA using $n_B=0$ and different values of $n_D$ ($=0,5,10$, 
the number of orders of derivatives of $\Psi$ that vanish at $t=0$).
Fig. \ref{synthmodomorianalysis2} shows that,
as $n_D$ increases, the power law scaling now holds for time scales larger than $10^{-2}$, so that we can provide a more reliable determination of $p$.

\subsubsection{Piecewise powerlaw scaling}

\begin{figure}
\begin{center}
\includegraphics[width=8cm]{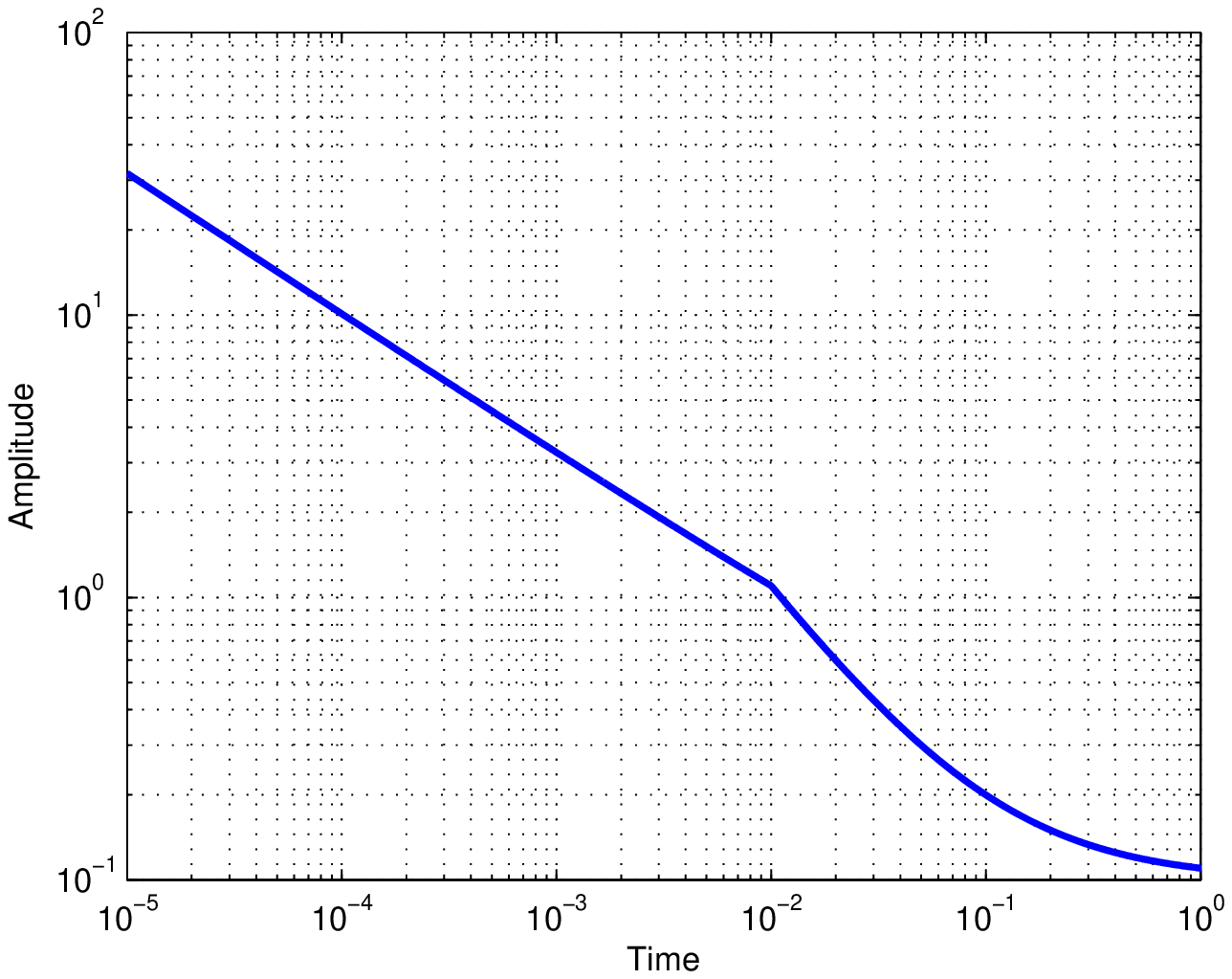}
\caption{\label{synthminpow} Piecewise power law with constant background term as defined by expression (\ref{jhjpefv}).}
\end{center}
\end{figure}

The last synthetic example we consider is the case of a piecewise powerlaw scaling with constant background,
\be
N(t) = min\left[\left(\frac{t}{10^{-2}} \right)^{-0.5} ; \left(\frac{t}{10^{-2}} \right)^{-1}\right] + 0.1~,
\label{jhjpefv}
\ee
which is plotted in Fig. \ref{synthminpow}. This function has a characteristic time scale
of $10^{-2}$.

\begin{figure}
\begin{center}
\includegraphics[width=8cm]{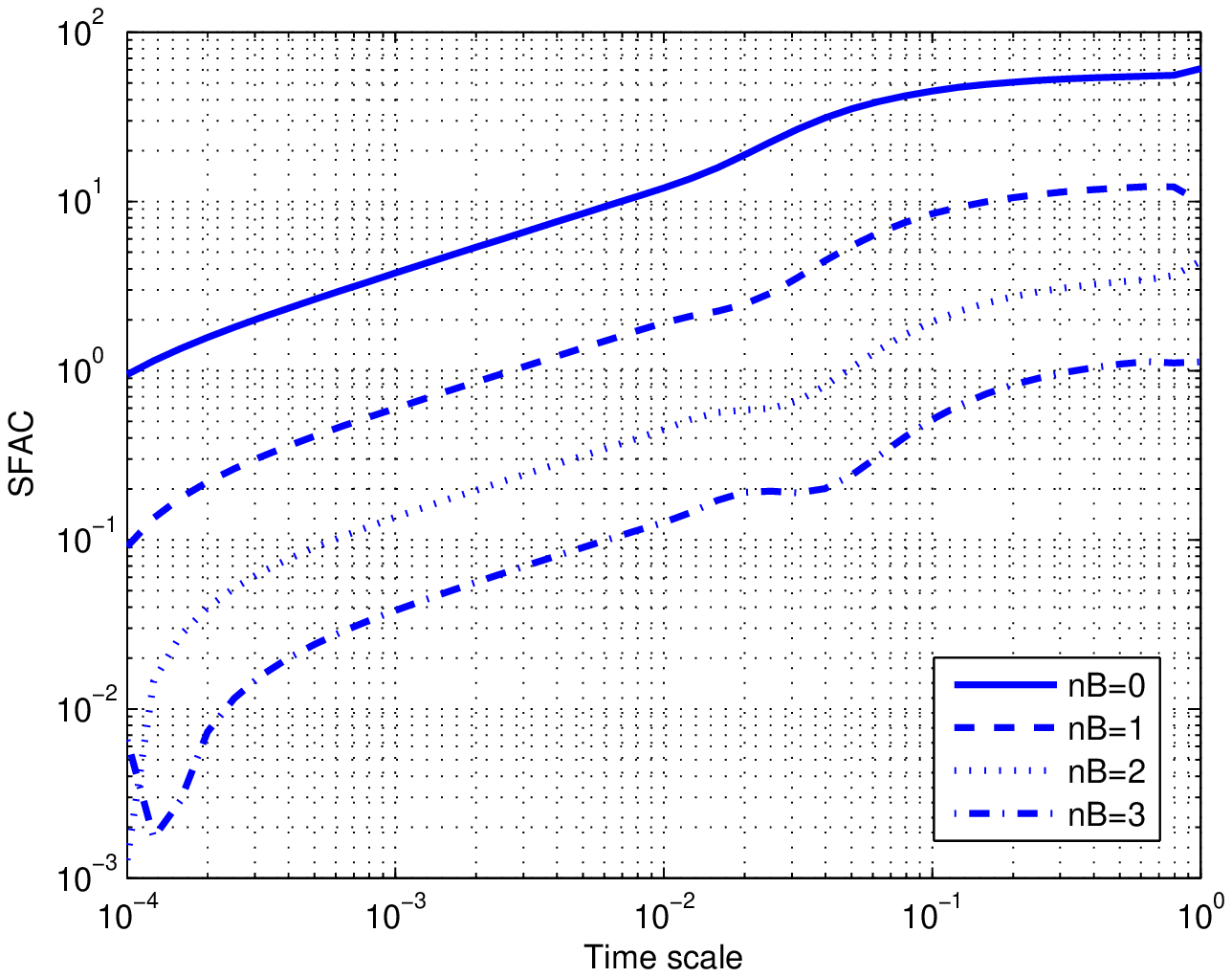}
\caption{\label{synthminpowanalysis} Scaling function analysis of the time series
shown in Fig. \ref{synthminpow}. Each curve corresponds to a given value of $n_B$ used to build the 
corresponding MSF.}
\end{center}
\end{figure}

The result of the scaling function analysis
is plotted in Fig. \ref{synthminpowanalysis}. As the time scale increases,
 two powerlaw scaling regimes are revealed, separated by a smooth step at a time scale of about
$2 \cdot 10^{-2}$, not too far from the built-in characteristic time scale of the process 
defined by expression (\ref{jhjpefv}).
The left part of the curves allows one to infer that the corresponding $p$-value
is close to $0.5$. The second right scaling range is not long enough to determine
the scaling exponent with sufficient accuracy, but it gives however a
rather good description of the change of exponent with scale/time.

\begin{figure}
\begin{center}
\includegraphics[width=8cm]{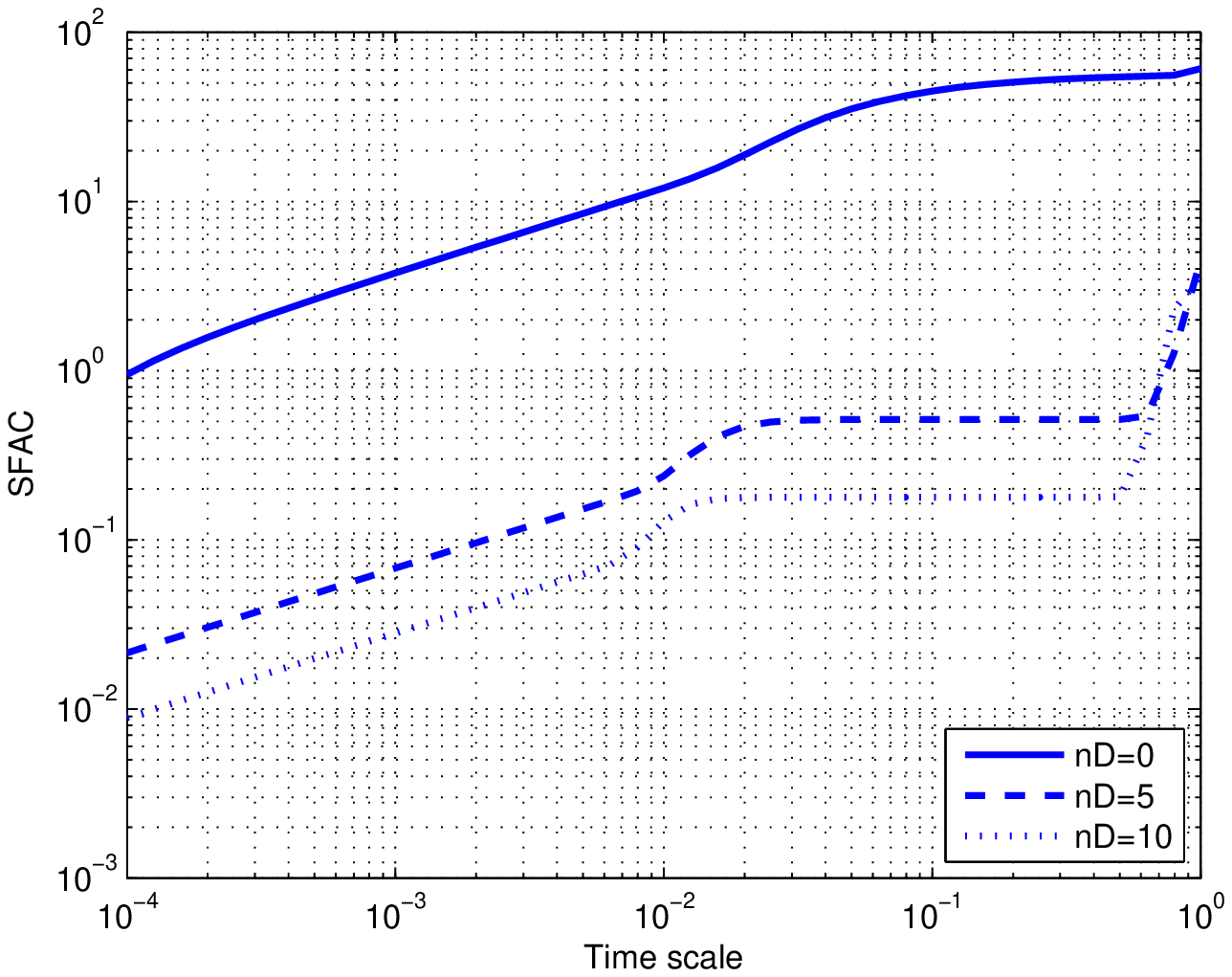}
\caption{\label{synthminpowanalysis2} Scaling function analysis of the time series
shown in Fig. \ref{synthminpow}. Each curve corresponds to a given value of $n_D$ used to build the 
corresponding MSF.}
\end{center}
\end{figure}

Now, setting $n_B=0$ and using non-zero values for $n_D$, one can get a better picture of the complex scaling  of $N(t)$.
Fig. \ref{synthminpowanalysis2} shows that increasing $n_D$ sharpens the transition at time scale $\simeq 10{-2}$, 
and that two different scaling ranges can clearly distinguished, over which the corresponding two values of the exponent $p$ can be determined with high accuracy. 

\clearpage

\clearpage

\clearpage

\clearpage

\clearpage

\clearpage

\clearpage

\begin{thebibliography}{}

\bibitem{Bacry} Bacry, E., J. Muzy, and A. Arneodo, 1993. Singularity spectrum
of fractal signals from wavelet analysis: exact results, 
{\it Journal of Statistical Physics}, {\bf 70} (3/4), 635-674.

\bibitem{Cili} Ciliberto, S., A. Guarino, and R. Scorretti, 2001.
The effect of disorder on the fracture nucleation process,
{\it Physica D}, {\bf 158}, 83-104.

\bibitem{Diete} Dieterich, J., 1994. A constitutive law for rate of earthquake production 
and its application to earthquake clustering,
{\it J. Geophys. Res.},{\bf 99}(B2), 2601-2618.

\bibitem{K1} Kagan, Y.Y., 2003. Accuracy of modern global earthquake catalogs,
{\it Phys. Earth \& Plan. Int.},{\bf 135} (2-3), 173-209.

\bibitem{K2} Kagan, Y.Y., 2004. Short-term properties of earthquake catalogs and models
of earthquake source, {\it Bull. Seism. Soc. Am.},{\bf 94} (4), 1207-1228.

\bibitem{KK1} Kagan, Y.Y., and L. Knopoff, 1981.
	Stochastic synthesis of earthquake catalogs,
	{\it  J. Geophys. Res.}, {\it 86}, 2853-2862.

\bibitem{King} King, G.C.P., R.S. Stein, and J. Lin, 1994.
Static stress changes and the triggering of earthquakes, 
{\it Bull. Seism. Soc. Am.},{\bf 84} (3), 935-953.

\bibitem{Miltenbergeretal93} Miltenberger, P., D. Sornette and C. Vanneste, 1993.
Fault self-organization as optimal random paths selected by critical spatio-temporal dynamics of
earthquakes, {\it Phys.Rev.Lett.},{\bf 71}, 3604-3607.

\bibitem{Ogata} Ogata, Y., 1988. Statistical models for earthquake occurrence
	and residual analysis for point processes, {\it J. Am. stat. Assoc.},
	{\bf  83}, 9-27.

\bibitem{OS} Ouillon, G. and D. Sornette, 2005.
Magnitude-Dependent Omori Law: Theory and Empirical Study,
{\it J. Geophys. Res.},{\bf 110}, B04306, doi:10.1029/2004JB003311.

\bibitem{SS1} Saichev, A. and D. Sornette, 2005.
Andrade, Omori and Time-to-failure Laws from Thermal Noise in Material Rupture,
{\it Phys. Rev. E},{\bf 71}, 016608.

\bibitem{SS2} Saichev, A. and D. Sornette, 2006.
Power law distribution of seismic rates: theory and data,
{\it Eur. Phys. J. B},{\bf 49}, 377-401.

\bibitem{Scholzbook} Scholz, C., 2002. The Mechanics of Earthquakes and Faulting,
2nd Ed., Cambridge University Press, Cambridge.

\bibitem{DS91} Sornette, D., 1991. Self-organized criticality in plate tectonics, in the proceedings of 
the NATO ASI ``Spontaneous formation of space-time structures and criticality,''
Geilo, Norway 2-12 april 1991, edited by T. Riste and D. Sherrington, Dordrecht, Boston,
Kluwer Academic Press (1991), {\bf 349}, 57-106.

\bibitem{SMV94} Sornette, D., P. Miltenberger and C. Vanneste, 1994.
 Statistical physics of fault patterns self-organized by repeated earthquakes, 
{\it Pure and Applied Geophysics},{\bf 142} (3/4), 491-527.

\bibitem{SO} Sornette, D. and G. Ouillon, 2005.
Multifractal Scaling of Thermally-Activated Rupture Processes,
{\it Phys. Rev. Lett.},{\bf 94}, 038501.

\bibitem{Stein}  Stein, R.S., Earthquake conversations, 2003.
{\it Scientific American}, {\bf 288} (1), 72-79.

\bibitem{wells} Wells, D.L., and K.J. Coppersmith, 1994.
New empirical relationships among
magnitude, rupture length, rupture width, rupture area, and surface displacement, 
{\it Bull. Seism. Soc. Am.},{\bf 84}(4), 974-1002.

\end{thebibliography}
\end{document}